\PassOptionsToPackage{svgnames}{xcolor}
\documentclass[twocolumn]{aastex62}
\usepackage{graphicx}
\usepackage[flushleft]{threeparttable}
\usepackage{blindtext}
\usepackage{amsmath}
\usepackage{mathtools}
\usepackage{multirow}
\usepackage{comment}
\usepackage{booktabs}
\usepackage{bm}

\makeatletter 
  \patchcmd{\NAT@citex}
    {\@citea\NAT@hyper@{%
      \NAT@nmfmt{\NAT@nm}%
      \hyper@natlinkbreak{\NAT@aysep\NAT@spacechar}{\@citeb\@extra@b@citeb}%
      \NAT@date}}
    {\@citea\NAT@nmfmt{\NAT@nm}%
    \NAT@aysep\NAT@spacechar\NAT@hyper@{\NAT@date}}{}{}

  \patchcmd{\NAT@citex}
    {\@citea\NAT@hyper@{%
      \NAT@nmfmt{\NAT@nm}%
      \hyper@natlinkbreak{\NAT@spacechar\NAT@@open\if*#1*\else#1\NAT@spacechar\fi}%
        {\@citeb\@extra@b@citeb}%
      \NAT@date}}
    {\@citea\NAT@nmfmt{\NAT@nm}%
    \NAT@spacechar\NAT@@open\if*#1*\else#1\NAT@spacechar\fi\NAT@hyper@{\NAT@date}}
    {}{}
\makeatother

\newcommand{\beq}{\begin{equation}}
\newcommand{\eeq}{\end{equation}}
\newcommand{\bea}{\begin{eqnarray}}
\newcommand{\eea}{\end{eqnarray}}

\begin{document}
\shortauthors{Park et al.}
\def\nar{New Astron.}
\def\na{New Astron.}
\title{\large \textbf{Double-Peaked Ly$\bm{\alpha}$ Emission during Reionization Requires \\ Nearby Voids and a Favorable Local Ionizing Background}}

\correspondingauthor{Hyunbae Park}
\email{hyunbae.park@gmail.com}

\author[0000-0002-7464-7857]{Hyunbae Park}
\affil{Center for Theoretical Physics of the Universe, Institute for Basic Science, Daejeon, 34126, Republic of Korea}
\affil{Center for Computational Sciences, Tsukuba, Ibaraki 305-8577, Japan}
\author[0000-0002-2838-9033]{Aaron Smith}
\affil{Department of Physics, The University of Texas at Dallas, Richardson, TX 75080, USA}
\author[0000-0003-1187-4240]{Intae Jung}
\affil{Department of Astronomy and Space Science, Chungbuk National University, Cheongju, 28644, Republic of Korea}
\author[0000-0002-1319-3433]{Hidenobu Yajima}
\affil{Center for Computational Sciences, Tsukuba, Ibaraki 305-8577, Japan}
\author{Pierre Ocvirk}
\affil{Observatoire Astronomique de Strasbourg, Université de Strasbourg, CNRS UMR 7550, 11 rue de l’Université, 67000 Strasbourg, France}
\author{Joseph S. W. Lewis}
\affil{Institut d’Astrophysique de Paris, UMR 7095, CNRS, UPMC Univ. Paris VI, 98 bis boulevard Arago, 75014 Paris, France}
\affil{Laboratoire AIM, CEA/DSM-CNRS-Université Paris Diderot, IRFU/Service d’Astrophysique, Bâtiment 709, CEA Saclay, F-91191 Gif-sur-Yvette Cedex, France}
\author[0000-0002-6580-7177]{Luke Conaboy}
\affil{School of Physics and Astronomy, The University of Nottingham, University Park, Nottingham NG7 2RD, UK}
\author[0000-0002-0410-3045]{Paul R. Shapiro}
\affil{Department of Astronomy, University of Texas, Austin, TX 78712-1083, USA}
\author[0000-0002-5174-1365]{Ilian T. Iliev}
\affil{Astronomy Centre, Department of Physics \& Astronomy, Pevensey III Building, University of Sussex, Falmer, Brighton, BN1 9QH, United Kingdom}
\author[0000-0003-3974-1239]{Kyungjin Ahn}
\affil{Department of Earth Sciences, Chosun University, Gwangju 61452, Republic of Korea}
\author[0000-0001-8593-8222]{Joohyun Lee}
\affil{Department of Astronomy, University of Texas, Austin, TX 78712-1083, USA}
\author{Jenny G. Sorce}
\affil{Univ. Lille, CNRS, Centrale Lille, UMR 9189 CRIStAL, F-59000 Lille, France}
\affil{Universit\'e Paris-Saclay, CNRS, Institut d'Astrophysique Spatiale, 91405, Orsay, France}
\author{Yohan Dubois}
\affil{Institut d’Astrophysique de Paris, UMR 7095, CNRS, UPMC Univ. Paris VI, 98 bis boulevard Arago, 75014 Paris, France}
\author{Dominique Aubert}
\affil{Observatoire Astronomique de Strasbourg, Université de Strasbourg, CNRS UMR 7550, 11 rue de l’Université, 67000 Strasbourg, France}

\begin{abstract} 
Several Lyman-alpha (Ly$\alpha$) emitters deep into the reionization era exhibit double-peaked Ly$\alpha$ emission profiles, raising the question of how the intergalactic medium can transmit photons blueward of the Ly$\alpha$ resonance at such high redshifts. To investigate this, we compute Ly$\alpha$ transmission along sightlines originating from galaxies in the Cosmic Dawn III simulation and identify cases that closely reproduce the observed double-peaked emission. In these cases, the sightlines intersect highly underdense voids located a few comoving megaparsecs from the source galaxy. These voids allow photons emitted blueward of Ly$\alpha$ to redshift through resonance without scattering while traversing them. The low opacity arises because the neutral hydrogen density scales with the square of the underlying gas density under ionization equilibrium, making sufficiently underdense regions with $\lesssim30~\%$ of cosmic mean density highly transmissive. Such voids naturally occur in the fluctuating cosmic density field, even in the vicinity of galaxies, and can also be associated with transmissive spikes in the Ly$\alpha$ forest. We find that the global probability of observing double-peaked emission is $\sim0.5~\%$ during reionization at an 80\% global ionization fraction, while no cases are found at 60\% ionization. We also find that this probability depends sensitively on the local ionizing background intensity, increasing by a factor of $\sim10^3$ for a tenfold increase in intensity. Recent accretion episodes of supermassive black holes can further increase this probability. These results suggest that double-peaked Ly$\alpha$ emission in high-$z$ galaxies can serve as a sensitive probe of the ionizing background during the late stages of cosmic reionization.\\
\vspace{1em}
\end{abstract}

\section{Introduction}

The first billion years of cosmic history is emerging as a promising laboratory for testing our understanding of astrophysics. During this epoch, the impact of complex baryonic processes on observables is relatively limited compared to later epochs, allowing the Universe to be simulated with high fidelity and compared directly with observations \citep{2001PhR...349..125B,2013fgu..book.....L,2013RPPh...76k2901B}.

One of the most distinctive phenomena of this early Universe is cosmic reionization, during which the intergalactic medium (IGM) is gradually photoionized by radiation from the first luminous sources. While radiation from star-forming galaxies is widely believed to have been the dominant driver of reionization \citep[e.g.,][]{2015ApJ...802L..19R,2018MNRAS.473..227H,2018MNRAS.474.2904P,2019ApJ...879...36F,2020ApJ...892..109N,Yeh2023}, radiation from non-stellar sources such as active galactic nuclei and compact binaries has also been considered to make a non-negligible contribution \citep[e.g.,][]{2010A&A...523A...4B,2013ApJ...764...41F,2013MNRAS.431..621M,2015ApJ...813L...8M,2020MNRAS.498.6083E,2022ARA&A..60..121R,2025A&A...697A.211D}.

Reionization is thought to have begun in the overdensities of early galaxies \citep{2022ApJ...933...87J,2022arXiv221209850J,2023A&A...678A..68S,2026ApJ...997..102M} and progressed outward into underdense regions \citep{2023ApJ...954L..14H}, eventually ionizing most of the neutral hydrogen in the IGM by $z \sim 5.5$ \citep{2015MNRAS.447..499M,2025ApJS..277...37U}. The evolving geometry of the ionization field encodes critical information about the nature of the ionizing sources, including their spatial distribution, spectral properties, and relative contributions across cosmic time \citep[e.g.,][]{2004ApJ...613....1F,2006MNRAS.369.1625I,2012MNRAS.423.2222I,2024MNRAS.531.2943N,2026OJAp....960756N}. Consequently, probing the ionization structure of the IGM is a central goal of modern astronomy.

Lyman-alpha (Ly$\alpha$) emission from star-forming galaxies provides a powerful probe of the IGM ionization state \citep[see e.g.,][for reviews]{1998ApJ...501...15M,2015MNRAS.446..566M,2020ARA&A..58..617O}. Extensive efforts have been devoted to constraining reionization using the damping-wing opacity imprinted redward of the Ly$\alpha$ resonance in galaxy spectra \citep[e.g.,][]{2020ApJ...904..144J,2022ApJ...927...36W,2022ApJ...930..104L,2022MNRAS.517.3263B,2023ApJ...947L..24M,2024ApJ...967...28N,2025ApJ...983...91P,2026A&A...705A.114M}. In this work, however, we focus on the resonant opacity experienced by photons emitted slightly blueward of the Ly$\alpha$ resonance. 

Observations at $z \lesssim 3$, where IGM opacity is negligible, show that a substantial fraction of intrinsic Ly$\alpha$ emission emerges on the blue side of the Ly$\alpha$ resonance \citep[e.g.,][]{2014ApJ...795...33E,2015ApJ...809...89T}. With increasing redshift, the fraction of Ly$\alpha$ emitters (LAEs) exhibiting blue-side emission steadily declines as the IGM progressively attenuates blueward photons, becoming nearly absent above $z \sim 5$ \citep{Stark2011, 2020ApJ...904..144J, Jones2024, Tang2024a, Tang2024b, Kageura2025, 2026PASA...43...21M}. Even though the IGM is mostly ionized after the end of reionization ($z\sim5.5$), trace amounts of residual neutral hydrogen within ionized regions can still produce significant opacity to blue-wing photons as they redshift through the Ly$\alpha$ resonance \citep{2011ApJ...728...52L}. This behavior is closely analogous to the steeply declining transmission observed in the Ly$\alpha$ forest toward higher redshifts \citep[e.g.,][]{2006AJ....132..117F,2015PASA...32...45B,2018ApJ...864...53E,2020ApJ...904...26Y,2022MNRAS.514...55B}.

Interestingly, several unusual cases exhibiting double-peaked Ly$\alpha$ emission profiles have been identified at $z\gtrsim6$. Two such objects have been discovered at $z=6.6$ \citep[COLA1 and NEPLA4;][]{2016ApJ...825L...7H,2018A&A...619A.136M, 2018ApJ...859...91S} and one at $z=6.8$ \citep[A370p\_z1;][]{2021MNRAS.500..558M}. None of these objects show evidence for active galactic nuclei in or around them or existence of highly clustered neighboring galaxies. Although the lack of precise systemic velocity measurements previously left open the possibility that both peaks lay on the red side of the Ly$\alpha$ resonance \citep{2018A&A...619A.136M}, recent observations have revealed that COLA1 and NEPLA4 have systemic redshifts located precisely between their two peaks. This confirms that the blue peak indeed originates from wavelengths shorter than the rest-frame Ly$\alpha$ line \citep{2024A&A...689A..44T,2026arXiv260500763M}.

These high-redshift double-peaked LAEs raise the question of how the IGM during the reionization era could transmit photons emitted on the blue side of the Ly$\alpha$ resonance without unusually strong radiation sources. The HII region surrounding the source galaxy must be sufficiently large ($\sim 0.7$ physical Mpc) for blueward photons to redshift into resonance before encountering neutral gas \citep{2024A&A...689A..44T}. This condition can be readily satisfied in the late stages of reionization, when ionized bubbles overlap to form giant HII regions up to $\sim 10$ physical Mpc, as seen in many previous studies \citep[e.g.,][]{2004ApJ...613....1F,2006MNRAS.369.1625I,2007MNRAS.377.1043M,2007ApJ...671....1T,2016MNRAS.463.1462O}.

However, having a large HII region alone is not sufficient to allow transmission of blue wing photons. During reionization, the IGM is expected to remain highly opaque at the Ly$\alpha$ resonance due to the high mean density of the Universe and the weaker ionizing background radiation. This is supported by declining Ly$\alpha$ forest transmission toward higher redshift from $z\sim 5$, prior to the onset of reionization. 

Notably, \cite{2008MNRAS.391...63I} reported a significant fraction of sightlines in their reionization simulation exhibiting double-peaked Ly$\alpha$ emission. However, their assumed ionizing efficiency of galaxies was much higher ($\gtrsim 10^{-12}~\rm s^{-1}$ at $z \sim 6$) than values typically adopted today ($\sim 10^{-13}~\rm s^{-1}$ at $z \sim 6$), based on recent observations \citep{2002AJ....123.1247F,2011MNRAS.412.2543C,2013MNRAS.436.1023B,2018MNRAS.473..560D,2021MNRAS.508.1853B,2022MNRAS.514...55B,2023MNRAS.525.4093G}. This difference arises from earlier constraints on the electron-scattering optical depth from the Wilkinson Microwave Anisotropy Probe, which allowed for significantly earlier reionization than is currently favored \citep{2007ApJS..170..335P}.

Recently, \cite{2026arXiv260500763M} reported that COLA1 and NEPLA4 host supermassive black holes (SMBHs) with masses of $\sim 2 \times 10^8\,\text{M}_\odot$, inferred from the H$\alpha$ and H$\beta$ line widths observed by the James Webb Space Telescope. They proposed that a recent AGN episode, during which these SMBHs accreted near the Eddington limit less than a million years ago, could have reduced the neutral hydrogen density of the surrounding IGM. Because the IGM would not yet have returned to ionization equilibrium, this scenario could enable the transmission of blue-wing photons.

In this work, we analyze the state-of-the-art reionization simulation Cosmic Dawn III (CoDaIII) to search for analogs of the observed double-peaked LAEs. We note that \citet{2021MNRAS.508.3697G} previously investigated this question using the predecessor simulation, Cosmic Dawn II. They showed that $\sim 1\%$ of sightlines exhibit significant transmission on the blue side of Ly$\alpha$, but did not identify any sightlines capable of reproducing the observed double-peaked profiles from a typical intrinsic emission line of a galaxy. By improving the Ly$\alpha$ transmission calculation by more accurately estimating the halo peculiar velocities and substantially increasing the number of sampled sightlines, we now identify a significant population of sightlines exhibiting blue-side transmission that resembles the observed double-peaked LAEs. Building on these results, we examine the physical conditions required for such transmission and discuss the implications for current and future observations.

The rest of the paper is as follows. In Section~\ref{sec:method}, we describe the simulation used in this study and how we find double-peaked LAE cases from the data. In Section~\ref{sec:results}, we present the main results of this paper. In Section~\ref{sec:summary}, we summarize our results with discussions. Throughout this paper, distances are given in comoving units and are expressed in ckpc or cMpc unless stated otherwise.

\section{Methods} \label{sec:method}

\subsection{Cosmic Dawn III simulation}

\begin{figure*}
  \begin{center}
    \includegraphics[scale=0.745]{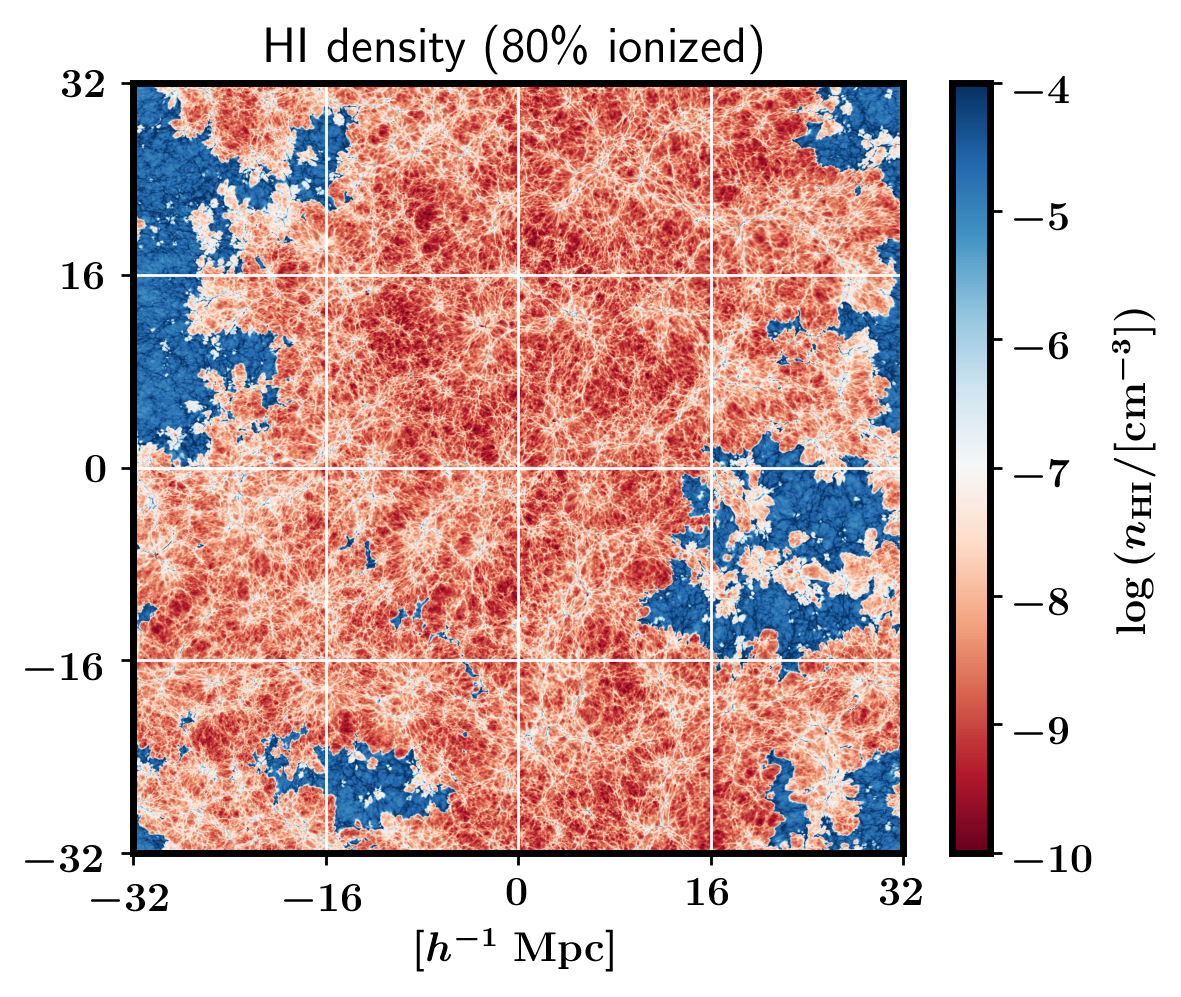}
    \includegraphics[scale=0.745]{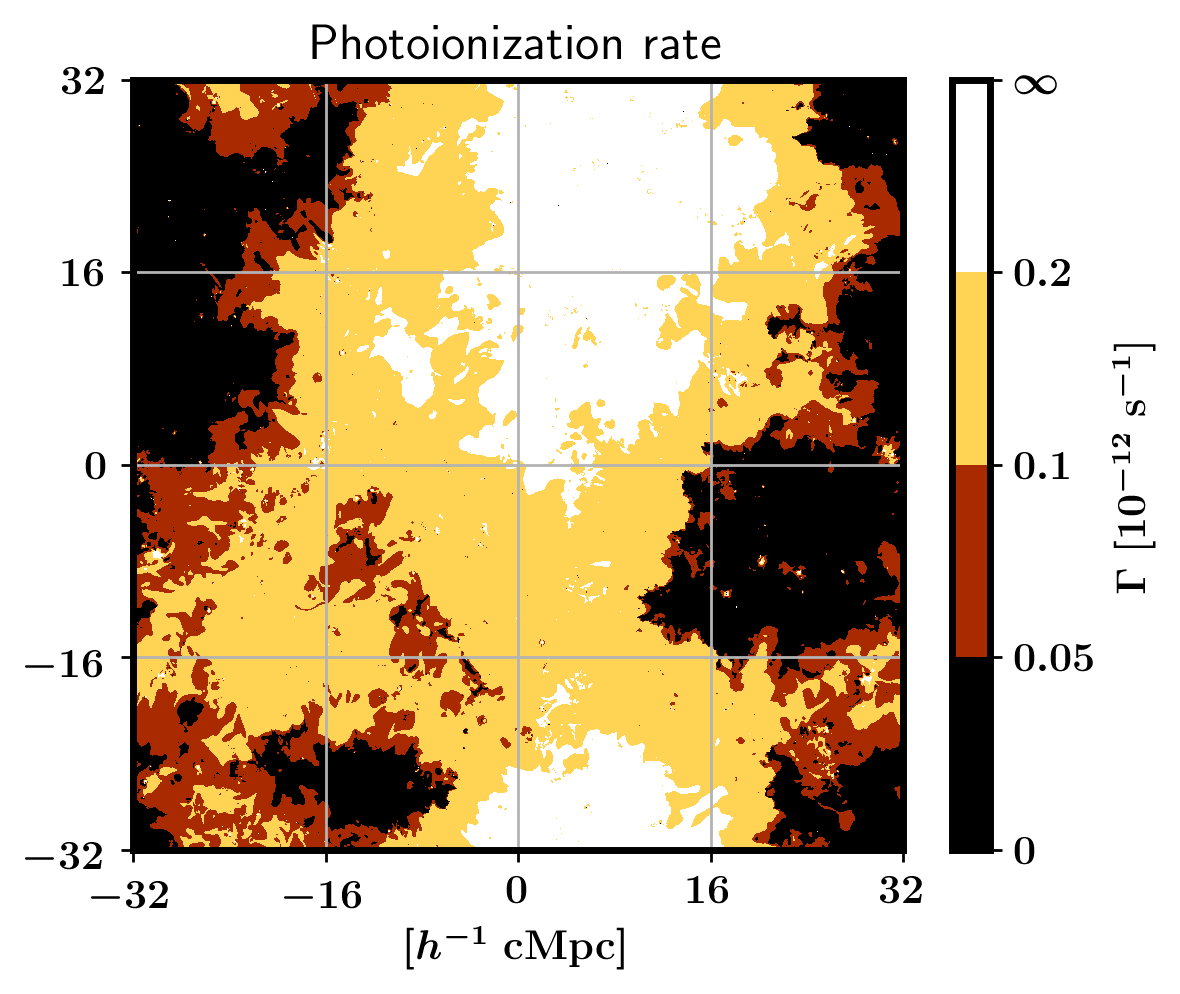}
  \caption{Maps of neutral hydrogen density (left) and photoionization rate due to ionizing background radiation (right) on a square slice across the entire simulation box ($64~h^{-1}~{\rm cMpc}$) at $z=6$. The global volume-averaged ionization fraction is 80\% in the snapshot, resulting in significant percolation of ionized bubbles into a single continuous structure. The intensity in the right panel declines abruptly on $\sim \rm cMpc$ scales near the edges of the ionized bubble, toward the surrounding neutral regions. The map is centred on one of the galaxies described in Section~\ref{sec:void}. }
   \label{fig:LSmap}
  \end{center}
\end{figure*}

A reliable calculation of Ly$\alpha$ opacity in the IGM requires accurate modeling of the relevant physical quantities with a spatial resolution of $\lesssim 20~{\rm ckpc}$ over a cosmological volume \citep{2015MNRAS.446.3697L}. The Cosmic Dawn III (CoDaIII) simulation provides a suitable mock Universe for the reionization era, in a cubic volume of size $L_{\rm box}=64~h^{-1}~{\rm cMpc}\approx94.4~{\rm cMpc}$ on a uniform grid of $8192^3$ cells, with an equal number of dark matter particles of mass $5.09\times10^4\,\text{M}_\odot$. This setup resolves baryonic structures with a uniform spatial resolution of $7.8~h^{-1}~{\rm ckpc}$ across the entire simulation volume. We use the baryon density, baryon velocity, hydrogen ionization fraction, temperature, and photoionization rate fields from the simulation in our analysis.

CoDaIII self-consistently models structure formation and radiative transfer using fully coupled radiation hydrodynamics implemented in the \textsc{RAMSES-CUDATON} code \citep{2016MNRAS.463.1462O}. \textsc{RAMSES-CUDATON} employs graphical processing units to track ionization fronts at the full speed of light, thereby avoiding reduced-speed-of-light artefacts \citep{2019A&A...626A..77O, 2019A&A...622A.142D}. Subgrid prescriptions are used to convert gas into stars, represented by collisionless star particles of mass $1.17\times10^4\,\text{M}_\odot$, and to estimate the ionizing radiation produced by these stars. The star-formation prescription is detailed in Section 2.2.1 of \cite{2020MNRAS.496.4087O} describing the predecessor simulation Cosmic Dawn II (CoDaII), and the modifications made in CoDaIII is described in Section 2.1.2 of \citet{2022MNRAS.516.3389L}.

These prescriptions are calibrated to reproduce several observables of the Epoch of Reionization: the observed galaxy ultraviolet luminosity function at $z\sim6$, the evolution of the neutral fraction of the IGM and the ionization rate, the Thomson scattering optical depth measured in the CMB. The simulation also produces an ionizing photon mean free path evolution in remarkable agreement with observations \citet{2022MNRAS.516.3389L}. Cosmic Dawn III’s dust model is also calibrated against REBELS and ALPINE dust masses \citep{2023MNRAS.519.5987L}, and the impact of dust on ultraviolet luminosity function and galaxy morphology is studied in \citet{2025A&A...703A..98O}.

Compared to its predecessor CoDaII, CoDaIII improves on spatial and mass resolution by a factor 2 and 8, respectively.  Moreover, a temperature criterion has been added to the star-formation prescription, which results in a greater sensitivity of low-mass galaxies to radiative feedback. This in turn moderates the growth of the ionizing background intensity and brings the simulation into agreement with observational constraints \citep{2021MNRAS.507.6108O,2022MNRAS.516.3389L}. 
Reionization in CoDaIII ends at $z=5.5$, instead of $z=6.2$ in CoDaII, providing better agreement with recent reionization constraints \citep{2026MNRAS.545f1862D}. We present double-peaked LAE cases identified in the $z = 6$ snapshot of CoDaIII and our corresponding analysis. In CoDaIII, $\approx80\%$ of the simulation volume is ionized at this redshift, marking a late stage of reionization. 

From the snapshot data, the HI density is computed using the gas density ($\rho$) and ionization mesh ($\chi$) from the simulation as $n_{\rm HI}=X\rho[1-\chi]/m_{\rm p}$, where $X=0.76$ is the hydrogen mass fraction in the Universe and $m_{\rm p}=1.67\times10^{-24}~\rm g$ is the proton mass. The CoDa simulations do not account for ionization of helium, which is typically expected to be singly ionized when hydrogen is ionized during the reionization era. To correct for the underestimated recombination rate due to the missing free electrons from helium, we increase $n_{\rm HI}$ by a factor of $0.82/0.76\approx1.08$ in highly ionized regions ($\chi>0.99$).

We visualize the HI density in a slice of this snapshot in the left panel of Figure~\ref{fig:LSmap}. The figure shows that ionized bubbles are merging into a single giant HII region (shown mostly red), leaving neutral regions (shown mostly blue) isolated. At this stage, most of the volume lies several cMpc away from the nearest neutral regions, and the damping-wing opacity therefore has limited contribution to opacity in most locations.

CoDaIII provides mesh data for the photoionization rate due to the ionizing background ($\Gamma$), as a natural outcome of solving the ionizing radiative transfer throughout the simulation. We visualize $\Gamma$ in the right panel of Figure~\ref{fig:LSmap} for the same slice shown in the left panel. The figure shows that $\Gamma$ varies smoothly over large ($\gtrsim \rm cMpc$) scales, consistent with results from other recent simulations \citep[e.g.,][]{2016MNRAS.460.1328D,2018MNRAS.473..560D}, before declining abruptly near neutral regions. A comparison of $n_{\rm HI}$ and $\Gamma$ shows that the typical value of $\Gamma$ lies in the range $(1$–$2)\times10^{-13}~\mathrm{s^{-1}}$ within ionized regions. $\Gamma$ exceeds $2\times10^{-13}~\mathrm{s^{-1}}$ in the central parts of the ionized regions, where a larger number of galaxies contribute to $\Gamma$, and falls below $1\times10^{-13}~\mathrm{s^{-1}}$ near neutral regions, where fewer galaxies contribute.

We have also analyzed the $z = 6.5$ snapshot, in which $\approx 60\%$ of the simulation volume is ionized. This redshift is closer to that of the highest-redshift observed double-peaked LAEs; however, we find no sightlines exhibiting double-peaked emission. This may indicate that such systems arise in extremely rare regions that are not statistically captured within the CoDaIII simulation volume, that the reionization history adopted in CoDaIII is delayed relative to reality, or that the simulated ionization topology differs substantially from that of the real Universe, which could also result in significantly different Ly$\alpha$ opacity \citep{2007MNRAS.377.1043M}. Nevertheless, we believe that our results at $z = 6$ provide useful insight into the observed double-peaked LAEs at $z \sim 6.5$.

Our galaxy identification method is described in detail in \cite{2025A&A...703A..98O}. We identify galaxies using a Friends-of-Friends (FoF) algorithm with a linking length of $0.15$ times the mean inter-particle separation. This choice yields a FoF mass function that better reproduces the halo mass function defined by a spherical overdensity of 200 at $z\sim6$ than the conventional linking length of $0.2$ \citep{2013MNRAS.433.1230W}. 

The intrinsic UV magnitude $M_{\rm UV,int}$ is first computed from the ages and metallicities of halo star particles using the BPASS V2.2.1 \citep{2022ARA&A..60..455E} stellar population models. We then apply a dust correction to galaxies with $M_{\rm UV,int}<-21$ by taking $M_{\rm UV}=M_{\rm UV,int}-(M_{\rm UV,int}+21)/2.5$, bringing the resulting UV luminosity function into agreement with observations. This prescription leaves galaxies with $M_{\rm UV,int}\ge-21$ unchanged and makes brighter galaxies ($M_{\rm UV,int}<-21$) fainter, with the correction increasing toward higher intrinsic UV luminosity. The prescription is based on the empirical calibration of dust attenuation in CoDaIII galaxies by \cite{2025A&A...703A..98O}, which was chosen so that the resulting bright end of the UV luminosity function agrees with observations. We use 938 identified galaxies with $M_{\rm UV}<-19$ in this snapshot for our analysis. The galaxies are numbered in decreasing order of UV brightness. For example, galaxy \#337 corresponds to the $337^\text{th}$ UV-brightest galaxy in our sample.

\subsection{Ly$\alpha$ Emission Line Calculation} 

While in real observations a galaxy can be viewed from only a single line of sight, simulated universes allow us to calculate emission along numerous hypothetical sightlines, giving a large statistical sample of mock observations. To identify sightlines that exhibit double-peaked emission profiles, we first compute the Ly$\alpha$ opacity along multiple directions and then apply the resulting transmission to an assumed intrinsic line profile to obtain the transmitted spectrum. We then apply a set of criteria to classify whether each spectrum is double-peaked.

We generally follow the procedure described in \cite{2021ApJ...922..263P} to calculate the IGM opacity. The optical depth, $\tau_\alpha$, is obtained by integrating the product of the neutral hydrogen number density, $n_{\rm HI}$, and the Ly$\alpha$ absorption cross section, $\sigma_\alpha$, along each sightline. For photons emitted by the source at an initial frequency $\nu_e$, the opacity is given by
\bea \label{eq:Tr}
\tau_\alpha(\nu_e) = \int_{r_{\rm min}}^{r_{\rm max}} n_{\rm HI}(\boldsymbol{r}) \sigma_\alpha \left(T(\boldsymbol{r}), \nu(\boldsymbol{r})\right) a(z) \, \text{d}r \, ,
\eea
where $\boldsymbol{r} = r\hat{\boldsymbol{n}}$ denotes a comoving position along a line-of-sight (LOS) direction $\hat{\boldsymbol{n}}$ and $a(z)$ is the scale factor to convert the comoving distance to a physical one. The cross section $\sigma_\alpha$ follows a Voigt profile as a function of photon frequency $\nu$, whose Gaussian core is determined by the local gas temperature $T$. Here, $\nu$ is measured in the rest frame of the IGM along the LOS and is given by
\bea
\nu(\boldsymbol{r}) = \nu_e - \nu_\alpha \frac{r a(z)H(z) + \hat{\boldsymbol{n}} \cdot \boldsymbol{v}_{\rm pe}(\boldsymbol{r})}{c},
\eea
where $\nu_\alpha=2.46607\times10^{15}~{\rm Hz}$ is the frequency of a Ly$\alpha$ photon, and $c$ is the speed of light.
The above equation accounts for the cosmological redshift through the Hubble expansion rate, $H(z)$, and the Doppler shift induced by the peculiar velocity of the IGM, $\boldsymbol{v}_{\rm pe}$. We evaluate the integral from the virial radius of the galaxy ($r_{\rm min} = r_{\rm vir}$) to half the box size ($r_{\rm max} = L_{\rm box}/2 = 32~h^{-1}~{\rm cMpc}$), thereby capturing the IGM opacity while excluding contributions from the interstellar medium (ISM) and circumgalactic medium (CGM). Damping-wing opacity from neutral regions beyond $r_{\rm max}$ is expected to contribute minimally at the late stage of reionization considered in this work \citep{2022MNRAS.512.3243S}. We compute $\tau_\alpha$ by the IGM for 360 sightlines separated by $1^\circ$ on the $xy$-plane for each galaxy, yielding in total of $360\times938 =  337,680$ Ly$\alpha$ transmission for sightlines\footnote{Ideally, the sightlines need to be distributed uniformly throughout the full 3D volume to maximize the statistical sampling of the data. However, for computational convenience, we restrict the sightlines to a 2D plane in this work. While our current sample size is sufficient to support the key conclusions of this study, we plan to improve this aspect in future work.}. 

The opacity in the ISM and CGM is generally very high ($\gg 1$), and Ly$\alpha$ photons typically undergo numerous scatterings with complex trajectories before escaping into the IGM \citep{2014PASA...31...40D, 2021MNRAS.504.1902G, 2022MNRAS.512.3243S}. These photons would still constitute emission from the target galaxies unless they are absorbed by dust because they have not traveled far away from the source galaxy. In the case of photons scattered in the IGM, dust absorption is highly unlikely; however, the photons have traveled far enough from the galaxy that change in their propagation direction would likely remove them from the direct sightline toward the galaxy, causing them to appear as diffuse emission outside the galaxies.

In this work, we focus on how the IGM opacity reshapes the emission that escapes the ISM and CGM by scattering photons off the sightline. We define the intrinsic emission as the radiation escaping the galaxy at the virial radius after being processed by the ISM and CGM. For the intrinsic spectral shape, we assume a Gaussian profile with a fixed standard deviation of $\sigma=200~{\rm km~s^{-1}}$ for all galaxies in the simulation:
\bea \label{eq:taualpha}
f_{\rm int} = \exp\left(-\frac{v_\alpha^2}{2\sigma^2}\right),
\eea
where $v_\alpha \equiv c(\lambda - \lambda_\alpha)/\lambda_\alpha$ is the wavelength offset relative to the Ly$\alpha$ wavelength $\lambda_\alpha = 1215.67~\text{\AA}$, expressed in units of velocity.
Then, the transmitted profile is given by
\bea
f_{\rm obs}(v_\alpha) = f_{\rm int}(v_\alpha) e^{-\tau_\alpha(v_\alpha)} \, .
\eea
The normalization of this profile is arbitrary, as the double-peak classification criteria applied in the subsequent step are independent of its intrinsic amplitude. 

The assumed profile is admittedly highly simplified. Observations show that intrinsic Ly$\alpha$ emission profiles can be double-peaked or exhibit more complex structures \citep{2018ApJ...855...96H,2023MNRAS.523.3749B}, and that the spectral extent of the emission correlates with the UV luminosity of the source galaxy \citep{2022MNRAS.517.5642E}. However, our current understanding of the connection between galactic properties and Ly$\alpha$ line formation remains insufficient to generate realistic intrinsic emission profiles directly from large-volume simulations with limited resolutions at ISM scales such as CoDaIII. Moreover, allowing the intrinsic profile to vary with galaxy properties would complicate the interpretation of our results, which will focus on the IGM physics. We therefore adopt this simplified prescription in the present analysis \citep[See e.g.,][for studies exploring alternative intrinsic emission profiles.]{2013MNRAS.428.1366J,2019MNRAS.485.1350W,2020A&A...642L..16B,2023MNRAS.521.4356X,2026OJAp....960756N}. Nevertheless, our choice of $\sigma = 200~{\rm km~s^{-1}}$ yields a Ly$\alpha$ emission profile whose wavelength extent is roughly comparable to that of green pea galaxies in the local Universe \citep{2016ApJ...820..130Y}.

We classify a sightline as exhibiting double-peaked emission based on the intrinsic and observed blue- and red-side fluxes, defined as
\bea
F_{\rm blue,obs} &\equiv \int_{-\infty}^{0} f_{\rm obs}(v_\alpha)\, dv_\alpha \, , \nonumber\\
F_{\rm red,obs}  &\equiv \int_{0}^{\infty} f_{\rm obs}(v_\alpha)\, dv_\alpha \, , \nonumber\\
F_{\rm red,int}  &\equiv \int_{0}^{\infty} f_{\rm int}(v_\alpha)\, dv_\alpha \, . \nonumber
\eea
We consider a sightline as showing double-peaked emission if the following conditions are satisfied:
\bea
F_{\rm red,obs}  &> 0.05\, F_{\rm red,int} \, ,  \nonumber \\
F_{\rm blue,obs} &> 0.1\, F_{\rm red,obs} \, .
\eea
The first criterion requires that at least 5\% of the intrinsic red-side flux be transmitted, while the second requires that the blue-side flux exceed 10\% of the red-side flux. 

We note that above criteria can falsely classify a single red peak with a broad blueward extension as a double peak in the case of single broad-peak intrinsic emission profile assumed here. To exclude such cases, we impose an additional requirement that the flux between the resonance and the red peak falls below 50\% of the peak flux:
\bea
\max_{0 < v_\alpha < v_{\rm peak,red}} f_{\rm obs}(v_\alpha)
\;<\; 0.5\, f_{\rm obs}(v_{\rm peak,red}),
\eea
where $v_{\rm peak,red}$ is the velocity offset of the red peak in the emergent emission. Visual inspection of the emergent profiles confirms that these criteria correctly identify genuine double-peaked Ly$\alpha$ emission.

\section{Results} \label{sec:results}

\begin{figure*}
  \begin{center}
    \includegraphics[scale=0.83]{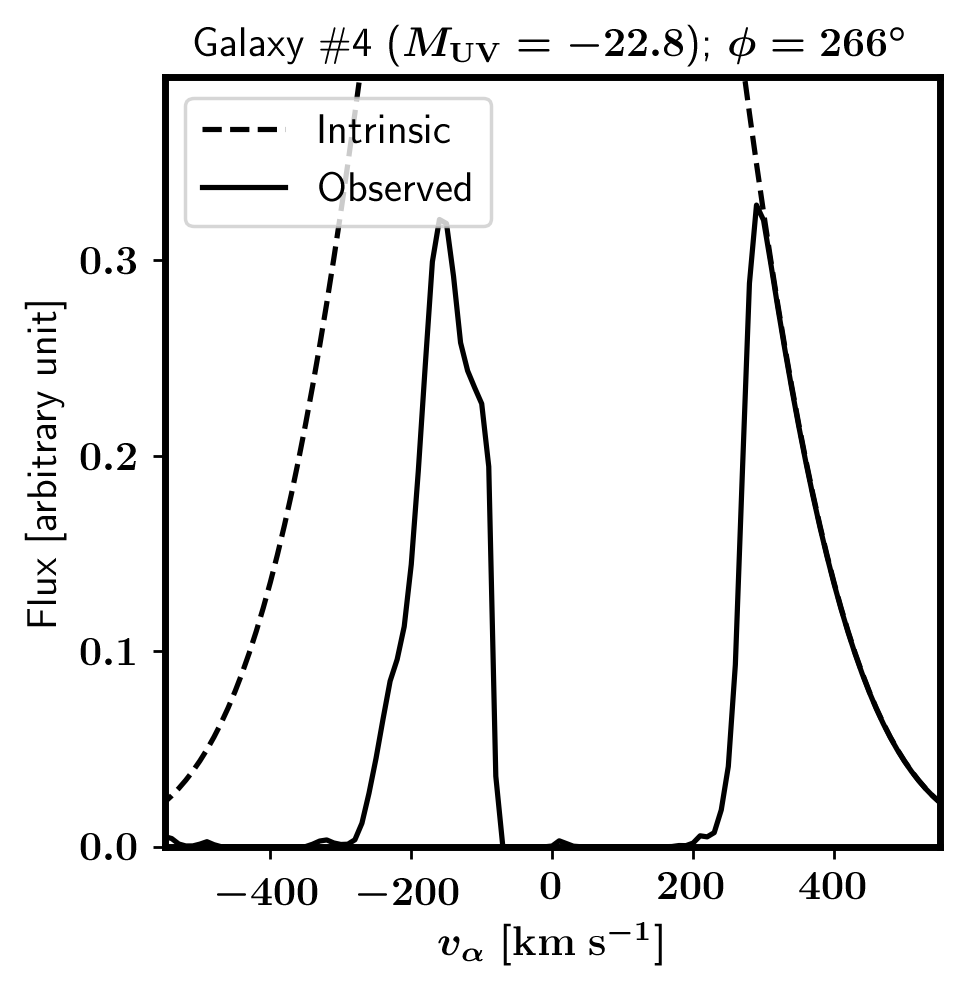}
    \includegraphics[scale=0.83]{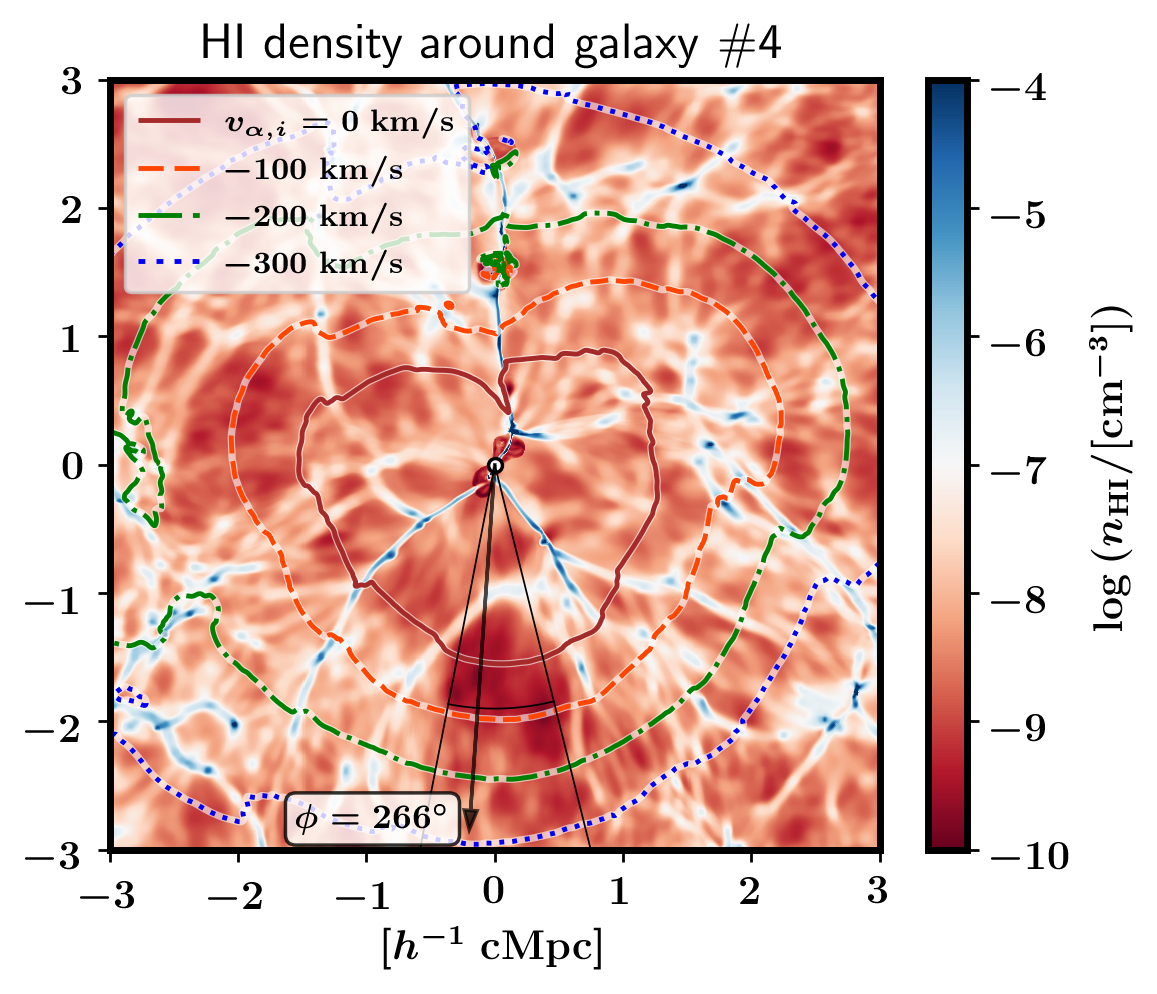}
  \caption{Left: Intrinsic ($F_{\rm in}$; dashed line) and observed Ly$\alpha$ emission profiles ($F_{\rm obs}$; solid line) as functions of $v_\alpha$ for one of the sightlines from galaxy \#4, with $M_{\rm UV} = -22.1$. Right: HI density map of a square slice centred on the source (circular symbol), illustrating a COLA1-like double-peaked Ly$\alpha$ emission profile corresponding to the left panel. Pairs of thin radial lines connected by arcs indicate the range of directions along which the emission appears double-peaked. The thick arrow extending from the centre marks the sightline with $\phi = 266^\circ$, whose emission profile is shown in the left panel. The solid, dashed, dot-dashed, and dotted contours mark the location where photons emitted at $v_\alpha=0,~-100,~-200,$ and $~-300~\rm km~s^{-1}$ redshift to the Ly$\alpha$ resonance according to Eq.~(\ref{eq:v}), respectively.}
   \label{fig:DP1}
  \end{center}
\end{figure*}

\begin{figure*}
  \begin{center}
    \includegraphics[scale=0.83]{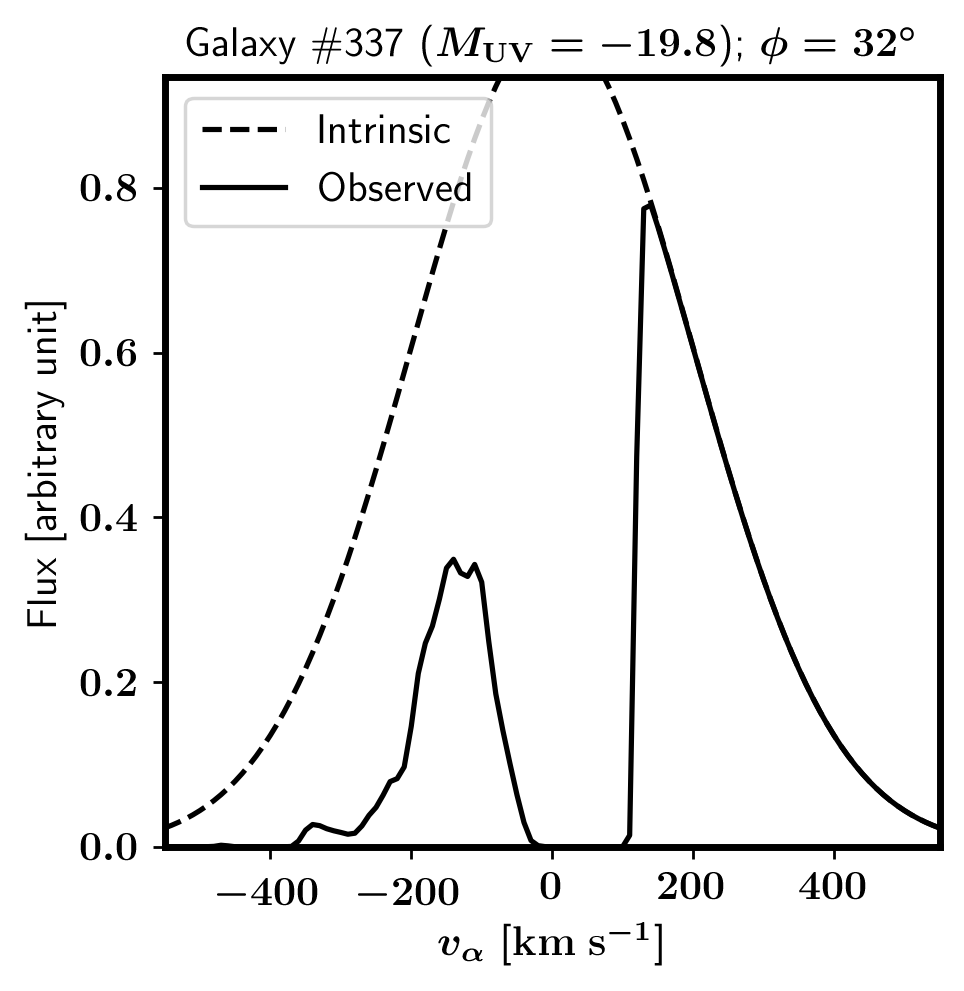}
    \includegraphics[scale=0.83]{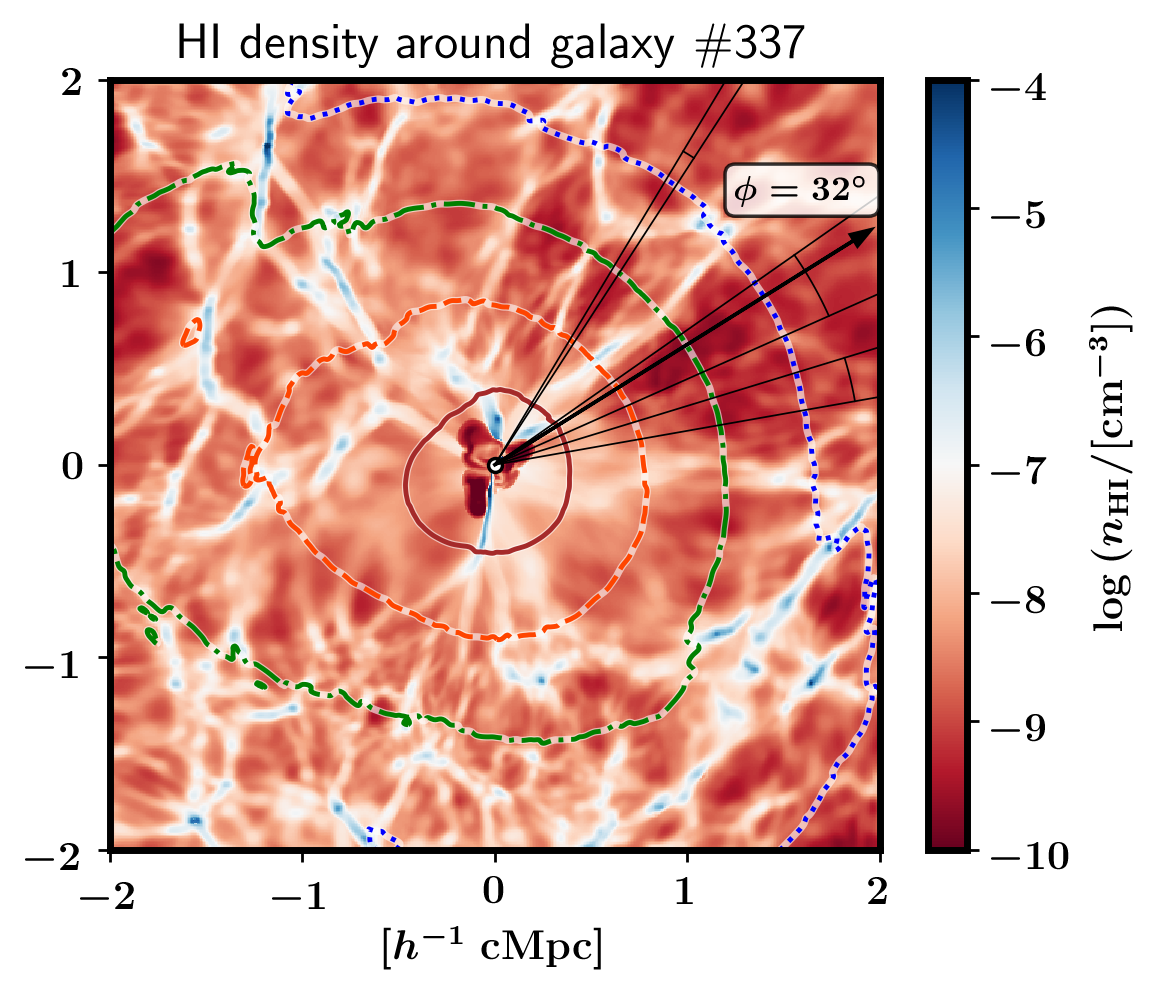}
  \caption{Same as Figure 2, but for another sightline from galaxy \#337, with $M_{\rm UV} = -19.8$ and $\phi = 32^\circ$, showing a double-peaked Ly$\alpha$ emission profile.}
   \label{fig:DP2}
  \end{center}
\end{figure*}

We find that 1858 out of 337,680 sightlines exhibit double-peaked emission, corresponding to a fraction of 0.55\%. This low occurrence rate is consistent with observations that double-peaked LAEs are highly rare cases. We also find that the number of double-peaked cases depends sensitively on the adopted selection criteria. For example, tightening the requirement of
$F_{\rm blue,obs} > 0.1~F_{\rm red,obs}$ to
$F_{\rm blue,obs} > 0.2~F_{\rm red,obs}$
decreases the number of double-peaked cases to 798 or 0.24\%. Consequently, we do not place strong emphasis on the absolute value of the probability from our calculation. We shall instead focus on how it depends on other environmental factors.

Figures~\ref{fig:DP1} and \ref{fig:DP2} show two examples among the double-peaked Ly$\alpha$ emission cases found in our calculation along with the HI density maps of the surrounding region. The case illustrated in Figure~\ref{fig:DP1} is associated with a bright galaxy with $M_{\rm UV}=-22.1$ similar to COLA1. In this case, the blue and red peaks are separated by nearly $400~\rm km~s^{-1}$. The case illustrated in Figure~\ref{fig:DP2} exhibits double peaks with smaller separation ($\sim200~\rm km~s^{-1}$) more similar to the double-peaked profile of COLA1. Instead, this sightline is associated with a much fainter galaxy ($M_{\rm UV}=-19.8$) than COLA1. 

We note that the blue side of the red peak appears sharper than in the observations. As for observed line profiles, the line-spread function of spectrographs introduces spectral smoothing on scales of $\sim 100~\rm km~s^{-1}$, broadening the observed profile. As for the theoretical line profiles calculated here, the Ly$\alpha$ opacity is computed using an integration that begins abruptly at $r = r_{\rm vir}$ (Eq.~(\ref{eq:Tr})), causing photons blueward of the red peak to be sharply attenuated by resonant opacity. In reality, some photons scattered near $r = r_{\rm vir}$ are not completely removed and may still contribute to the observed line flux after undergoing additional scatterings within $r_{\rm vir}$, resulting in a less sharply defined red peak than predicted here. However, we expect that these issues have negligible impact on the integrated flux, $F_{\rm red,obs}$.

The cases shown in Figures~\ref{fig:DP1} and \ref{fig:DP2} are representative of the double-peaked profiles in our sample and share a common physical origin. We therefore focus on these examples throughout this paper. In Section~\ref{sec:void}, we describe how underdense regions surrounding the source enable blue-side transmission. In Section~\ref{sec:IBR}, we examine the relationship between the local $\Gamma$ and the occurrence of the blue peak. Finally, in Section~\ref{sec:MUV}, we show that the blue-peak occurrence exhibits little dependence on either the UV magnitude or the local density.

\subsection{Underdense Void in the Vicinity of the Source} \label{sec:void}

\begin{figure*}
  \begin{center}
    \includegraphics[scale=1.1]{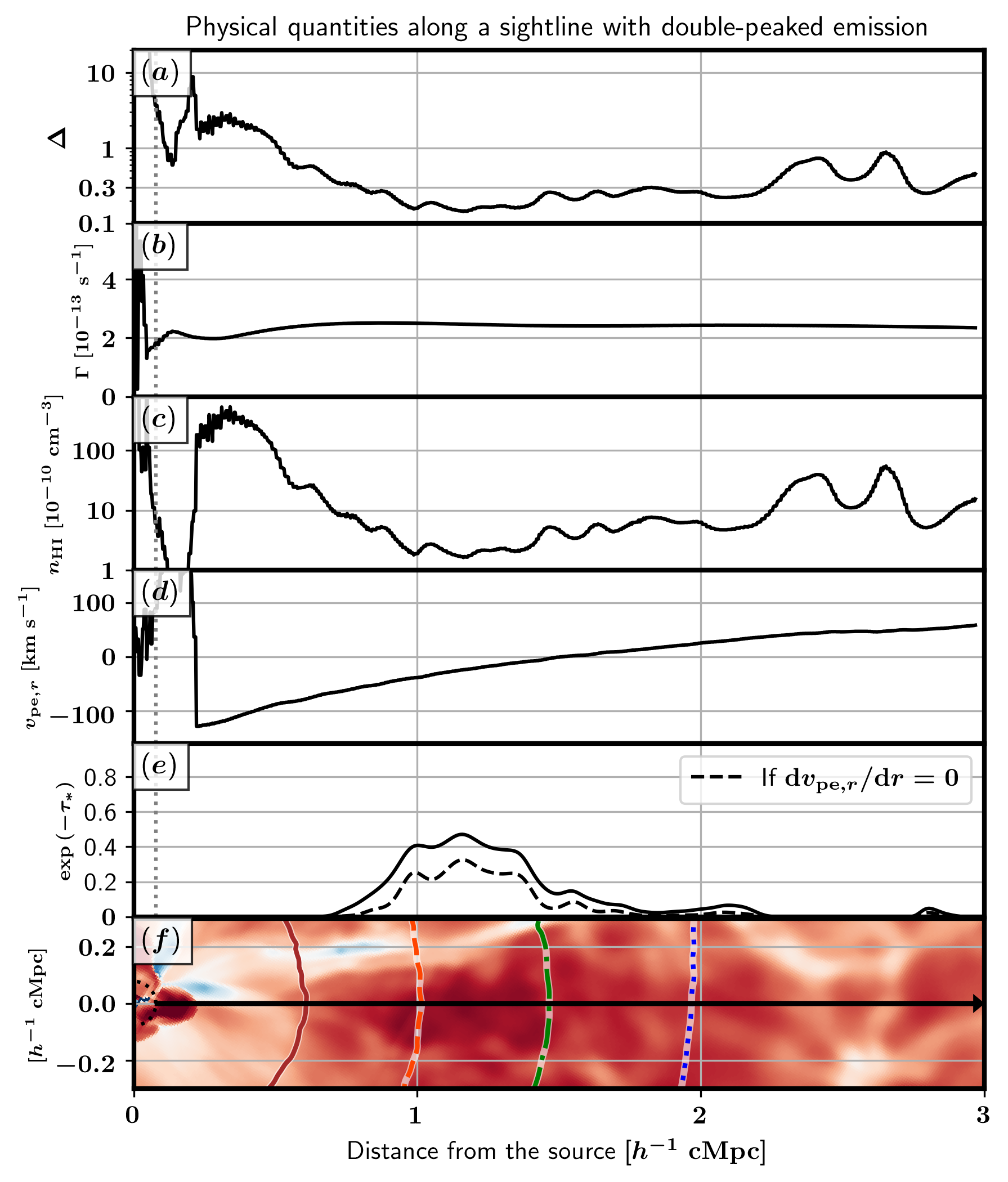}
  \caption{Physical quantities of the IGM along the sightline exhibiting double-peaked emission in Figure~\ref{fig:DP2}, shown as a function of distance from the centre of the source halo. The black dotted line marks $r_{200}$ of the halo. Panel $a$: gas density normalized by its cosmic mean. Panel $b$: Photoionization rate in the unit of $10^{-13}~{\rm s}^{-1}$. Panel $c$: HI density in the unit of $10^{-10}~{\rm cm}^{-3}$. Panel $d$: radial peculiar velocity in the unit of $\rm km~s^{-1}$. Panel $e$: transmission of photons entering the Ly$\alpha$ resonance from Eq.~(\ref{eq:tau}). The dashed line shows the result when the radial peculiar velocity gradient is ignored (i.e., $\text{d}v_{{\rm pe},r}/\text{d}r=0$). Panel $f$: HI density map, with the same color scheme and line contours as in the right panel of Figure~\ref{fig:DP2}. The black arrow marks the sightline of interest.}
   \label{fig:DP2_LOS}
  \end{center}
\end{figure*}

In the right panels of Figures~\ref{fig:DP1} and \ref{fig:DP2}, the blue shading extending outward from the galaxy marks the directions along which the Ly$\alpha$ emission appears double-peaked to hypothetical observers. In both cases, the blue-shaded directions intersect regions of exceptionally low HI density ($n_{\rm HI}\lesssim 10^{-9}~{\rm cm^{-3}}$), shown in dark red. These highly underdense voids are the primary cause of the double-peaked profiles, which we examine in detail in this section. In Figure~\ref{fig:DP2_LOS}, we show various physical quantities along the sample sightline from Figure~\ref{fig:DP2} for detailed investigation.

In the typical IGM with $T=10^4~\rm K$ at $z=6$, photons traverse the thermally broadened Ly$\alpha$ resonance line within approximately the Sobolev length,
\bea
l_s=v_{\rm th}/H(z)=0.13~h^{-1}~\rm cMpc \, .
\eea
Because $l_s$ is much shorter than a cMpc, the resonance opacity can be estimated from the HI density at the location where photons redshift into resonance on large scales in the IGM \citep{2025MNRAS.544.4246P}. Equation~(A6) of \cite{2025MNRAS.544.4246P} gives the opacity as
\bea\label{eq:tau}
&&\tau_* (\mathbf{r}) \approx \\
&&\left[\frac{n_{\rm HI}(\mathbf{r})}{0.75 \times 10^{-10}~{\rm cm^{-3}}}\right]
\left[\frac{H(z)+[1+z]\text{d}v_{{\rm pe}, r}/\text{d}r}{3\times10^2~\rm km~s^{-1}~pMpc^{-1}} \right]^{-1} \, , \nonumber
\eea
where $\mathbf{r}$ denotes the location relative to the Ly$\alpha$ source. Here $v_{{\rm pe}, r}\equiv \hat{\mathbf{n}}\cdot\mathbf{v}_{\rm pe}(\mathbf{r})$ is the radial peculiar velocity along the sightline direction $\hat{\mathbf{n}}$, and pMpc denotes the physical megaparsec. Using $n_{\rm HI}$ and $v_{{\rm pe},r}$ along the sightline, gaussian-smoothed over $l_s$, we identify transmissive locations (panel~$e$ of Figure~\ref{fig:DP2_LOS}) that are responsible for the blue-side emission in the left panel of Figure~\ref{fig:DP2}. Panel $d$ of Figure~\ref{fig:DP2_LOS} shows that the radial velocity gradient $\text{d}v_{{\rm pe}, r}/\text{d}r$ is typically positive near the source galaxy because the gravitational infall velocity decreases with increasing distance from the source\footnote{Panel~$d$ of Figure~\ref{fig:DP2_LOS} also shows a strong outflow within $0.15~h^{-1}~{\rm cMpc}$ due to pressurized gas from supernovae, and a mild outflow beyond $2~h^{-1}~{\rm cMpc}$ due to large-scale expansion around the void. The former does not affect the transmission of the blueward photons of interest, while the latter further increases $\mathrm{d}v_{{\rm pe}, r}/\mathrm{d}r$, thereby enhancing their transmission.} \citep[see Sec.~3.1 of][]{2021ApJ...922..263P}. This velocity gradient effect produces an order-unity increase in the transmission, as illustrated in panel~$e$ of Figure~\ref{fig:DP2_LOS}.

For simplicity in the explanation, we reduce Equation~(\ref{eq:tau}) by dropping the $\text{d}v_{{\rm pe}, r}/\text{d}r$ term and substituting the cosmological parameters, while ignoring the dark energy contribution $\Omega_\Lambda = 1 - \Omega_{\rm m}$ (where $\Omega_{\rm m}$ is the matter density today) in $H(z)$ (valid for $z \gtrsim 3$):
\bea\label{eq:tau2}
\tau_*(\mathbf{r}) \approx \left[\frac{n_{\rm HI}(\mathbf{r})}{1.7 \times 10^{-10}~{\rm cm^{-3}}}\right]z_6^{-1.5},
\eea
where $z_6\equiv[1+z]/7$. This expression indicates that the IGM permits significant transmission when $n_{\rm HI} \lesssim 2\times10^{-10}~{\rm cm^{-3}}$. This is consistent with our finding in Figure~\ref{fig:DP2_LOS} that regions with significant transmission ($\tau_* \lesssim1$) coincide with locations where $n_{\rm HI}\lesssim2\times10^{-10}~{\rm cm^{-3}}$.

In addition to low density, the spatial location of the void relative to the galaxy is also important. For our assumed intrinsic emission profile, most blueward photons are emitted at $-300\lesssim v_{\alpha,i} \lesssim 0~{\rm km~s^{-1}}$. The underdense region must therefore lie within the distance over which such photons redshift into the Ly$\alpha$ resonance. This distance can be obtained by solving for $r$ along each direction $\hat{\mathbf{n}}$ that satisfies
\bea \label{eq:v}
v_{\alpha,i} + ra(z)H(z) + v_{{\rm pe},r}(\mathbf{r}) = 0 \, .
\eea
We show these locations for $v_{\alpha,i} = 0$, $-100$, $-200$, and $-300~{\rm km~s^{-1}}$ as contours on the HI density maps in Figures~\ref{fig:DP1}–\ref{fig:DP2_LOS}. The contours are approximately circular and increase in radius for more negative $v_{\alpha,i}$, as a larger distance is required to redshift the photons into the Ly$\alpha$ resonance. Deviations from perfect circular shapes arise from spatial fluctuations in the peculiar velocity field associated with the inhomogeneous large-scale structure.\footnote{As illustrated by the red contour, photons emitted at the Ly$\alpha$ resonance in the galaxy frame ($v_{\alpha,i}=0$) must still travel a finite distance before reaching resonance in the IGM frame. This offset arises because gravitational infall motions around galaxies blueshift the photons in the IGM frame \citep{2004MNRAS.349.1137S,2008MNRAS.391...63I,2011MNRAS.414.2139D,2021ApJ...922..263P,2022ApJ...931..126P}.} Panel $f$ of Figure~\ref{fig:DP2_LOS} shows that the sightlines of interest intersect the underdense voids with $n_{\rm HI}\lesssim3\times10^{-10}~{\rm cm^{-3}}$ shown in dark red between the red contour corresponding to $v_{\alpha,i} = 0~{\rm km~s^{-1}}$ and the blue contour corresponding to $v_{\alpha,i} = -300~{\rm km~s^{-1}}$, allowing significant blueward Ly$\alpha$ flux to escape.

How can regions with such low HI density exist at $z=6$? In HII regions, ionization equilibrium gives the HI density as
\bea \label{eq:nHI}
n_{\rm HI} &=& \frac{\alpha_{\rm B} n_{\rm HII}n_e}{\Gamma}
\approx 1.08 \frac{\alpha_{\rm B} \bar{n}^2_{\rm H}}{\Gamma} \Delta^2 \nonumber\\
&=& \left[1.21\times 10^{-8}~{\rm cm^{-3}}\right]
\left[\frac{10^{-13}~{\rm s^{-1}}}{\Gamma}\right]
T_4^{0.7}
z_6^6 \Delta^2 ,
\eea
where $T_4\equiv T/[10^4~\rm K]$, $\alpha_{\rm B} = 2.7\times 10^{-13}~T_4^{0.7}~{\rm cm^3~s^{-1}}$ is the case B recombination coefficient, and $\Delta \equiv n_{\rm H}/\bar{n}_{\rm H}$ is the hydrogen density normalized by its cosmic mean $\bar{n}_{\rm H}=[6.45\times10^{-5}]z_6^3~{\rm cm^{-3}}$. In HII regions we assume $n_{\rm HII}\approx n_{\rm H}$ and that helium is singly ionized giving $n_e\approx 1.08~n_{\rm H}$.

As shown in panel~$b$ of Figure~\ref{fig:DP2_LOS}, $\Gamma$ quickly approaches the background value ($\sim 2\times10^{-13}~{\rm s^{-1}}$) outside the virial radius, making $n_{\rm HI}$ the primary factor controlling $\tau_*$. The scaling $n_{\rm HI}\propto \Delta^2$ implies that baryon density fluctuations are strongly amplified in the neutral hydrogen distribution. Substituting Equation~(\ref{eq:nHI}) into Equation~(\ref{eq:tau2}) yields
\bea \label{eq:tau3}
\tau_* \approx 71
\left[ \frac{10^{-13}~{\rm s^{-1}}}{\Gamma}\right]
\left[\frac{T}{10^4~{\rm K}}\right]^{0.7}
z_6^{4.5} \Delta^2 .
\eea
Thus, the ionized IGM at the cosmic mean density ($\Delta = 1$) remains highly opaque at the Ly$\alpha$ resonance ($\tau_* \approx 35$) for a typical ionized gas temperature of $T = 10^4~{\rm K}$ and a moderately strong ionizing background with $\Gamma = 2 \times 10^{-13}~{\rm s^{-1}}$. However, in sufficiently underdense regions with $\Delta \lesssim 0.25$ (see panel~$a$), $\tau_*$ falls to a few or below, with an additional boost in transmission due to a positive radial peculiar velocity gradient ($\mathrm{d}v_{{\rm pe}, r}/\mathrm{d}r>0$), permitting significant Ly$\alpha$ transmission, as shown in panel~$e$.

\begin{figure}
  \begin{center}
    \includegraphics[scale=0.85]{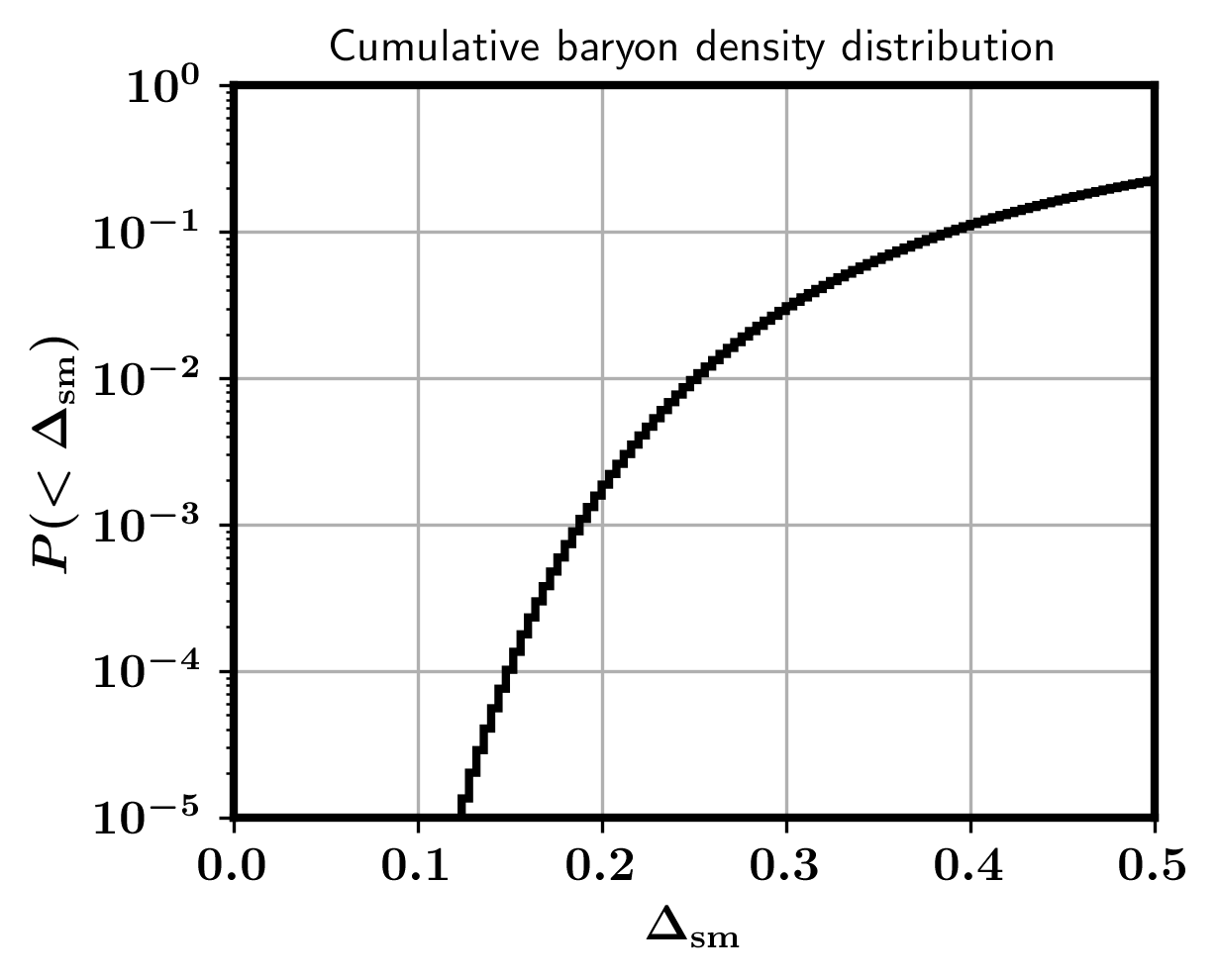}
  \caption{Cumulative probability distribution function of the normalized density, Gaussian-smoothed over the Sobolev length $l_s = 0.13~h^{-1}~\rm cMpc$, denoted as $\Delta_{\rm sm}$.}
   \label{fig:DeltDist}
  \end{center}
\end{figure}

Therefore, a small ($\sim0.5~\rm cMpc$) pocket of such underdensity located at $\sim1$–$2~{\rm cMpc}$ from the galaxy is sufficient to produce significant blue-side transmission; the surrounding environment need not be globally underdense. Because density fluctuations on $\lesssim{\rm cMpc}$ scales are already nonlinear by $z=6$, such voids are expected to be common in the IGM. The cumulative distribution of the normalized density, Gaussian-smoothed over the Sobolev length $l_s = 0.13~h^{-1}~\rm cMpc$ (Figure~\ref{fig:DeltDist}), shows that $\sim 1\%$ of the simulation volume is expected to host such transmissive regions with densities below $\sim25\%$ of the cosmic mean. Although galaxies reside in overdense nodes of the cosmic web—where such underdense regions are less likely to be—voids can still occur between sheets and filaments at these distances, as in the cases demonstrated in Figures~\ref{fig:DP1} and \ref{fig:DP2}. 

\begin{figure*}
  \begin{center}
    \includegraphics[scale=0.77]{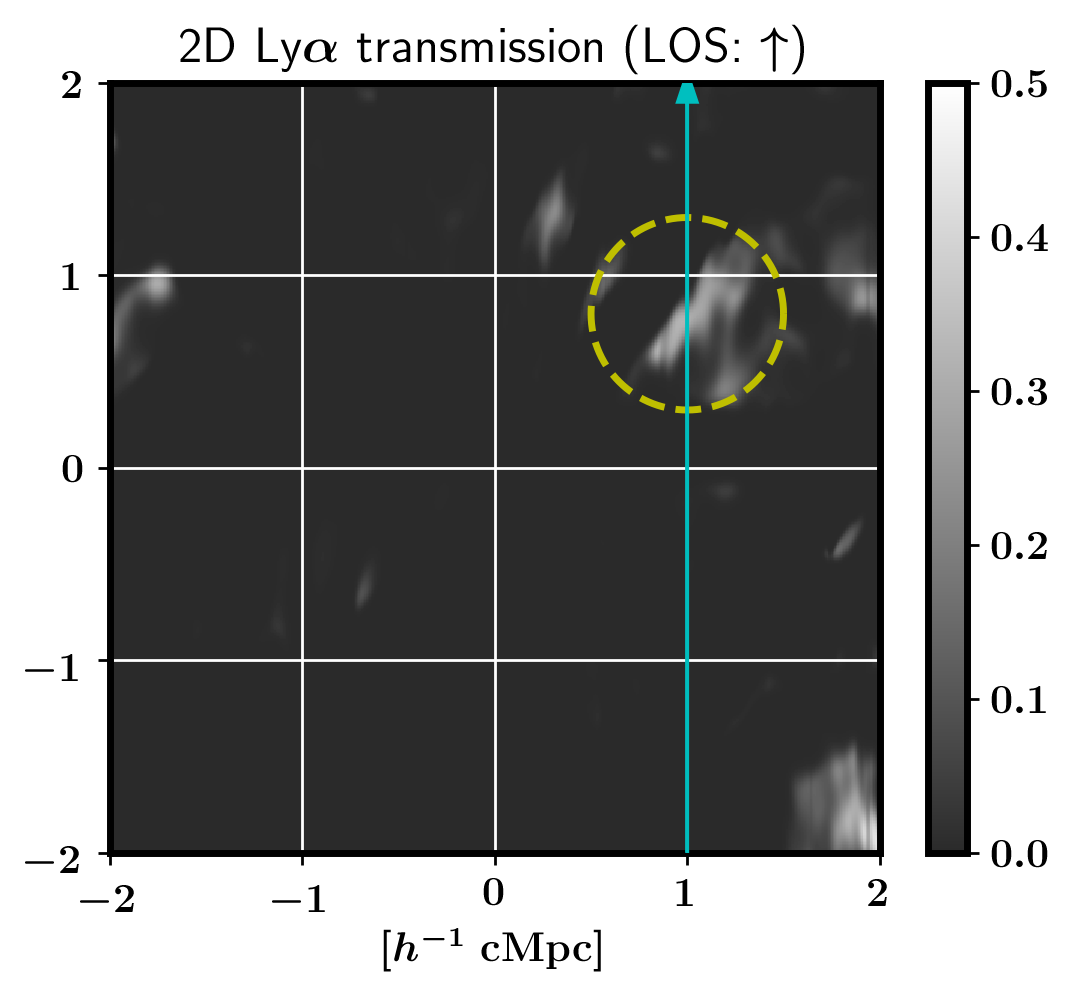}
    \includegraphics[scale=0.77]{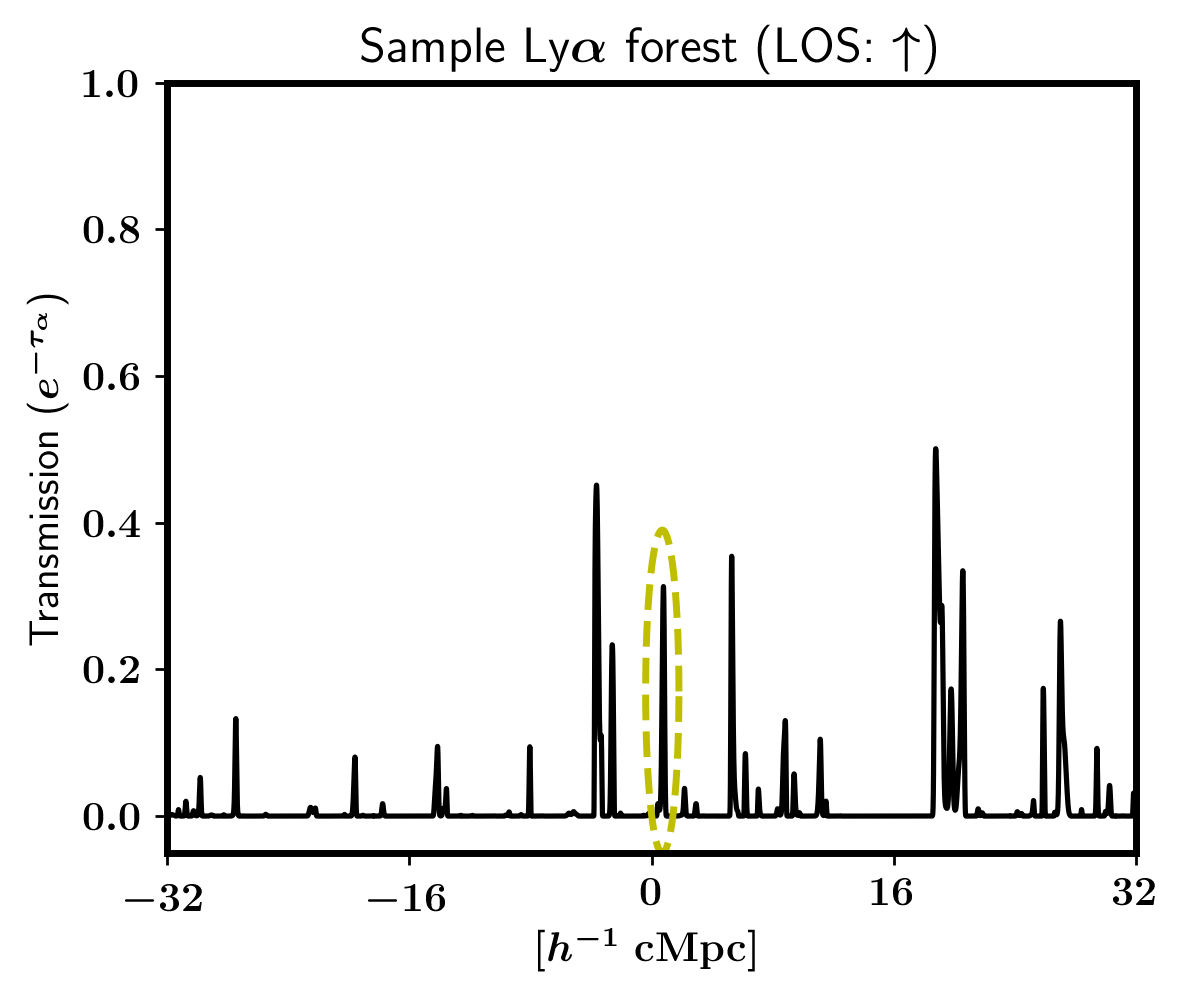}
    \includegraphics[scale=0.77]{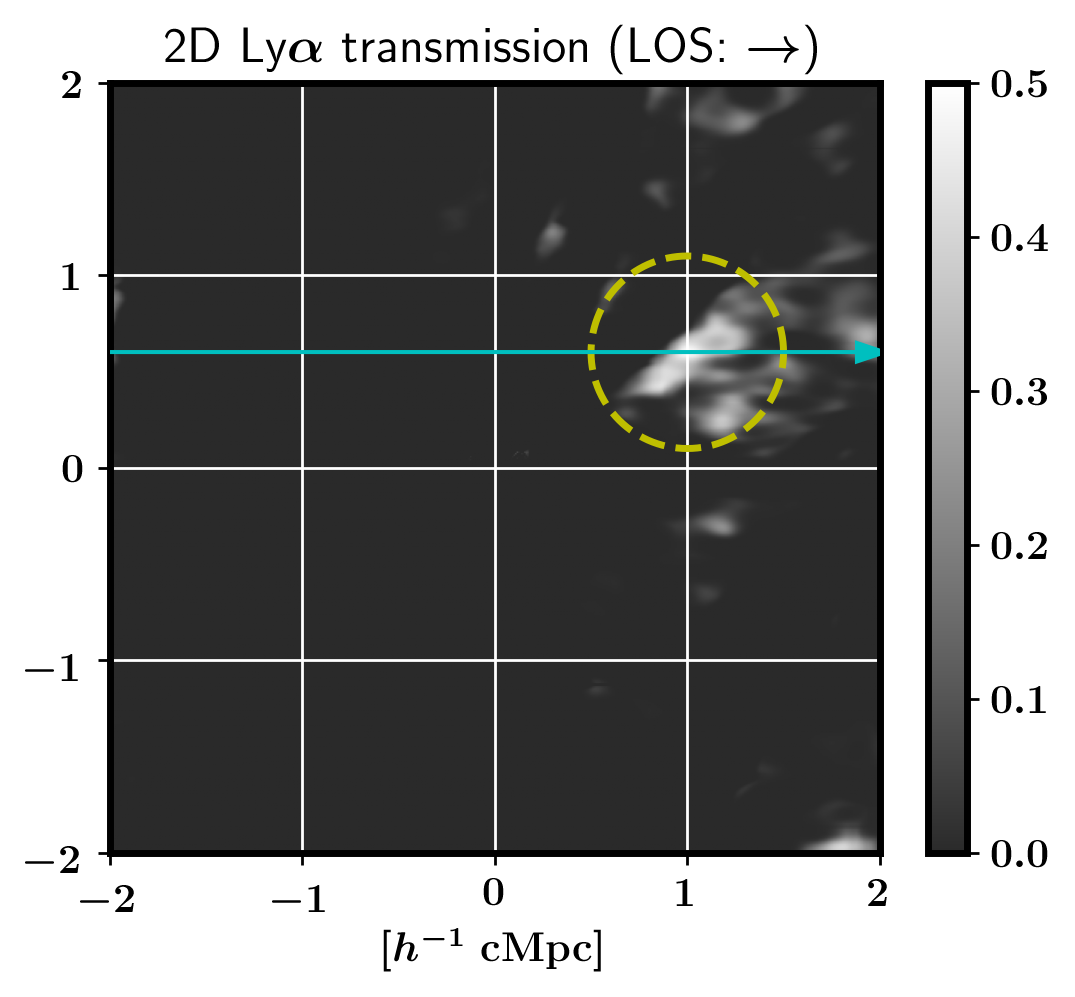}
    \includegraphics[scale=0.77]{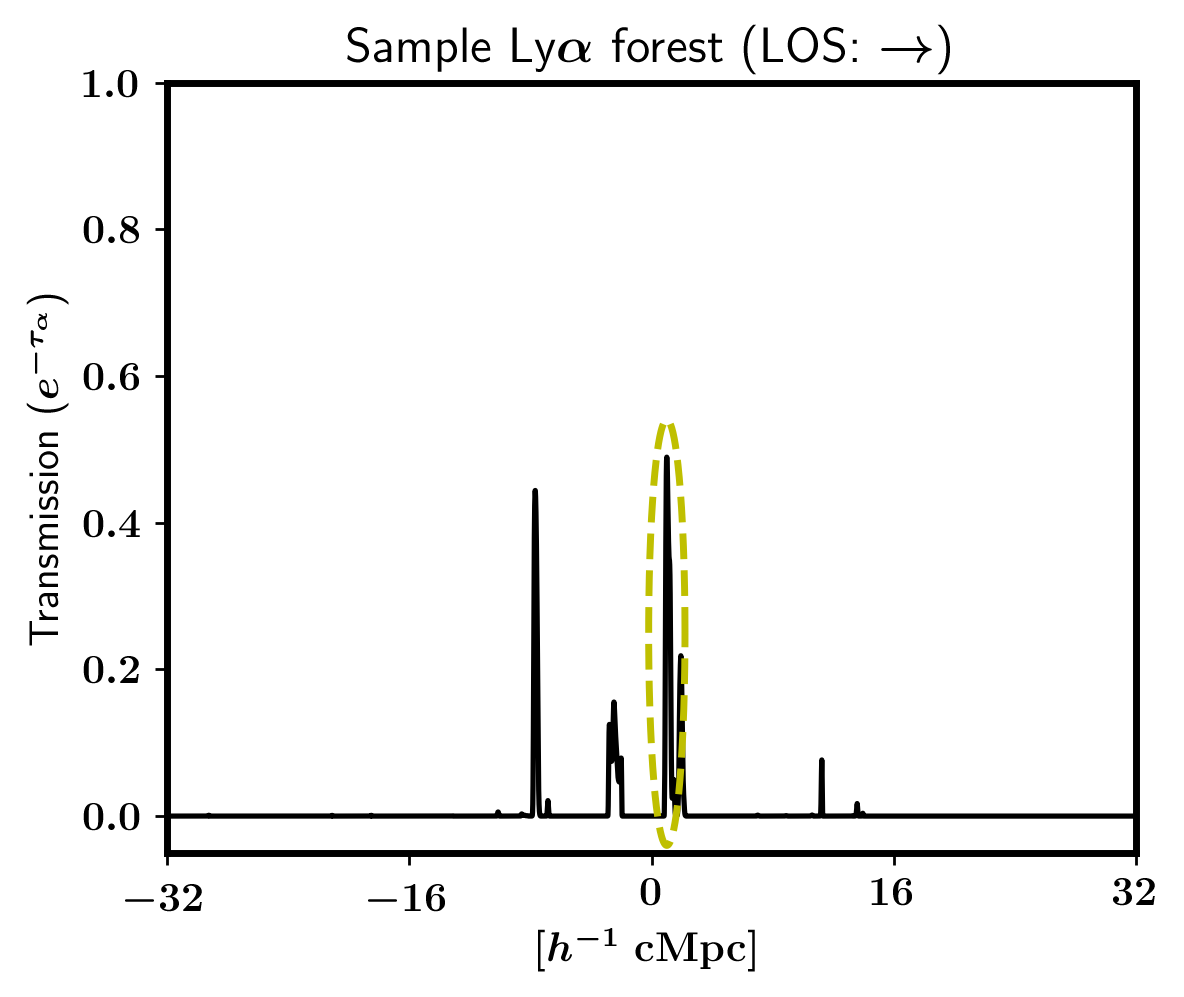}
  \caption{Left panels: two-dimensional Ly$\alpha$ transmission maps for vertical (upper panel) and horizontal (lower panel) sightlines through a $4~h^{-1}~{\rm cMpc}$ square slice centred on the source shown in Figure~\ref{fig:DP2}. The yellow dashed circles highlight transmission pockets corresponding to the underdense void in the right panel of Figure~\ref{fig:DP2}, which enables double-peaked transmission. The cyan arrows mark example sightlines intersecting the highlighted transmission regions. Right panels: Ly$\alpha$ forest transmission along the vertical (upper panel) and horizontal (lower panel) sightlines indicated by the cyan arrows in the left panels, extending across the entire simulation box. The yellow dashed circles mark the transmission spike corresponding to the highlighted transmission pocket in the left panel.}
   \label{fig:tauMap}
  \end{center}
\end{figure*}

Clearly, the location of the underdense void required for significant blueward transmission depends on the emergent line profile from the source, $f_{\rm int}(v_\alpha)$. For example, if the blueward flux of the emergent profile were distributed mostly between $v_\alpha=-300$ and $-600~\rm km~s^{-1}$, as found in some galaxies in the low-redshift Universe, the underdense void shown in this section would not allow meaningful transmission of the blueward flux, because most of the flux would redshift into resonance only after passing through that void. Sightlines from such galaxies would therefore require underdense voids at greater distances to allow a transmission of blue-wing photons.

\subsubsection{Analogy to the Ly$\alpha$ Forest}

The physical mechanism that allows blue-side Ly$\alpha$ peaks to be transmitted through the IGM is the same as that responsible for transmission in the Ly$\alpha$ forest. Thus, the underdense voids that give rise to double-peaked LAEs are also expected to produce Ly$\alpha$ forest transmission spikes along sightlines that intersect them. Although transmission becomes nearly zero above $z\sim5$, some sightlines have been observed to exhibit transmission spikes up to $z\sim6$ \citep[e.g.,][]{2022MNRAS.514...55B}.

In the left panels of Figure~\ref{fig:tauMap}, we show two-dimensional Ly$\alpha$ transmission maps for the same $4~h^{-1}~{\rm cMpc}$ slice shown in the HI density map of Figure~\ref{fig:DP2}, assuming that the LOS direction is parallel to the $x$- and $y$- axes for each map. That is, any sightline sampled parallel to these axes corresponds to a hypothetical normalized Ly$\alpha$ forest spectrum along that sightline. As expected at $z=6$, the transmission is nearly zero throughout most of the volume, with only small, localized pockets of nonzero transmission. The locations of these transmissive pockets coincide with low-HI-density regions ($n_{\rm HI}\lesssim2\times 10^{-10}~{\rm cm^{-3}}$), which appear dark red in the HI density map.

The extended transmission pockets marked by the dashed circles in the left panels of Figure~\ref{fig:tauMap} are associated with the same void that gives rise to the blue peak in Figure~\ref{fig:DP2}. The transmissive regions appear slightly different for different LOS directions due to the change in redshift-space distortion along the line of sight. The right panels of Figure~\ref{fig:tauMap} show the Ly$\alpha$ forest along two sightlines intersecting this void in the $x$ and $y$ directions, extending across the entire simulation box, with the transmission spikes associated with the underdense void marked by dashed lines. Our results here clearly show that underdense voids result in both rare double-peaked emission in LAEs and rare transmission spikes in the high-$z$ Ly$\alpha$ forest. It is, therefore, expected that these two observables would spatially correlate with each other. 

The statistical properties of the Ly$\alpha$ forest at these redshifts are an important and extensive topic in their own right; however, they are beyond the scope of this paper. A full analysis based on calculations performed on the complete $8192^3$ mesh for multiple CoDaIII snapshots will be presented in a forthcoming paper.

\subsection{Ionizing Background Intensity} \label{sec:IBR}

\begin{figure}
  \begin{center}
    \includegraphics[scale=0.8]{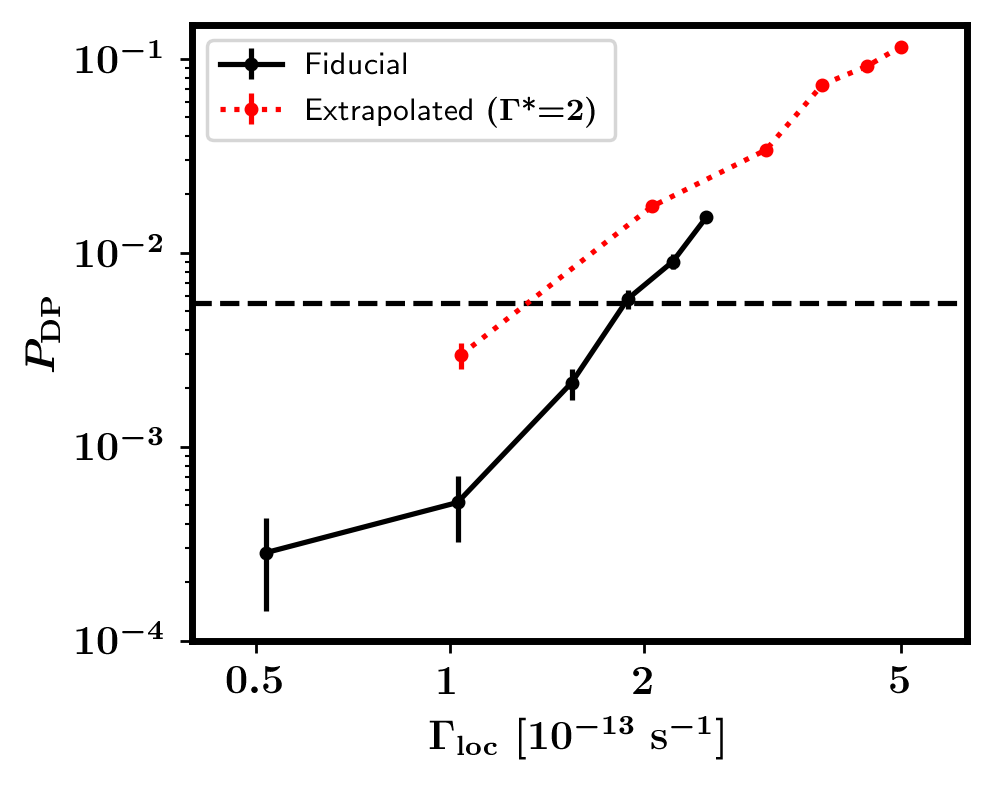}
  \caption{ Probability of exhibiting double-peaked Ly$\alpha$ emission, $P_{\rm DP}$, as a function of $\Gamma_{\rm loc}$. Each data point represents a binned average of $\approx 43{,}000$ sightlines from $\approx 120$ galaxies, sorted according to $\Gamma_{\rm loc}$. The error bars indicate the $2\sigma$ uncertainty. The solid black line shows the fiducial case, while the red dotted line corresponds to the case in which $n_{\rm HI}$ is artificially reduced by a factor of two in highly ($\gtrsim99\%$) ionized regions estimate $P_{\rm DP}$ under a two times stronger $\Gamma_{\rm loc}$. The horizontal dashed line marks the average $P_{\rm DP}$ over all sightlines in the fiducial case, $0.28\%$.}
   \label{fig:PdpVsGamma}
  \end{center}
\end{figure}

The ionizing background intensity, quantified by $\Gamma$, is another key factor governing the IGM opacity. As described in Equation~(\ref{eq:tau}), the Ly$\alpha$ opacity, $\tau_\alpha$, scales as $\Delta^2/\Gamma$, such that a higher (lower) $\Gamma$ leads to a higher (lower) threshold density for the transmission of blue-side Ly$\alpha$ photons. As shown in Figure~\ref{fig:DeltDist}, the probability that the density falls below this threshold varies steeply around $\Delta = 0.25$. This steep dependence leads to a substantial difference in the probability of observing double-peaked emission ($P_{\rm DP}$, hereafter) for a moderate change in $\Gamma$.

The COLA1-like case shown in Figure~\ref{fig:DP2} supports this intuition. The galaxy lies near the centre of a large ionized region, where the ionizing intensity is relatively high, with $\Gamma\approx2\times10^{-13}~{\rm s^{-1}}$ (see Figure~\ref{fig:LSmap}). In such environments, even moderately underdense regions with $\Delta\approx0.3$ can become sufficiently transparent to allow blue-side Ly$\alpha$ transmission. In contrast, a region with $\Gamma \approx 10^{-13}~{\rm s^{-1}}$, for instance, would require more extreme underdensities, such as $\Delta \approx 0.2$, which are $\sim 20$ times less likely to occur (see Figure~\ref{fig:DeltDist}).

For a statistical analysis, we compute the local average of the ionizing intensity, $\Gamma_{\rm loc}$, for each galaxy by taking the median value of $\Gamma$ between the virial radius and $3~h^{-1}~{\rm cMpc}$. This quantity serves as a proxy for the characteristic ionizing background that regulates blue-side Ly$\alpha$ transmission. Figure~\ref{fig:PdpVsGamma} shows a strong dependence of $P_{\rm DP}$ on $\Gamma_{\rm loc}$: $P_{\rm DP}$ increases by nearly an order of magnitude as $\Gamma_{\rm loc}$ increases from $1$ to $2\times10^{-13}~\rm s^{-1}$. 

To further illustrate this sensitivity, we show in Figure~\ref{fig:PdpVsGamma} as a dotted line an approximate estimate for a twofold increase in $\Gamma_{\rm loc}$ of the sample galaxies with the same ionization morphology. This estimate is obtained by recalculating $\tau_\alpha$ after reducing the HI density by a factor of two at highly ionized regions ($\chi>0.99$), motivated by the ionization-equilibrium relation $n_{\rm HI} \propto \Gamma^{-1}$. The resulting curve provides an approximate extrapolation of the $P_{\rm DP}$–$\Gamma_{\rm loc}$ relation toward higher $\Gamma_{\rm loc}$. 

The extrapolated $P_{\rm DP}$--$\Gamma_{\rm loc}$ relation remains broadly consistent with the fiducial result. It overshoots the fiducial result in the lowest $\Gamma_{\rm loc}$ bin, at $10^{-13}~{\rm s^{-1}}$, by nearly a factor of ten. However, the discrepancy decreases to a factor of two at $\Gamma_{\rm loc}=2\times10^{-13}~{\rm s^{-1}}$, indicating convergence toward higher $\Gamma_{\rm loc}$, where the extrapolation is useful. The larger discrepancy in the lowest $\Gamma_{\rm loc}$ bin arises because the sample galaxies lie near neutral regions, where $\Gamma$ can be highly inhomogeneous within $3~h^{-1}~\rm cMpc$ (see the brown regions in the right panel of Figure~\ref{fig:LSmap}). These galaxies can have $\Gamma\approx0$ in the neutral regions but much higher $\Gamma$ than $\Gamma_{\rm loc}$ in the ionized regions, yielding a much elevated $P_{\rm DP}$ compared to the fiducial case at the same $\Gamma_{\rm loc}=10^{-13}~{\rm s^{-1}}$, where $\Gamma$ is more homogeneous within $3~h^{-1}~\rm cMpc$.

Combining the extrapolated and fiducial results, we find that a tenfold increase in $\Gamma_{\rm loc}$, from $0.5$ to $5 \times 10^{-13}~{\rm s^{-1}}$, increases $P_{\rm DP}$ by nearly three orders of magnitude, from $\sim10^{-4}$ to $\sim 10^{-1}$. Although the exact values of $P_{\rm DP}$ depend on the assumptions adopted in the calculation, this result demonstrates that $P_{\rm DP}$ is extremely sensitive to $\Gamma_{\rm loc}$.

\subsubsection{Transmissivity Boost by recent AGN activity}

\begin{figure}
  \begin{center}
    \includegraphics[scale=0.8]{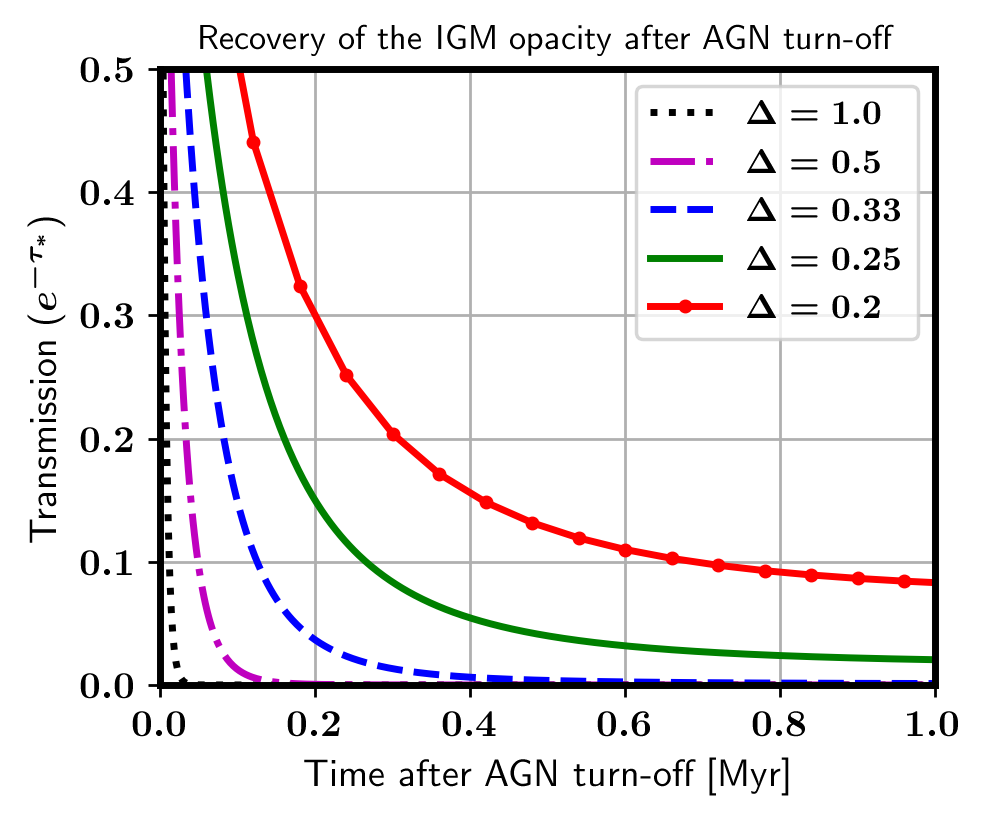}
  \caption{Evolution of the Ly$\alpha$ optical-depth parameter $\tau_*$ following the end of an AGN episode at $z=6$. The AGN is assumed to have reduced $\tau_*$ to nearly zero prior to $t=0$. After the AGN turns off, the photoionization rate is set to the background value of $\Gamma = 10^{-13}\ {\rm s}^{-1}$. The curves (dotted, dot-dashed, dashed, solid and solid with dots) correspond to normalized gas densities of $\Delta \equiv \rho/\bar\rho = 1.0$, 0.5, 0.33, 0.25, and 0.2, respectively.}
   \label{fig:AfterAGN}
  \end{center}
\end{figure}

AGNs are extremely powerful sources of ionizing radiation that can reduce the HI density in the IGM to levels that are highly transmissive at the Ly$\alpha$ resonance within their proximity zones. Most quasar proximity zones at $z\gtrsim6$ extend well beyond the distance required for blue-wing photons emitted a few hundred $\rm~km~s^{-1}$ blueward of the Ly$\alpha$ resonance to redshift into resonance. However, quasars become increasingly rare at $z\gtrsim6$, with the number density of objects brighter than $M_{\rm UV}<-25.5$ falling below one per cubic gigaparsec \citep{2019A&A...632A..45R}, and no such sources have been observed in the vicinity of the two known double-peaked LAEs at $z\approx6.6$, COLA1 and NEPLA4. Nevertheless, the recent discovery of $\sim2\times10^8\,\text{M}_\odot$ SMBHs in these galaxies by \citet{2026arXiv260500763M} suggests that the surrounding IGM may have been affected by strong AGN radiation, which was turned off recently. If these SMBHs experienced an AGN phase sufficiently recently, the IGM may not yet have returned to its ionization-equilibrium HI density and could therefore exhibit enhanced transmissivity to Ly$\alpha$ photons.

Assuming that $\Gamma$ was sufficiently high during a recent AGN phase to suppress $\tau_*$ to nearly zero, and that the AGN ceased its activity a time $t$ ago, after which the photoionization rate returned to $\Gamma_{\rm now}$, the subsequent evolution of $\tau_*$ can be described by substituting $\Gamma_{\rm eff} = \Gamma_{\rm now}/[1-\exp(-t\Gamma_{\rm now})]$ for $\Gamma$ in Equation~(\ref{eq:taualpha}). Based on this, Figure~\ref{fig:AfterAGN} shows how the IGM transmissivity within HII regions, $e^{-\tau_*}$, evolves at different gas densities ($\Delta=1$, 0.5, 0.33, 0.25, and 0.2) at $z=6$, assuming $\Gamma_{\rm now}=10^{-13}~\rm s^{-1}$.

The figure shows that underdense regions ($\Delta<1$) can remain transmissive for substantially longer than regions at the mean density ($\Delta=1$). For example, if we adopt $e^{-\tau_*}=0.1$ as the threshold above which the IGM transmits a significant fraction of blue-wing photons, gas at the mean density remains transmissive for only 0.012 megayears (Myr, hereafter), as it approaches its equilibrium value of $\tau_*=71$ on a timescale of $1/\Gamma_{\rm now}\approx0.3$ Myr. In contrast, this transmissive phase lasts 0.05, 0.13, 0.27, and 0.69 Myr for lower gas densities of $\Delta=$ 0.5, 0.33, 0.25, and 0.2, respectively, as the equilibrium value of $\tau_*$ decreases to 36, 8, 4.5, and 2.8 (Eq.~(\ref{eq:taualpha})).

Given that regions with $\Delta=1$ is much more common than those with $\Delta\lesssim0.5$, the short-lived ($\sim10^4~\rm yr$) Ly$\alpha$ transmissivity boost following AGN activity can still have a substantial impact on the transmission of blue peaks in mean density regions. However, it is also notable that AGN activity can enable transmission through mildly underdense regions, whose equilibrium opacity can marginally suppress blue peaks, for even longer periods. This analysis suggests that the recent AGN activity and underdense voids can act as complementary effects in producing double-peaked LAEs.

\subsection{UV Magnitude and Local Density Around the Source} \label{sec:MUV}

\begin{figure*}
  \begin{center}
    \includegraphics[scale=0.84]{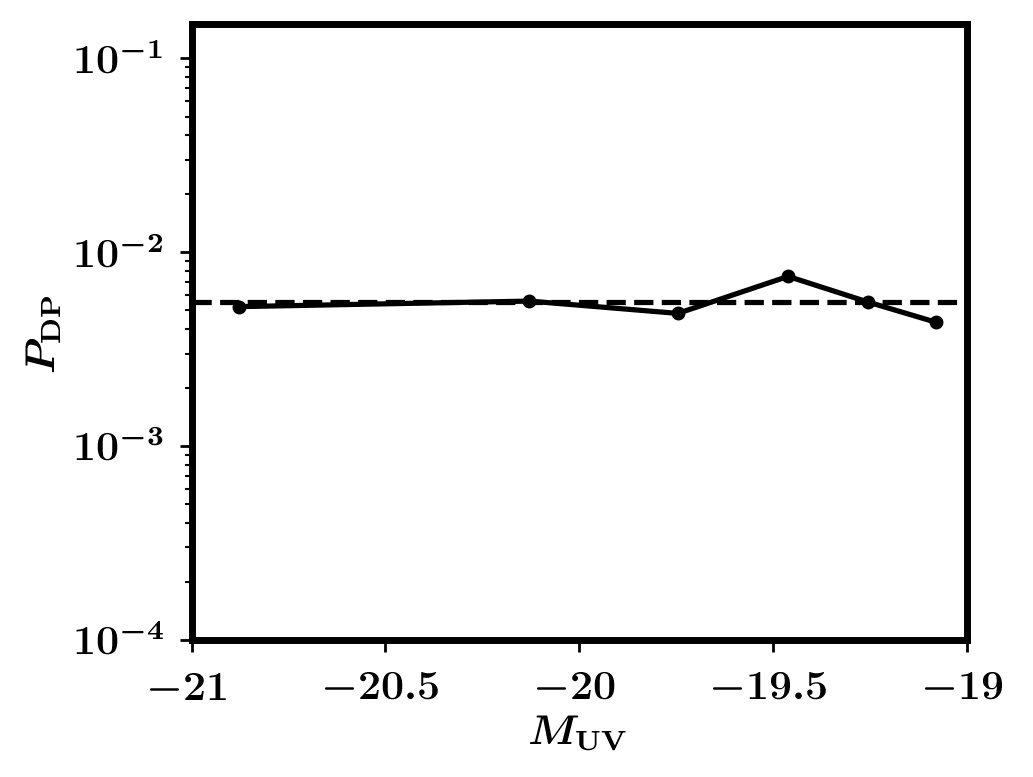}
    \includegraphics[scale=0.84]{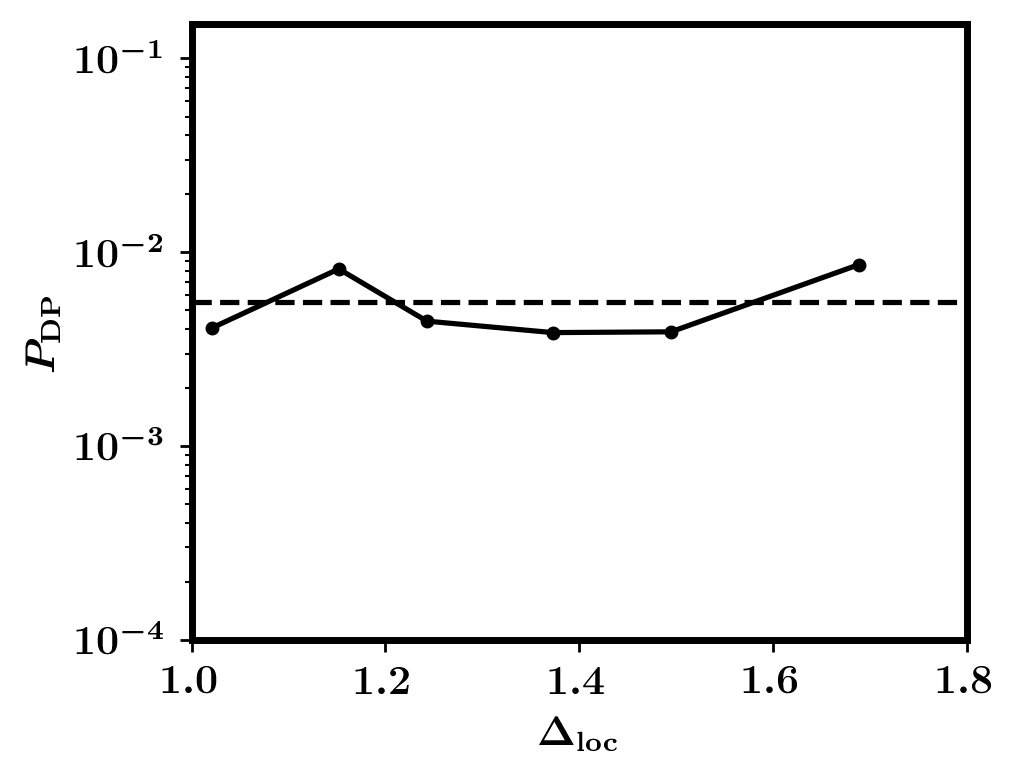}
\caption{ Probability of exhibiting double-peaked Ly$\alpha$ emission $P_{\rm DP}$ for different $M_{\rm UV}$'s (left) and local density $\Delta_{\rm loc}$ (right) of source galaxies. $\Delta_{\rm loc}$ is obtained by averaging the gas density within 0.5 and $3~h^{-1}~\rm cMpc$ from the source galaxy. Other details are same as in Figure~\ref{fig:PdpVsGamma}.}
   \label{fig:PdpVsMuv}
  \end{center}
\end{figure*}

In contrast to $\Gamma_{\rm loc}$, neither $M_{\rm UV}$ nor the local density $\Delta_{\rm loc}$ appears to be an important factor for double-peaked Ly$\alpha$ emission. We compute $\Delta_{\rm loc}$ by averaging the normalized density $\Delta$ between the virial radius and $3~h^{-1}~{\rm cMpc}$ from the source galaxy. Figure~\ref{fig:PdpVsMuv} shows that neither $M_{\rm UV}$ nor $\Delta_{\rm loc}$ exhibits a clear correlation with $P_{\rm DP}$.

At first glance, this result may seem counterintuitive, as brighter galaxies in higher-density regions would naturally produce a stronger $\Gamma$ and a lower $n_{\rm HI}$. However, we note that the IGM responsible for producing double-peaked emission is located at least $1~h^{-1}~{\rm cMpc}$ away from the source galaxy, where the radiation from the galaxy itself contributes only minimally to $\Gamma$, as shown in panel $b$ of Figure~\ref{fig:DP2_LOS}. Additionally, higher local densities make the presence of underdense voids less likely, potentially offsetting the effects of an increased galaxy density.

We note that the size of our simulation may not be sufficient to capture a large number of bright galaxies to obtain reliable statistics for $M_{\rm UV} \lesssim -21$. For example, the number of galaxies decreases steeply toward higher UV brightness in our sample, and $P_{\rm DP}$ for the UV-brightest bin in the left panel of Figure~\ref{fig:PdpVsMuv} is computed by averaging over 120 galaxies with $-22 \lesssim M_{\rm UV} \lesssim -20.5$. A much larger simulation volume is therefore required to obtain robust statistics for these galaxies, as double-peaked emission is intrinsically rare.

Finally, any dependence of the intrinsic Ly$\alpha$ line profile on $M_{\rm UV}$—which is not included in this work—could introduce a corresponding dependence in $P_{\rm DP}$. For example, UV-brighter galaxies are generally more massive and are known to produce more extended emergent profiles both in the blue and red directions. A more extended blueward profile might increase the likelihood of flux transmission because the photons enter the Ly$\alpha$ resonance at larger distances from the overdense environments of the source galaxies. Since we use the flux ratio between the blue and red peaks in our double-peak criteria in this work, changes in the redward profile would also affect $P_{\rm DP}$ in our calculation. However, these effects are unlikely to produce variations comparable to the strong dependence on $\Gamma$ seen in Figure~\ref{fig:PdpVsGamma}.

\section{Summary and Discussion}\label{sec:summary}

In this paper, we analyzed Ly$\alpha$ opacity along sightlines originating from simulated galaxies in the CoDaIII simulation to investigate the physical conditions that give rise to double-peaked Ly$\alpha$ emitters observed in recent surveys. We identify several cases that closely resemble the observed systems and use them to isolate the key conditions required for double-peaked Ly$\alpha$ emission.

\textit{Physical conditions for double-peaked LAEs:}
Double-peaked Ly$\alpha$ emission requires the presence of a highly underdense pocket, with a characteristic size of a few hundred ckpc, located at the distance where the blueward flux from the source galaxy redshifts into resonance, typically one to a few cMpc. These small voids enable a significant fraction of photons emitted blueward of Ly$\alpha$ to be transmitted through the IGM as they redshift through resonance. Although such underdense regions are rare, they can arise naturally in the vicinity of galaxies, forming in the gaps between sheets and filaments of the cosmic web as a consequence of inhomogeneous structure formation.

The intensity of local ionizing background is another critical factor enabling double-peaked emission. The threshold density required for voids to transmit blue-side photons lies in the extreme low-density tail of the density distribution. Increasing (decreasing) the intensity therefore leads to a steep increase (decrease) in the number of voids capable of transmitting blue-side photons. We find that a factor-of-two change in the ionizing background intensity results in nearly an order-of-magnitude change in the probability of seeing double-peaked emission ($P_{\rm DP}$). This strong dependence implies that even small variations in the ionizing background during the late stages of reionization can produce large changes in the abundance of observable double-peaked LAEs.

In contrast, $P_{\rm DP}$ shows no significant dependence on the UV luminosity or local density of the source galaxy. This indicates that the dominant factor governing double-peaked emission is the UV radiation from other galaxies within the mean free path of ionizing photons, rather than radiation from the source galaxy itself. This interpretation is consistent with observations of COLA1 by \cite{2024A&A...689A..44T}, where the source galaxy is neither particularly UV-bright nor located in an extreme galaxy overdensity.

Another possibility is a supermassive black hole that was accreting until recently ($\lesssim\rm Myr$ ago). Extreme ionization in the proximity zone can leave the region transmissive even after the active galactic nucleus (AGN) has turned off, as recently suggested by \cite{2026arXiv260500763M}. In this case, mildly underdense pockets whose equilibrium opacity is only marginally high enough to scatter blue-wing photons can remain transmissive for $\rm (a~few)\times10^{5}~yr$, while the mean-density IGM would become opaque within $\rm (a~few)\times10^{4}~yr$. Thus, recent AGN activity and underdense pockets can act as complementary factors in explaining double-peaked LAEs.

\textit{Implications for reionization scenarios:}
The steep dependence of $P_{\rm DP}$ on photoionization rate ($\Gamma$) suggests that double-peaked LAEs can serve as a sensitive probe of reionization models. For example, the absence of double-peaked emission in the $z=6.5$ snapshot of CoDaIII may indicate that reionization was more advanced at $z=6.5$ than assumed in the simulation, which adopts a global ionized fraction of $60\%$. In our analysis, double-peaked emission is predominantly found near the centres of ionized regions, where $\Gamma \gtrsim 10^{-13}~\mathrm{s^{-1}}$, at $z \approx 6$, when the Universe is approximately $80\%$ ionized. At this stage, ionized regions grow to tens of comoving megaparsecs through the overlap of smaller bubbles, leading to enhanced ionizing backgrounds near their centres. Equation~(\ref{eq:tau3}) shows that the IGM opacity increases by $\approx 30\%$ for a fixed $\Gamma$ and $\Delta$ when going from $z = 6$ to $6.5$, and the probability of finding highly underdense voids also decreases slightly. However, regions with $\Gamma \gtrsim 10^{-13}~\mathrm{s^{-1}}$ could still allow a small fraction of sightlines to exhibit double-peaked emission. The observation of double-peaked LAEs at $z = 6.5$ may therefore indicate that reionization had already entered this late-stage percolation phase by that redshift.

Depending on the mean free path of ionizing photons and the halo-mass dependence of ionizing efficiency, reionization may be dominated either by numerous small HII bubbles or by rare, large bubbles, resulting in different ionization geometries \citep[see, e.g.,][]{2004ApJ...613....1F,2007MNRAS.377.1043M,2022LRCA....8....3G}. For a fixed global ionization fraction, the latter (i.e., the large HII bubble scenario) would be more favorable for double-peaked LAEs, as $\Gamma$ would be higher on average in HII regions. 

\textit{Relation to the Ly$\alpha$ forest:}
Double-peaked LAEs are closely related to transmission spikes in the high-redshift Ly$\alpha$ forest, as the physical conditions required for both phenomena—underdense pockets and a strong ionizing background—are essentially the same \citep{2024ApJ...975..115Z,2026ApJ..1002...93Z}. Because the ionizing background radiation field traces the ionization topology of the IGM (see the right panel of Figure~\ref{fig:LSmap}), large-scale spatial correlations between these two observables are expected. This implies that environments favorable for double-peaked Ly$\alpha$ emission should coincide with those that give rise to unusually Ly$\alpha$-transmissive sightlines near the end of reionization.

\textit{Caveats:}
The sightlines identified as exhibiting double-peaked Ly$\alpha$ emission are subject to several assumptions adopted in this work. In particular, we assume a fixed intrinsic Ly$\alpha$ emission profile for all galaxies, owing to the limited understanding of the detailed Ly$\alpha$ radiative transfer processes within simulated galaxies. In addition, the criteria used to identify double-peaked emission is somewhat arbitrary, as no standardized definition currently exists in the literature. For instance, one could instead employ topological persistence in the line shape, rather than the integrated flux used in this work. Consequently, the inferred number of double-peaked sightlines depends on the specific definition adopted. Despite these limitations, our central result—that double-peaked Ly$\alpha$ emission serves as a sensitive probe of the ionizing background intensity $\Gamma$—remains robust, as all assumptions are applied consistently throughout the analysis.

The volume of CoDaIII is not large enough to sample large HII bubbles spanning tens of cMpc that form from the percolation of smaller bubbles toward the end of reionization (see Figure~\ref{fig:LSmap}). Since most double-peaked LAEs are found in the central regions of these bubbles, the limited volume likely imposes statistical uncertainties on our results. 

CoDaIII also does not include AGNs, which can substantially reduce the Ly$\alpha$ opacity in their surroundings. The probability $P_{\rm DP}$ is therefore likely to be higher in regions near AGNs. However, observed double-peaked LAEs at $z \gtrsim 6.5$ do not show evidence for AGN activity within or around their host galaxies, suggesting that their surrounding environments may be similar to those represented in CoDaIII. As mentioned above, recently ceased AGN activity could be another factor that further increases the occurrence of double-peaked LAEs.

\textit{Future prospects:}
Upcoming spectroscopic surveys with \textit{JWST}, existing facilities (e.g., X-shooter on the Very Large Telescope, \citealt{2011A&A...536A.105V}; FOCAS on the Subaru telescope, \citealt{2022arXiv220614908G}), and next-generation ground-based telescopes (e.g., GMACS on the Giant Magellan Telescope, \citealt{10.1117/12.2313940}; MOSAIC on the Extremely Large Telescope, \citealt{10.1117/12.3019047}) will substantially increase the samples of high-redshift LAEs and Ly$\alpha$ forest spectra with well-resolved line profiles. Joint analyses of LAEs and the Ly$\alpha$ forest—such as measurements of their two-point correlation—may provide additional constraints on the ionizing background intensity and the reionization history during its late stages. Ultimately, these probes could complement traditional approaches, such as Ly$\alpha$ equivalent-width measurements and Ly$\alpha$ damping-wing analyses, providing new insights into the ionizing sources during the reionization era.

\section*{Acknowledgments}
We thank the referee for constructive feedback that has improved the quality of this work. This work used resources of the Oak Ridge Leadership Computing Facility, in particular the Frontier supercomputer, under project AST031. H.P. thanks the organizers of the Cosmic Dawn at High Latitude conference held in Stockholm, Sweden between June 24--28, 2024, which allowed helpful discussion with the participants. H.P. was supported in part by NSF grant PHY-2309135 to the Kavli Institute for Theoretical Physics (KITP). H.P. is supported by IBS under the project code IBS-R018-D3. H.Y. is supported by MEXT/JSPS KAKENHI grant No. 21H04489, 26H02061 and JST FOREST Program, grant No. JP-MJFR202Z. L.C. is supported by STFC consolidated grant ST/X000982/1. J.S. is supported by the University of Lille via the Welcoming Internationals to Lille initiative for the UNIVERSITWINS project. KA is supported by the National Research Foundation of Korea (NRF) RS-2021-NR058956, RS-2025-16302968 and the Korea Astronomy and Space Science Institute under the R\&D program (Project No. 2025-9-844-00) supervised by the Korea AeroSpace Administration.

\section*{Data Availability}
The CoDaIII simulation data used in this study are available from P. Ocvirk upon reasonable request. Derived data products generated in this work are available from the first author, H. Park, upon reasonable request.

\bibliographystyle{apj}
\bibliography{reference}

@ARTICLE{2022arXiv220614908G,
       author = {{Greene}, Jenny and {Bezanson}, Rachel and {Ouchi}, Masami and {Silverman}, John and {the PFS Galaxy Evolution Working Group}},
        title = "{The Prime Focus Spectrograph Galaxy Evolution Survey}",
      journal = {arXiv e-prints},
         year = 2022,
        month = jun,
          eid = {arXiv:2206.14908},
        pages = {arXiv:2206.14908},
          doi = {10.48550/arXiv.2206.14908},
archivePrefix = {arXiv},
       eprint = {2206.14908},
 primaryClass = {astro-ph.GA},
       adsurl = {https://ui.adsabs.harvard.edu/abs/2022arXiv220614908G}
}

@ARTICLE{2019A&A...632A..45R,
       author = {{Romano}, M. and {Grazian}, A. and {Giallongo}, E. and {Cristiani}, S. and {Fontanot}, F. and {Boutsia}, K. and {Fiore}, F. and {Menci}, N.},
        title = "{Lyman continuum escape fraction and mean free path of hydrogen ionizing photons for bright z {\ensuremath{\sim}} 4 QSOs from SDSS DR14}",
      journal = {\aap},
         year = 2019,
        month = dec,
       volume = {632},
          eid = {A45},
        pages = {A45},
          doi = {10.1051/0004-6361/201935550},
archivePrefix = {arXiv},
       eprint = {1910.02775},
 primaryClass = {astro-ph.GA},
       adsurl = {https://ui.adsabs.harvard.edu/abs/2019A&A...632A..45R}
}

@ARTICLE{2022ApJ...927...36W,
       author = {{Wold}, Isak G.~B. and {Malhotra}, Sangeeta and {Rhoads}, James and {Wang}, Junxian and {Hu}, Weida and {Perez}, Lucia A. and {Zheng}, Zhen-Ya and {Khostovan}, Ali Ahmad and {Walker}, Alistair R. and {Barrientos}, L. Felipe and {Gonz{\'a}lez-L{\'o}pez}, Jorge and {Harish}, Santosh and {Infante}, Leopoldo and {Jiang}, Chunyan and {Pharo}, John and {Moya-Sierralta}, Crist{\'o}bal and {Bauer}, Franz E. and {Galaz}, Gaspar and {Valdes}, Francisco and {Yang}, Huan},
        title = "{LAGER Ly{\ensuremath{\alpha}} Luminosity Function at z   7: Implications for Reionization}",
      journal = {\apj},
         year = 2022,
        month = mar,
       volume = {927},
       number = {1},
          eid = {36},
        pages = {36},
          doi = {10.3847/1538-4357/ac4997},
archivePrefix = {arXiv},
       eprint = {2105.12191},
 primaryClass = {astro-ph.GA},
       adsurl = {https://ui.adsabs.harvard.edu/abs/2022ApJ...927...36W}
}

@ARTICLE{2022ApJ...930..104L,
       author = {{Larson}, Rebecca L. and {Finkelstein}, Steven L. and {Hutchison}, Taylor A. and {Papovich}, Casey and {Bagley}, Micaela and {Dickinson}, Mark and {Rojas-Ruiz}, Sof{\'\i}a and {Ferguson}, Harry C. and {Jung}, Intae and {Giavalisco}, Mauro and {Grazian}, Andrea and {Pentericci}, Laura and {Tacchella}, Sandro},
        title = "{Searching for Islands of Reionization: A Potential Ionized Bubble Powered by a Spectroscopic Overdensity at z = 8.7}",
      journal = {\apj},
         year = 2022,
        month = may,
       volume = {930},
       number = {2},
          eid = {104},
        pages = {104},
          doi = {10.3847/1538-4357/ac5dbd},
archivePrefix = {arXiv},
       eprint = {2203.08461},
 primaryClass = {astro-ph.GA},
       adsurl = {https://ui.adsabs.harvard.edu/abs/2022ApJ...930..104L}
}

@ARTICLE{2022MNRAS.517.3263B,
       author = {{Bolan}, Patricia and {Lemaux}, Brian C. and {Mason}, Charlotte and {Brada{\v{c}}}, Maru{\v{s}}a and {Treu}, Tommaso and {Strait}, Victoria and {Pelliccia}, Debora and {Pentericci}, Laura and {Malkan}, Matthew},
        title = "{Inferring the intergalactic medium neutral fraction at z   6-8 with low-luminosity Lyman break galaxies}",
      journal = {\mnras},
         year = 2022,
        month = dec,
       volume = {517},
       number = {3},
        pages = {3263-3274},
          doi = {10.1093/mnras/stac1963},
archivePrefix = {arXiv},
       eprint = {2111.14912},
 primaryClass = {astro-ph.GA},
       adsurl = {https://ui.adsabs.harvard.edu/abs/2022MNRAS.517.3263B}
}

@ARTICLE{2023ApJ...947L..24M,
       author = {{Morishita}, Takahiro and {Roberts-Borsani}, Guido and {Treu}, Tommaso and {Brammer}, Gabriel and {Mason}, Charlotte A. and {Trenti}, Michele and {Vulcani}, Benedetta and {Wang}, Xin and {Acebron}, Ana and {Bah{\'e}}, Yannick and {Bergamini}, Pietro and {Boyett}, Kristan and {Bradac}, Marusa and {Calabr{\`o}}, Antonello and {Castellano}, Marco and {Chen}, Wenlei and {De Lucia}, Gabriella and {Filippenko}, Alexei V. and {Fontana}, Adriano and {Glazebrook}, Karl and {Grillo}, Claudio and {Henry}, Alaina and {Jones}, Tucker and {Kelly}, Patrick L. and {Koekemoer}, Anton M. and {Leethochawalit}, Nicha and {Lu}, Ting-Yi and {Marchesini}, Danilo and {Mascia}, Sara and {Mercurio}, Amata and {Merlin}, Emiliano and {Metha}, Benjamin and {Nanayakkara}, Themiya and {Nonino}, Mario and {Paris}, Diego and {Pentericci}, Laura and {Rosati}, Piero and {Santini}, Paola and {Strait}, Victoria and {Vanzella}, Eros and {Windhorst}, Rogier A. and {Xie}, Lizhi},
        title = "{Early Results from GLASS-JWST. XIV. A Spectroscopically Confirmed Protocluster 650 Million Years after the Big Bang}",
      journal = {\apjl},
         year = 2023,
        month = apr,
       volume = {947},
       number = {2},
          eid = {L24},
        pages = {L24},
          doi = {10.3847/2041-8213/acb99e},
archivePrefix = {arXiv},
       eprint = {2211.09097},
 primaryClass = {astro-ph.GA},
       adsurl = {https://ui.adsabs.harvard.edu/abs/2023ApJ...947L..24M}
}

@ARTICLE{2024ApJ...967...28N,
       author = {{Nakane}, Minami and {Ouchi}, Masami and {Nakajima}, Kimihiko and {Harikane}, Yuichi and {Ono}, Yoshiaki and {Umeda}, Hiroya and {Isobe}, Yuki and {Zhang}, Yechi and {Xu}, Yi},
        title = "{Ly{\ensuremath{\alpha}} Emission at z = 7─13: Clear Evolution of Ly{\ensuremath{\alpha}} Equivalent Width Indicating a Late Cosmic Reionization History}",
      journal = {\apj},
         year = 2024,
        month = may,
       volume = {967},
       number = {1},
          eid = {28},
        pages = {28},
          doi = {10.3847/1538-4357/ad38c2},
archivePrefix = {arXiv},
       eprint = {2312.06804},
 primaryClass = {astro-ph.GA},
       adsurl = {https://ui.adsabs.harvard.edu/abs/2024ApJ...967...28N}
}

@ARTICLE{2026ApJ...997..102M,
       author = {{Martin}, Crystal L. and {Hu}, Weida and {Wold}, Isak G.~B. and {Faisst}, Andreas and {Moya-Sierralta}, Crist{\'o}bal and {Malhotra}, Sangeeta and {Rhoads}, James E. and {Barrientos}, Luis Felipe and {Harikane}, Yuichi and {Infante}, Leopoldo and {Koekemoer}, Anton M. and {Gonzalez Lopez}, Jorge and {Ouchi}, Masami and {Xu}, Junyan and {Yang}, Jiayang and {Yung}, L.~Y. Aaron and {Weaver}, John R. and {McCracken}, Henry and {Zheng}, Zhenya and {Wang}, Junxian},
        title = "{Galaxy Protoclusters as Drivers of Cosmic Reionization: I. Bubble Overlap at Redshift z {\ensuremath{\sim}} 7 in LAGER-z7OD1}",
      journal = {\apj},
         year = 2026,
        month = jan,
       volume = {997},
       number = {1},
          eid = {102},
        pages = {102},
          doi = {10.3847/1538-4357/ae1d6b},
archivePrefix = {arXiv},
       eprint = {2510.13140},
 primaryClass = {astro-ph.GA},
       adsurl = {https://ui.adsabs.harvard.edu/abs/2026ApJ...997..102M}
}

@ARTICLE{2020ARA&A..58..617O,
       author = {{Ouchi}, Masami and {Ono}, Yoshiaki and {Shibuya}, Takatoshi},
        title = "{Observations of the Lyman-{\ensuremath{\alpha}} Universe}",
      journal = {\araa},
         year = 2020,
        month = aug,
       volume = {58},
        pages = {617-659},
          doi = {10.1146/annurev-astro-032620-021859},
archivePrefix = {arXiv},
       eprint = {2012.07960},
 primaryClass = {astro-ph.GA},
       adsurl = {https://ui.adsabs.harvard.edu/abs/2020ARA&A..58..617O}
}

@ARTICLE{2025ApJ...983...91P,
       author = {{Park}, Hyunbae and {Jung}, Intae and {Yajima}, Hidenobu and {Sorce}, Jenny G. and {Shapiro}, Paul R. and {Ahn}, Kyungjin and {Ocvirk}, Pierre and {Teyssier}, Romain and {Yepes}, Gustavo and {Iliev}, Ilian T. and {Lewis}, Joseph S.~W.},
        title = "{Constraining Reionization with Ly{\ensuremath{\alpha}} Damping-wing Absorption in Galaxy Spectra: A Machine Learning Model Based on Reionization Simulations}",
      journal = {\apj},
         year = 2025,
        month = apr,
       volume = {983},
       number = {2},
          eid = {91},
        pages = {91},
          doi = {10.3847/1538-4357/adc001},
archivePrefix = {arXiv},
       eprint = {2410.07377},
 primaryClass = {astro-ph.CO},
       adsurl = {https://ui.adsabs.harvard.edu/abs/2025ApJ...983...91P}
}

@ARTICLE{2026A&A...705A.114M,
       author = {{Mason}, Charlotte A. and {Chen}, Zuyi and {Stark}, Daniel P. and {Yi Lu}, Ting and {Topping}, Michael and {Tang}, Mengtao},
        title = "{Constraints on the z {\ensuremath{\sim}} 6{\ensuremath{-}}13 intergalactic medium from JWST spectroscopy of Lyman-alpha damping wings in galaxies}",
      journal = {\aap},
         year = 2026,
        month = jan,
       volume = {705},
          eid = {A114},
        pages = {A114},
          doi = {10.1051/0004-6361/202553820},
archivePrefix = {arXiv},
       eprint = {2501.11702},
 primaryClass = {astro-ph.GA},
       adsurl = {https://ui.adsabs.harvard.edu/abs/2026A&A...705A.114M}
}

@ARTICLE{2025ApJS..277...37U,
       author = {{Umeda}, Hiroya and {Ouchi}, Masami and {Kikuta}, Satoshi and {Harikane}, Yuichi and {Ono}, Yoshiaki and {Shibuya}, Takatoshi and {Inoue}, Akio K. and {Shimasaku}, Kazuhiro and {Liang}, Yongming and {Matsumoto}, Akinori and {Saito}, Shun and {Kusakabe}, Haruka and {Kageura}, Yuta and {Nakane}, Minami},
        title = "{SILVERRUSH. XIV. Ly{\ensuremath{\alpha}} Luminosity Functions and Angular Correlation Functions from 20,000 Ly{\ensuremath{\alpha}} Emitters at z {\ensuremath{\sim}} 2.2─7.3 from up to 24 deg$^{2}$ HSC-SSP and CHORUS Surveys: Linking the Postreionization Epoch to the Heart of Reionization}",
      journal = {\apjs},
         year = 2025,
        month = apr,
       volume = {277},
       number = {2},
          eid = {37},
        pages = {37},
          doi = {10.3847/1538-4365/adb1c0},
archivePrefix = {arXiv},
       eprint = {2411.15495},
 primaryClass = {astro-ph.GA},
       adsurl = {https://ui.adsabs.harvard.edu/abs/2025ApJS..277...37U}
}

@ARTICLE{2023ApJ...954L..14H,
       author = {{Hayes}, Matthew J. and {Scarlata}, Claudia},
        title = "{On the Sizes of Ionized Bubbles Around Galaxies During the Reionization Epoch. The Spectral Shapes of the Ly{\ensuremath{\alpha}} Emission from Galaxies}",
      journal = {\apjl},
         year = 2023,
        month = sep,
       volume = {954},
       number = {1},
          eid = {L14},
        pages = {L14},
          doi = {10.3847/2041-8213/acee6a},
archivePrefix = {arXiv},
       eprint = {2303.03160},
 primaryClass = {astro-ph.GA},
       adsurl = {https://ui.adsabs.harvard.edu/abs/2023ApJ...954L..14H}
}

@ARTICLE{2022arXiv221209850J,
       author = {{Jung}, Intae and {Finkelstein}, Steven L. and {Larson}, Rebecca L. and {Hutchison}, Taylor A. and {Straughn}, Amber N. and {Bagley}, Micaela B. and {Castellano}, Marco and {Cleri}, Nikko J. and {Cooper}, M.~C. and {Dickinson}, Mark and {Ferguson}, Henry C. and {Holwerda}, Benne W. and {Kartaltepe}, Jeyhan S. and {Kim}, Seonwoo and {Koekemoer}, Anton M. and {Papovich}, Casey and {Park}, Hyunbae and {Pentericci}, Laura and {Perez-Gonzalez}, Pablo G. and {Song}, Mimi and {Tacchella}, Sandro and {Weiner}, Benjamin J. and {Willmer}, Christopher N.~A. and {Zavala}, Jorge A.},
        title = "{New $z > 7$ Lyman-alpha Emitters in EGS: Evidence of an Extended Ionized Structure at $z \sim 7.7$}",
      journal = {arXiv e-prints},
         year = 2022,
        month = dec,
          eid = {arXiv:2212.09850},
        pages = {arXiv:2212.09850},
          doi = {10.48550/arXiv.2212.09850},
archivePrefix = {arXiv},
       eprint = {2212.09850},
 primaryClass = {astro-ph.GA},
       adsurl = {https://ui.adsabs.harvard.edu/abs/2022arXiv221209850J}
}

@ARTICLE{2023A&A...678A..68S,
       author = {{Saxena}, Aayush and {Robertson}, Brant E. and {Bunker}, Andrew J. and {Endsley}, Ryan and {Cameron}, Alex J. and {Charlot}, Stephane and {Simmonds}, Charlotte and {Tacchella}, Sandro and {Witstok}, Joris and {Willott}, Chris and {Carniani}, Stefano and {Curtis-Lake}, Emma and {Ferruit}, Pierre and {Jakobsen}, Peter and {Arribas}, Santiago and {Chevallard}, Jacopo and {Curti}, Mirko and {D'Eugenio}, Francesco and {De Graaff}, Anna and {Jones}, Gareth C. and {Looser}, Tobias J. and {Maseda}, Michael V. and {Rawle}, Tim and {Rix}, Hans-Walter and {Del Pino}, Bruno Rodr{\'\i}guez and {Smit}, Renske and {{\"U}bler}, Hannah and {Eisenstein}, Daniel J. and {Hainline}, Kevin and {Hausen}, Ryan and {Johnson}, Benjamin D. and {Rieke}, Marcia and {Williams}, Christina C. and {Willmer}, Christopher N.~A. and {Baker}, William M. and {Bhatawdekar}, Rachana and {Bowler}, Rebecca and {Boyett}, Kristan and {Chen}, Zuyi and {Egami}, Eiichi and {Ji}, Zhiyuan and {Kumari}, Nimisha and {Nelson}, Erica and {Perna}, Michele and {Sandles}, Lester and {Scholtz}, Jan and {Shivaei}, Irene},
        title = "{JADES: Discovery of extremely high equivalent width Lyman-{\ensuremath{\alpha}} emission from a faint galaxy within an ionized bubble at z = 7.3}",
      journal = {\aap},
         year = 2023,
        month = oct,
       volume = {678},
          eid = {A68},
        pages = {A68},
          doi = {10.1051/0004-6361/202346245},
archivePrefix = {arXiv},
       eprint = {2302.12805},
 primaryClass = {astro-ph.GA},
       adsurl = {https://ui.adsabs.harvard.edu/abs/2023A&A...678A..68S}
}

@ARTICLE{2026ApJ..1002...93Z,
       author = {{Zhu}, Yongda and {Becker}, George D. and {D'Aloisio}, Anson and {Endsley}, Ryan and {Gangolli}, Nakul and {Cain}, Christopher and {Mason}, Charlotte A. and {Hashemi}, Seyedazim and {Hong}, Hui},
        title = "{Galaxy Underdensities Host the Clearest Intergalactic Medium Ly{\ensuremath{\alpha}} Transmission and Indicate Anisotropic Reionization}",
      journal = {\apj},
         year = 2026,
        month = may,
       volume = {1002},
       number = {1},
          eid = {93},
        pages = {93},
          doi = {10.3847/1538-4357/ae5bbb},
archivePrefix = {arXiv},
       eprint = {2510.09568},
 primaryClass = {astro-ph.GA},
       adsurl = {https://ui.adsabs.harvard.edu/abs/2026ApJ..1002...93Z}
}

@ARTICLE{2022ApJ...931..126P,
       author = {{Park}, Hyunbae and {Kim}, Hyo Jeong and {Ahn}, Kyungjin and {Song}, Hyunmi and {Jung}, Intae and {Ocvirk}, Pierre and {Shapiro}, Paul R. and {Dawoodbhoy}, Taha and {Sorce}, Jenny G. and {Iliev}, Ilian T.},
        title = "{Scattering of Ly{\ensuremath{\alpha}} Photons through the Reionizing Intergalactic Medium: I. Spectral Energy Distribution}",
      journal = {\apj},
         year = 2022,
        month = jun,
       volume = {931},
       number = {2},
          eid = {126},
        pages = {126},
          doi = {10.3847/1538-4357/ac69e4},
archivePrefix = {arXiv},
       eprint = {2202.06277},
 primaryClass = {astro-ph.CO},
       adsurl = {https://ui.adsabs.harvard.edu/abs/2022ApJ...931..126P}
}

@ARTICLE{2023MNRAS.523.3749B,
       author = {{Blaizot}, J{\'e}r{\'e}my and {Garel}, Thibault and {Verhamme}, Anne and {Katz}, Harley and {Kimm}, Taysun and {Michel-Dansac}, L{\'e}o and {Mitchell}, Peter D. and {Rosdahl}, Joakim and {Trebitsch}, Maxime},
        title = "{Simulating the diversity of shapes of the Lyman-{\ensuremath{\alpha}} line}",
      journal = {\mnras},
         year = 2023,
        month = aug,
       volume = {523},
       number = {3},
        pages = {3749-3772},
          doi = {10.1093/mnras/stad1523},
archivePrefix = {arXiv},
       eprint = {2305.10047},
 primaryClass = {astro-ph.GA},
       adsurl = {https://ui.adsabs.harvard.edu/abs/2023MNRAS.523.3749B}
}

@ARTICLE{2023MNRAS.525.4093G,
       author = {{Gaikwad}, Prakash and {Haehnelt}, Martin G. and {Davies}, Fredrick B. and {Bosman}, Sarah E.~I. and {Molaro}, Margherita and {Kulkarni}, Girish and {D'Odorico}, Valentina and {Becker}, George D. and {Davies}, Rebecca L. and {Nasir}, Fahad and {Bolton}, James S. and {Keating}, Laura C. and {Ir{\v{s}}i{\v{c}}}, Vid and {Puchwein}, Ewald and {Zhu}, Yongda and {Asthana}, Shikhar and {Yang}, Jinyi and {Lai}, Samuel and {Eilers}, Anna-Christina},
        title = "{Measuring the photoionization rate, neutral fraction, and mean free path of H I ionizing photons at 4.9 {\ensuremath{\leq}} z {\ensuremath{\leq}} 6.0 from a large sample of XShooter and ESI spectra}",
      journal = {\mnras},
         year = 2023,
        month = nov,
       volume = {525},
       number = {3},
        pages = {4093-4120},
          doi = {10.1093/mnras/stad2566},
archivePrefix = {arXiv},
       eprint = {2304.02038},
 primaryClass = {astro-ph.CO},
       adsurl = {https://ui.adsabs.harvard.edu/abs/2023MNRAS.525.4093G}
}

@ARTICLE{2024ApJ...975..115Z,
       author = {{Zhu}, Hanjue and {Gnedin}, Nickolay Y. and {Avestruz}, Camille},
        title = "{On the Physical Nature of Ly{\ensuremath{\alpha}} Transmission Spikes in High-redshift Quasar Spectra}",
      journal = {\apj},
         year = 2024,
        month = nov,
       volume = {975},
       number = {1},
          eid = {115},
        pages = {115},
          doi = {10.3847/1538-4357/ad793c},
archivePrefix = {arXiv},
       eprint = {2401.04762},
 primaryClass = {astro-ph.CO},
       adsurl = {https://ui.adsabs.harvard.edu/abs/2024ApJ...975..115Z}
}

@ARTICLE{2013MNRAS.436.1023B,
       author = {{Becker}, George D. and {Bolton}, James S.},
        title = "{New measurements of the ionizing ultraviolet background over 2 < z < 5 and implications for hydrogen reionization}",
      journal = {\mnras},
         year = 2013,
        month = dec,
       volume = {436},
       number = {2},
        pages = {1023-1039},
          doi = {10.1093/mnras/stt1610},
archivePrefix = {arXiv},
       eprint = {1307.2259},
 primaryClass = {astro-ph.CO},
       adsurl = {https://ui.adsabs.harvard.edu/abs/2013MNRAS.436.1023B}
}

@ARTICLE{2002AJ....123.1247F,
       author = {{Fan}, Xiaohui and {Narayanan}, Vijay K. and {Strauss}, Michael A. and {White}, Richard L. and {Becker}, Robert H. and {Pentericci}, Laura and {Rix}, Hans-Walter},
        title = "{Evolution of the Ionizing Background and the Epoch of Reionization from the Spectra of z\raisebox{-0.5ex}\textasciitilde6 Quasars}",
      journal = {\aj},
         year = 2002,
        month = mar,
       volume = {123},
       number = {3},
        pages = {1247-1257},
          doi = {10.1086/339030},
archivePrefix = {arXiv},
       eprint = {astro-ph/0111184},
 primaryClass = {astro-ph},
       adsurl = {https://ui.adsabs.harvard.edu/abs/2002AJ....123.1247F}
}

@ARTICLE{2018MNRAS.473..560D,
       author = {{D'Aloisio}, Anson and {McQuinn}, Matthew and {Davies}, Frederick B. and {Furlanetto}, Steven R.},
        title = "{Large fluctuations in the high-redshift metagalactic ionizing background}",
      journal = {\mnras},
         year = 2018,
        month = jan,
       volume = {473},
       number = {1},
        pages = {560-575},
          doi = {10.1093/mnras/stx2341},
archivePrefix = {arXiv},
       eprint = {1611.02711},
 primaryClass = {astro-ph.CO},
       adsurl = {https://ui.adsabs.harvard.edu/abs/2018MNRAS.473..560D}
}

@ARTICLE{2019A&A...622A.142D,
       author = {{Deparis}, Nicolas and {Aubert}, Dominique and {Ocvirk}, Pierre and {Chardin}, Jonathan and {Lewis}, Joseph},
        title = "{Impact of the reduced speed of light approximation on ionization front velocities in cosmological simulations of the epoch of reionization}",
      journal = {\aap},
         year = 2019,
        month = feb,
       volume = {622},
          eid = {A142},
        pages = {A142},
          doi = {10.1051/0004-6361/201832889},
archivePrefix = {arXiv},
       eprint = {1803.01634},
 primaryClass = {astro-ph.CO},
       adsurl = {https://ui.adsabs.harvard.edu/abs/2019A&A...622A.142D}
}

@ARTICLE{2019A&A...626A..77O,
       author = {{Ocvirk}, P. and {Aubert}, D. and {Chardin}, J. and {Deparis}, N. and {Lewis}, J.},
        title = "{Impact of the reduced speed of light approximation on the post-overlap neutral hydrogen fraction in numerical simulations of the epoch of reionization}",
      journal = {\aap},
         year = 2019,
        month = jun,
       volume = {626},
          eid = {A77},
        pages = {A77},
          doi = {10.1051/0004-6361/201832923},
archivePrefix = {arXiv},
       eprint = {1803.02434},
 primaryClass = {astro-ph.CO},
       adsurl = {https://ui.adsabs.harvard.edu/abs/2019A&A...626A..77O}
}

@ARTICLE{2026arXiv260500763M,
       author = {{Meyer}, Romain A. and {Oesch}, Pascal A. and {Witten}, Callum and {Elllis}, Richard S. and {Bosman}, Sarah E.~I. and {Davies}, Fred and {Drake}, Alyssa B. and {Laporte}, Nicolas and {Matthee}, Jorryt and {Walter}, Fabian},
        title = "{Life After the Quasar: Overmassive Black Holes and Remnant Ionised Bubbles in and Around Two z\raisebox{-0.5ex}\textasciitilde6.6 Galaxies}",
      journal = {arXiv e-prints},
         year = 2026,
        month = may,
          eid = {arXiv:2605.00763},
        pages = {arXiv:2605.00763},
          doi = {10.48550/arXiv.2605.00763},
archivePrefix = {arXiv},
       eprint = {2605.00763},
 primaryClass = {astro-ph.GA},
       adsurl = {https://ui.adsabs.harvard.edu/abs/2026arXiv260500763M}
}

@ARTICLE{2023MNRAS.519.5987L,
       author = {{Lewis}, Joseph S.~W. and {Ocvirk}, Pierre and {Dubois}, Yohan and {Aubert}, Dominique and {Chardin}, Jonathan and {Gillet}, Nicolas and {Th{\'e}lie}, {\'E}milie},
        title = "{DUSTiER (DUST in the Epoch of Reionization): dusty galaxies in cosmological radiation-hydrodynamical simulations of the Epoch of Reionization with RAMSES-CUDATON}",
      journal = {\mnras},
         year = 2023,
        month = mar,
       volume = {519},
       number = {4},
        pages = {5987-6007},
          doi = {10.1093/mnras/stad081},
archivePrefix = {arXiv},
       eprint = {2204.03949},
 primaryClass = {astro-ph.GA},
       adsurl = {https://ui.adsabs.harvard.edu/abs/2023MNRAS.519.5987L}
}

@ARTICLE{2013MNRAS.433.1230W,
       author = {{Watson}, William A. and {Iliev}, Ilian T. and {D'Aloisio}, Anson and {Knebe}, Alexander and {Shapiro}, Paul R. and {Yepes}, Gustavo},
        title = "{The halo mass function through the cosmic ages}",
      journal = {\mnras},
         year = 2013,
        month = aug,
       volume = {433},
       number = {2},
        pages = {1230-1245},
          doi = {10.1093/mnras/stt791},
archivePrefix = {arXiv},
       eprint = {1212.0095},
 primaryClass = {astro-ph.CO},
       adsurl = {https://ui.adsabs.harvard.edu/abs/2013MNRAS.433.1230W}
}

@inproceedings{10.1117/12.3019047,
author = {Roser Pell{\'o} and Mathieu Puech and {\'E}ric Prieto and Myriam Rodrigues and Rub{\'e}n Sanchez-Janssen and Gavin B. Dalton and Franck Ducret and Kacem El Hadi and Mar{\'i}a L. Garc{\'i}a-Vargas and Jeff Lynn and Nazim A. Bharmal and Diane Chapuis and Michel Dupieux and Cl{\'e}ment Hottier and Marie Larrieu and Laurent Martin and Meghna Mohamed and Tim Morris and Ana P{\'e}rez and Walter Seifert and Wenli Xu and Simon Morris and Lex Kaper and Jes{\'u}s Gallego and Jose Afonso and Beatriz Barbuy and Thierry Contini and Alexis Finoguenov and Susan Kassin and Christopher Miller and G{\"o}ran Ostlin and Laura Pentericci and Daniel Schaerer and Matthias Steinmetz and Bodo Ziegler and Ricardo Araujo and Joar Brynnel and Bruno Castilho and Christopher J. Conselice and Nick Cvetojevic and Christopher Davison and Julien Dejonghe and Mirka Dessauges-Zavadsky and Kjetil Dohlen and D{\'e}cio Ferreira and Armando Gil de Paz and Thiago S. Gon{\c{c}}alves and Isabelle Guinouard and Matthew J. Hayes and Derek Ives and Annemieke Janssen and Carol Kehrig and Andreas Kelz and Davor Krajnović and Audrey A. Lanotte and Nicolas Laporte and Philippe Laporte and Soren Larsen and Bertran Lemasle and Ian Lewis and Jiang-Tao Li and Elena Pancino and Matthew M. Pieri and Christian Surace and Markus Thurneysen and Susanna Vergani and Fran{\c{c}}ois Wildi and Fernando {\'A}lvarez Moreno and Raziye Artan and Mathilde Beaulieu and Eva Besada and Arjan Bik and Charlotte Bond and Mohamed Bouri and J{\'e}r{\'e}mie Boy and David Bramall and Sarah Brands and Antonio Braulio and Tim Butterley and Cristina Cabello and Marina Calero de Ory and Rocio Calvo and Africa Castillo Morales and Zalpha Challita and Stephen Chittik and Andr{\'e}s Curto Maldonado and F{\'a}tima De Frontat and Elfi Dijkstra and Eddy Elswijk and Gilles Fasola and Carmen Feiz and Fabio Fialho and Johan Floriot and Paolo Franzetti and Marco Fumana and Luis Gabarra and Lia Garcia and Adriana Gargiulo and Julien Gaudemard and Domenico Giannone and Polly Gill and Alicia Gomez-Gutierrez and Carole Gouvret and Alan Guenther and Douglas Harvey and Jos{\'e} Miguel Ib{\'a}{\~n}ez Mengual and Jorge Iglesias and Yevgeniy Ivanisenko and Peter Kunst and Sander Kwast and Kieran Leschinski and Gianluca Licausi and Sebastiano Ligori and Juan Antonio L{\'o}pez Orozco and Adam Lowe and Mike Macintosh and H{\'e}ctor Magan and Manuel Maldonado and Thomas Marquart and Lucimara Martins and Marisole Melara and Jens Melinder and Jeannet Molema and David Montgomery and Mar{\'i}a Morales and Francisco Najarro and Nicolas Nardetto and Ramon Navarro and Antoine Ottomani and Tony Pamplona and Cyril Pannetier and Phil Parr-Burman and Sergio Pascual and Mar{\'i}a Pe{\~n}ataro and Isabel P{\'e}rez Grande and Jan Rinze Peterzon and Javier Piqueras and Nicolai Piskunov and Ramon Rodriguez Cardoso and Sergio Rodriguez Venzal and Rick Romp and Hossein Rostami and Fr{\'e}d{\'e}ric Royer and Daniel Sablowski and Ainhoa Sanchez and Ernesto S{\'a}nchez Blanco and Ellen Schalling and Jurgen Schmoll and Noah Schwartz and Jay Stephan and Sylvestre Taburet and David Terrett and Ignacio Torralbo and Niels Tromp and Gerardo Veredas and Yanbin Yang and Alec York and Werner Zeilinger},
title = {{MOSAIC at the ELT: a unique instrument for the largest ground-based telescope}},
volume = {13096},
booktitle = {Ground-based and Airborne Instrumentation for Astronomy X},
editor = {Julia J. Bryant and Kentaro Motohara and Jo{\"e}l R. D. Vernet},
organization = {International Society for Optics and Photonics},
publisher = {SPIE},
pages = {1309615},
year = {2024},
doi = {10.1117/12.3019047},
URL = {https://doi.org/10.1117/12.3019047}
}

@inproceedings{10.1117/12.2313940,
author = {D. L. DePoy and Luke M. Schmidt and Rafael Ribeiro and Keith Taylor and Damien Jones and Travis Prochaska and Jennifer L. Marshall and Erika Cook and Cynthia Froning and Tae-Geun Ji and Hye-In Lee and D. M. Faes and A. Souza and D. Bortoletto and Claudia Mendes de Oliveira and Soojong Pak and Casey Papovich},
title = {{GMACS: a wide-field, moderate-resolution spectrograph for the Giant Magellan Telescope}},
volume = {10702},
booktitle = {Ground-based and Airborne Instrumentation for Astronomy VII},
editor = {Christopher J. Evans and Luc Simard and Hideki Takami},
organization = {International Society for Optics and Photonics},
publisher = {SPIE},
pages = {107021X},
year = {2018},
doi = {10.1117/12.2313940},
URL = {https://doi.org/10.1117/12.2313940}
}

@ARTICLE{2011A&A...536A.105V,
       author = {{Vernet}, J. and {Dekker}, H. and {D'Odorico}, S. and {Kaper}, L. and {Kjaergaard}, P. and {Hammer}, F. and {Randich}, S. and {Zerbi}, F. and {Groot}, P.~J. and {Hjorth}, J. and {Guinouard}, I. and {Navarro}, R. and {Adolfse}, T. and {Albers}, P.~W. and {Amans}, J.-P. and {Andersen}, J.~J. and {Andersen}, M.~I. and {Binetruy}, P. and {Bristow}, P. and {Castillo}, R. and {Chemla}, F. and {Christensen}, L. and {Conconi}, P. and {Conzelmann}, R. and {Dam}, J. and {de Caprio}, V. and {de Ugarte Postigo}, A. and {Delabre}, B. and {di Marcantonio}, P. and {Downing}, M. and {Elswijk}, E. and {Finger}, G. and {Fischer}, G. and {Flores}, H. and {Fran{\c{c}}ois}, P. and {Goldoni}, P. and {Guglielmi}, L. and {Haigron}, R. and {Hanenburg}, H. and {Hendriks}, I. and {Horrobin}, M. and {Horville}, D. and {Jessen}, N.~C. and {Kerber}, F. and {Kern}, L. and {Kiekebusch}, M. and {Kleszcz}, P. and {Klougart}, J. and {Kragt}, J. and {Larsen}, H.~H. and {Lizon}, J.-L. and {Lucuix}, C. and {Mainieri}, V. and {Manuputy}, R. and {Martayan}, C. and {Mason}, E. and {Mazzoleni}, R. and {Michaelsen}, N. and {Modigliani}, A. and {Moehler}, S. and {M{\o}ller}, P. and {Norup S{\o}rensen}, A. and {N{\o}rregaard}, P. and {P{\'e}roux}, C. and {Patat}, F. and {Pena}, E. and {Pragt}, J. and {Reinero}, C. and {Rigal}, F. and {Riva}, M. and {Roelfsema}, R. and {Royer}, F. and {Sacco}, G. and {Santin}, P. and {Schoenmaker}, T. and {Spano}, P. and {Sweers}, E. and {Ter Horst}, R. and {Tintori}, M. and {Tromp}, N. and {van Dael}, P. and {van der Vliet}, H. and {Venema}, L. and {Vidali}, M. and {Vinther}, J. and {Vola}, P. and {Winters}, R. and {Wistisen}, D. and {Wulterkens}, G. and {Zacchei}, A.},
        title = "{X-shooter, the new wide band intermediate resolution spectrograph at the ESO Very Large Telescope}",
      journal = {\aap},
         year = 2011,
        month = dec,
       volume = {536},
          eid = {A105},
        pages = {A105},
          doi = {10.1051/0004-6361/201117752},
archivePrefix = {arXiv},
       eprint = {1110.1944},
 primaryClass = {astro-ph.IM},
       adsurl = {https://ui.adsabs.harvard.edu/abs/2011A&A...536A.105V}
}

@ARTICLE{2019MNRAS.485.1350W,
       author = {{Weinberger}, Lewis H. and {Haehnelt}, Martin G. and {Kulkarni}, Girish},
        title = "{Modelling the observed luminosity function and clustering evolution of Ly {\ensuremath{\alpha}} emitters: growing evidence for late reionization}",
      journal = {\mnras},
         year = 2019,
        month = may,
       volume = {485},
       number = {1},
        pages = {1350-1366},
          doi = {10.1093/mnras/stz481},
archivePrefix = {arXiv},
       eprint = {1902.05077},
 primaryClass = {astro-ph.GA},
       adsurl = {https://ui.adsabs.harvard.edu/abs/2019MNRAS.485.1350W}
}

@ARTICLE{2016MNRAS.460.1328D,
       author = {{Davies}, Frederick B. and {Furlanetto}, Steven R.},
        title = "{Large fluctuations in the hydrogen-ionizing background and mean free path following the epoch of reionization}",
      journal = {\mnras},
         year = 2016,
        month = aug,
       volume = {460},
       number = {2},
        pages = {1328-1339},
          doi = {10.1093/mnras/stw931},
archivePrefix = {arXiv},
       eprint = {1509.07131},
 primaryClass = {astro-ph.CO},
       adsurl = {https://ui.adsabs.harvard.edu/abs/2016MNRAS.460.1328D}
}

@ARTICLE{1998ApJ...501...15M,
       author = {{Miralda-Escud{\'e}}, Jordi},
        title = "{Reionization of the Intergalactic Medium and the Damping Wing of the Gunn-Peterson Trough}",
      journal = {\apj},
         year = 1998,
        month = jul,
       volume = {501},
       number = {1},
        pages = {15-22},
          doi = {10.1086/305799},
archivePrefix = {arXiv},
       eprint = {astro-ph/9708253},
 primaryClass = {astro-ph},
       adsurl = {https://ui.adsabs.harvard.edu/abs/1998ApJ...501...15M}
}

@ARTICLE{2011ApJ...728...52L,
       author = {{Laursen}, Peter and {Sommer-Larsen}, Jesper and {Razoumov}, Alexei O.},
        title = "{Intergalactic Transmission and Its Impact on the Ly{\ensuremath{\alpha}} Line}",
      journal = {\apj},
         year = 2011,
        month = feb,
       volume = {728},
       number = {1},
          eid = {52},
        pages = {52},
          doi = {10.1088/0004-637X/728/1/52},
archivePrefix = {arXiv},
       eprint = {1009.1384},
 primaryClass = {astro-ph.CO},
       adsurl = {https://ui.adsabs.harvard.edu/abs/2011ApJ...728...52L}
}

@ARTICLE{2016ApJ...820..130Y,
       author = {{Yang}, Huan and {Malhotra}, Sangeeta and {Gronke}, Max and {Rhoads}, James E. and {Dijkstra}, Mark and {Jaskot}, Anne and {Zheng}, Zhenya and {Wang}, Junxian},
        title = "{Green Pea Galaxies Reveal Secrets of Ly{\ensuremath{\alpha}} Escape}",
      journal = {\apj},
         year = 2016,
        month = apr,
       volume = {820},
       number = {2},
          eid = {130},
        pages = {130},
          doi = {10.3847/0004-637X/820/2/130},
archivePrefix = {arXiv},
       eprint = {1506.02885},
 primaryClass = {astro-ph.GA},
       adsurl = {https://ui.adsabs.harvard.edu/abs/2016ApJ...820..130Y}
}

@ARTICLE{2022ARA&A..60..455E,
       author = {{Eldridge}, Jan J. and {Stanway}, Elizabeth R.},
        title = "{New Insights into the Evolution of Massive Stars and Their Effects on Our Understanding of Early Galaxies}",
      journal = {\araa},
         year = 2022,
        month = aug,
       volume = {60},
        pages = {455-494},
          doi = {10.1146/annurev-astro-052920-100646},
archivePrefix = {arXiv},
       eprint = {2202.01413},
 primaryClass = {astro-ph.GA},
       adsurl = {https://ui.adsabs.harvard.edu/abs/2022ARA&A..60..455E}
}

@ARTICLE{2020A&A...642L..16B,
       author = {{Byrohl}, C. and {Gronke}, M.},
        title = "{Variations in shape among observed Lyman-{\ensuremath{\alpha}} spectra due to intergalactic absorption}",
      journal = {\aap},
         year = 2020,
        month = oct,
       volume = {642},
          eid = {L16},
        pages = {L16},
          doi = {10.1051/0004-6361/202038685},
archivePrefix = {arXiv},
       eprint = {2006.10041},
 primaryClass = {astro-ph.GA},
       adsurl = {https://ui.adsabs.harvard.edu/abs/2020A&A...642L..16B}
}

@ARTICLE{2013MNRAS.428.1366J,
       author = {{Jensen}, Hannes and {Laursen}, Peter and {Mellema}, Garrelt and {Iliev}, Ilian T. and {Sommer-Larsen}, Jesper and {Shapiro}, Paul R.},
        title = "{On the use of Ly{\ensuremath{\alpha}} emitters as probes of reionization}",
      journal = {\mnras},
         year = 2013,
        month = jan,
       volume = {428},
       number = {2},
        pages = {1366-1381},
          doi = {10.1093/mnras/sts116},
archivePrefix = {arXiv},
       eprint = {1206.4028},
 primaryClass = {astro-ph.CO},
       adsurl = {https://ui.adsabs.harvard.edu/abs/2013MNRAS.428.1366J}
}

@ARTICLE{2023MNRAS.521.4356X,
       author = {{Xu}, Clara and {Smith}, Aaron and {Borrow}, Josh and {Garaldi}, Enrico and {Kannan}, Rahul and {Vogelsberger}, Mark and {Pakmor}, R{\"u}diger and {Springel}, Volker and {Hernquist}, Lars},
        title = "{The THESAN project: Lyman-{\ensuremath{\alpha}} emitter luminosity function calibration}",
      journal = {\mnras},
         year = 2023,
        month = may,
       volume = {521},
       number = {3},
        pages = {4356-4374},
          doi = {10.1093/mnras/stad789},
archivePrefix = {arXiv},
       eprint = {2210.16275},
 primaryClass = {astro-ph.GA},
       adsurl = {https://ui.adsabs.harvard.edu/abs/2023MNRAS.521.4356X}
}

@ARTICLE{2026OJAp....960756N,
       author = {{Neyer}, Meredith and {Smith}, Aaron and {Vogelsberger}, Mark and {Garc{\'\i}a}, Luz {\'A}ngela and {Kannan}, Rahul and {Garaldi}, Enrico and {Keating}, Laura},
        title = "{The THESAN project: Lyman-alpha emitters as probes of ionized bubble sizes}",
      journal = {The Open Journal of Astrophysics},
         year = 2026,
        month = apr,
       volume = {9},
        pages = {60756},
          doi = {10.33232/001c.160756},
       adsurl = {https://ui.adsabs.harvard.edu/abs/2026OJAp....960756N}
}

@ARTICLE{2020ApJ...904..144J,
       author = {{Jung}, Intae and {Finkelstein}, Steven L. and {Dickinson}, Mark and {Hutchison}, Taylor A. and {Larson}, Rebecca L. and {Papovich}, Casey and {Pentericci}, Laura and {Straughn}, Amber N. and {Guo}, Yicheng and {Malhotra}, Sangeeta and {Rhoads}, James and {Song}, Mimi and {Tilvi}, Vithal and {Wold}, Isak},
        title = "{Texas Spectroscopic Search for Ly{\ensuremath{\alpha}} Emission at the End of Reionization. III. The Ly{\ensuremath{\alpha}} Equivalent-width Distribution and Ionized Structures at z > 7}",
      journal = {\apj},
         year = 2020,
        month = dec,
       volume = {904},
       number = {2},
          eid = {144},
        pages = {144},
          doi = {10.3847/1538-4357/abbd44},
archivePrefix = {arXiv},
       eprint = {2009.10092},
 primaryClass = {astro-ph.GA},
       adsurl = {https://ui.adsabs.harvard.edu/abs/2020ApJ...904..144J}
}

@ARTICLE{2022ApJ...933...87J,
       author = {{Jung}, Intae and {Papovich}, Casey and {Finkelstein}, Steven L. and {Simons}, Raymond C. and {Estrada-Carpenter}, Vicente and {Backhaus}, Bren E. and {Cleri}, Nikko J. and {Finlator}, Kristian and {Giavalisco}, Mauro and {Ji}, Zhiyuan and {Matharu}, Jasleen and {Momcheva}, Ivelina and {Straughn}, Amber N. and {Trump}, Jonathan R.},
        title = "{CLEAR: Boosted Ly{\ensuremath{\alpha}} Transmission of the Intergalactic Medium in UV-bright Galaxies}",
      journal = {\apj},
         year = 2022,
        month = jul,
       volume = {933},
       number = {1},
          eid = {87},
        pages = {87},
          doi = {10.3847/1538-4357/ac6fe7},
archivePrefix = {arXiv},
       eprint = {2111.14863},
 primaryClass = {astro-ph.GA},
       adsurl = {https://ui.adsabs.harvard.edu/abs/2022ApJ...933...87J}
}

@ARTICLE{2004MNRAS.349.1137S,
       author = {{Santos}, Michael R.},
        title = "{Probing reionization with Lyman {\ensuremath{\alpha}} emission lines}",
      journal = {\mnras},
         year = 2004,
        month = apr,
       volume = {349},
       number = {3},
        pages = {1137-1152},
          doi = {10.1111/j.1365-2966.2004.07594.x},
archivePrefix = {arXiv},
       eprint = {astro-ph/0308196},
 primaryClass = {astro-ph},
       adsurl = {https://ui.adsabs.harvard.edu/abs/2004MNRAS.349.1137S}
}

@ARTICLE{2021MNRAS.504.1902G,
       author = {{Garel}, Thibault and {Blaizot}, J{\'e}r{\'e}my and {Rosdahl}, Joakim and {Michel-Dansac}, L{\'e}o and {Haehnelt}, Martin G. and {Katz}, Harley and {Kimm}, Taysun and {Verhamme}, Anne},
        title = "{Ly {\ensuremath{\alpha}} as a tracer of cosmic reionization in the SPHINX radiation-hydrodynamics cosmological simulation}",
      journal = {\mnras},
         year = 2021,
        month = jun,
       volume = {504},
       number = {2},
        pages = {1902-1926},
          doi = {10.1093/mnras/stab990},
archivePrefix = {arXiv},
       eprint = {2104.03339},
 primaryClass = {astro-ph.GA},
       adsurl = {https://ui.adsabs.harvard.edu/abs/2021MNRAS.504.1902G}
}

@ARTICLE{2014PASA...31...40D,
       author = {{Dijkstra}, Mark},
        title = "{Ly{\ensuremath{\alpha}} Emitting Galaxies as a Probe of Reionisation}",
      journal = {\pasa},
         year = 2014,
        month = oct,
       volume = {31},
          eid = {e040},
        pages = {e040},
          doi = {10.1017/pasa.2014.33},
archivePrefix = {arXiv},
       eprint = {1406.7292},
 primaryClass = {astro-ph.CO},
       adsurl = {https://ui.adsabs.harvard.edu/abs/2014PASA...31...40D}
}

@ARTICLE{2025A&A...703A..98O,
       author = {{Ocvirk}, Pierre and {Lewis}, Joseph S.~W. and {Conaboy}, Luke and {Dubois}, Yohan and {Bethermin}, Matthieu and {Sorce}, Jenny G. and {Aubert}, Dominique and {Shapiro}, Paul R. and {Dawoodbhoy}, Taha and {Lee}, Joohyun and {Teyssier}, Romain and {Yepes}, Gustavo and {Gottl{\"o}ber}, Stefan and {Iliev}, Ilian T. and {Ahn}, Kyungjin and {Park}, Hyunbae and {Palanque}, Mei},
        title = "{Dust-UV offsets in high-redshift galaxies in the Cosmic Dawn III simulation}",
      journal = {\aap},
         year = 2025,
        month = nov,
       volume = {703},
          eid = {A98},
        pages = {A98},
          doi = {10.1051/0004-6361/202452098},
archivePrefix = {arXiv},
       eprint = {2409.05946},
 primaryClass = {astro-ph.GA},
       adsurl = {https://ui.adsabs.harvard.edu/abs/2025A&A...703A..98O}
}

@ARTICLE{2015PASA...32...45B,
       author = {{Becker}, George D. and {Bolton}, James S. and {Lidz}, Adam},
        title = "{Reionisation and High-Redshift Galaxies: The View from Quasar Absorption Lines}",
      journal = {\pasa},
         year = 2015,
        month = dec,
       volume = {32},
          eid = {e045},
        pages = {e045},
          doi = {10.1017/pasa.2015.45},
archivePrefix = {arXiv},
       eprint = {1510.03368},
 primaryClass = {astro-ph.CO},
       adsurl = {https://ui.adsabs.harvard.edu/abs/2015PASA...32...45B}
}

@ARTICLE{2018ApJ...864...53E,
       author = {{Eilers}, Anna-Christina and {Davies}, Frederick B. and {Hennawi}, Joseph F.},
        title = "{The Opacity of the Intergalactic Medium Measured along Quasar Sightlines at z {\ensuremath{\sim}} 6}",
      journal = {\apj},
         year = 2018,
        month = sep,
       volume = {864},
       number = {1},
          eid = {53},
        pages = {53},
          doi = {10.3847/1538-4357/aad4fd},
archivePrefix = {arXiv},
       eprint = {1807.04229},
 primaryClass = {astro-ph.GA},
       adsurl = {https://ui.adsabs.harvard.edu/abs/2018ApJ...864...53E}
}

@ARTICLE{2022ARA&A..60..121R,
       author = {{Robertson}, Brant E.},
        title = "{Galaxy Formation and Reionization: Key Unknowns and Expected Breakthroughs by the James Webb Space Telescope}",
      journal = {\araa},
         year = 2022,
        month = aug,
       volume = {60},
        pages = {121-158},
          doi = {10.1146/annurev-astro-120221-044656},
archivePrefix = {arXiv},
       eprint = {2110.13160},
 primaryClass = {astro-ph.CO},
       adsurl = {https://ui.adsabs.harvard.edu/abs/2022ARA&A..60..121R}
}

@ARTICLE{2025A&A...697A.211D,
       author = {{Dayal}, Pratika and {Volonteri}, Marta and {Greene}, Jenny E. and {Kokorev}, Vasily and {Goulding}, Andy D. and {Williams}, Christina C. and {Furtak}, Lukas J. and {Zitrin}, Adi and {Atek}, Hakim and {Bezanson}, Rachel and {Chemerynska}, Iryna and {Feldmann}, Robert and {Glazebrook}, Karl and {Labbe}, Ivo and {Nanayakkara}, Themiya and {Oesch}, Pascal A. and {Weaver}, John R.},
        title = "{UNCOVERing the contribution of black holes to reionization}",
      journal = {\aap},
         year = 2025,
        month = may,
       volume = {697},
          eid = {A211},
        pages = {A211},
          doi = {10.1051/0004-6361/202449331},
archivePrefix = {arXiv},
       eprint = {2401.11242},
 primaryClass = {astro-ph.GA},
       adsurl = {https://ui.adsabs.harvard.edu/abs/2025A&A...697A.211D}
}

@ARTICLE{2026PASA...43...21M,
       author = {{Mukherjee}, Tamal and {Zafar}, Tayyaba and {Nanayakkara}, Themiya and {Gurung-L{\'o}pez}, Siddhartha and {Gupta}, Anshu and {Croom}, Scott and {Battisti}, Andrew and {Glazebrook}, Karl and {Papaderos}, Polychronis and {Riggs}, Melissa and {Wisnioski}, Emily and {Foster}, Caroline and {Harborne}, Katherine and {Lagos}, Claudia and {Mendel}, Jon Trevor and {Prathap}, Jahang and {Barsanti}, Stefania and {Sweet}, Sarah and {Valenzuela}, Lucas and {Mailvaganam}, Anilkumar},
        title = "{A census of double-peaked Lyman-{\ensuremath{\alpha}} emitters in MAGPI: Classification, global characteristics, and spatially resolved properties}",
      journal = {\pasa},
         year = 2026,
        month = jan,
       volume = {43},
          eid = {e021},
        pages = {e021},
          doi = {10.1017/pasa.2026.10145},
archivePrefix = {arXiv},
       eprint = {2510.18398},
 primaryClass = {astro-ph.GA},
       adsurl = {https://ui.adsabs.harvard.edu/abs/2026PASA...43...21M}
}

@ARTICLE{2024A&A...689A..44T,
       author = {{Torralba}, Alberto and {Matthee}, Jorryt and {Naidu}, Rohan P. and {Mackenzie}, Ruari and {Pezzulli}, Gabriele and {Hutter}, Anne and {Arnalte-Mur}, Pablo and {Gurung-L{\'o}pez}, Siddhartha and {Tacchella}, Sandro and {Oesch}, Pascal and {Kashino}, Daichi and {Conroy}, Charlie and {Sobral}, David},
        title = "{Anatomy of an ionized bubble: NIRCam grism spectroscopy of the z = 6.6 double-peaked Lyman-{\ensuremath{\alpha}} emitter COLA1 and its environment}",
      journal = {\aap},
         year = 2024,
        month = sep,
       volume = {689},
          eid = {A44},
        pages = {A44},
          doi = {10.1051/0004-6361/202450318},
archivePrefix = {arXiv},
       eprint = {2404.10040},
 primaryClass = {astro-ph.GA},
       adsurl = {https://ui.adsabs.harvard.edu/abs/2024A&A...689A..44T}
}

@ARTICLE{2015MNRAS.446..566M,
       author = {{Mesinger}, Andrei and {Aykutalp}, Aycin and {Vanzella}, Eros and {Pentericci}, Laura and {Ferrara}, Andrea and {Dijkstra}, Mark},
        title = "{Can the intergalactic medium cause a rapid drop in Ly{\ensuremath{\alpha}} emission at z > 6?}",
      journal = {\mnras},
         year = 2015,
        month = jan,
       volume = {446},
       number = {1},
        pages = {566-577},
          doi = {10.1093/mnras/stu2089},
archivePrefix = {arXiv},
       eprint = {1406.6373},
 primaryClass = {astro-ph.CO},
       adsurl = {https://ui.adsabs.harvard.edu/abs/2015MNRAS.446..566M}
}

@ARTICLE{2012MNRAS.423.2222I,
       author = {{Iliev}, Ilian T. and {Mellema}, Garrelt and {Shapiro}, Paul R. and {Pen}, Ue-Li and {Mao}, Yi and {Koda}, Jun and {Ahn}, Kyungjin},
        title = "{Can 21-cm observations discriminate between high-mass and low-mass galaxies as reionization sources?}",
      journal = {\mnras},
         year = 2012,
        month = jul,
       volume = {423},
       number = {3},
        pages = {2222-2253},
          doi = {10.1111/j.1365-2966.2012.21032.x},
archivePrefix = {arXiv},
       eprint = {1107.4772},
 primaryClass = {astro-ph.CO},
       adsurl = {https://ui.adsabs.harvard.edu/abs/2012MNRAS.423.2222I}
}

@ARTICLE{2018MNRAS.473..227H,
       author = {{Hassan}, Sultan and {Dav{\'e}}, Romeel and {Mitra}, Sourav and {Finlator}, Kristian and {Ciardi}, Benedetta and {Santos}, Mario G.},
        title = "{Constraining the contribution of active galactic nuclei to reionization}",
      journal = {\mnras},
         year = 2018,
        month = jan,
       volume = {473},
       number = {1},
        pages = {227-240},
          doi = {10.1093/mnras/stx2194},
archivePrefix = {arXiv},
       eprint = {1705.05398},
 primaryClass = {astro-ph.CO},
       adsurl = {https://ui.adsabs.harvard.edu/abs/2018MNRAS.473..227H}
}

@ARTICLE{2024MNRAS.531.2943N,
       author = {{Neyer}, Meredith and {Smith}, Aaron and {Kannan}, Rahul and {Vogelsberger}, Mark and {Garaldi}, Enrico and {Gal{\'a}rraga-Espinosa}, Daniela and {Borrow}, Josh and {Hernquist}, Lars and {Pakmor}, R{\"u}diger and {Springel}, Volker},
        title = "{The THESAN project: connecting ionized bubble sizes to their local environments during the Epoch of Reionization}",
      journal = {\mnras},
         year = 2024,
        month = jul,
       volume = {531},
       number = {3},
        pages = {2943-2957},
          doi = {10.1093/mnras/stae1325},
archivePrefix = {arXiv},
       eprint = {2310.03783},
 primaryClass = {astro-ph.GA},
       adsurl = {https://ui.adsabs.harvard.edu/abs/2024MNRAS.531.2943N}
}

@ARTICLE{2004ApJ...613....1F,
       author = {{Furlanetto}, Steven R. and {Zaldarriaga}, Matias and {Hernquist}, Lars},
        title = "{The Growth of H II Regions During Reionization}",
      journal = {\apj},
         year = 2004,
        month = sep,
       volume = {613},
       number = {1},
        pages = {1-15},
          doi = {10.1086/423025},
archivePrefix = {arXiv},
       eprint = {astro-ph/0403697},
 primaryClass = {astro-ph},
       adsurl = {https://ui.adsabs.harvard.edu/abs/2004ApJ...613....1F}
}

@ARTICLE{2015MNRAS.447..499M,
       author = {{McGreer}, Ian D. and {Mesinger}, Andrei and {D'Odorico}, Valentina},
        title = "{Model-independent evidence in favour of an end to reionization by z {\ensuremath{\approx}} 6}",
      journal = {\mnras},
         year = 2015,
        month = feb,
       volume = {447},
       number = {1},
        pages = {499-505},
          doi = {10.1093/mnras/stu2449},
archivePrefix = {arXiv},
       eprint = {1411.5375},
 primaryClass = {astro-ph.CO},
       adsurl = {https://ui.adsabs.harvard.edu/abs/2015MNRAS.447..499M}
}

@ARTICLE{2018MNRAS.474.2904P,
       author = {{Parsa}, Shaghayegh and {Dunlop}, James S. and {McLure}, Ross J.},
        title = "{No evidence for a significant AGN contribution to cosmic hydrogen reionization}",
      journal = {\mnras},
         year = 2018,
        month = mar,
       volume = {474},
       number = {3},
        pages = {2904-2923},
          doi = {10.1093/mnras/stx2887},
archivePrefix = {arXiv},
       eprint = {1704.07750},
 primaryClass = {astro-ph.GA},
       adsurl = {https://ui.adsabs.harvard.edu/abs/2018MNRAS.474.2904P}
}

@ARTICLE{2020ApJ...892..109N,
       author = {{Naidu}, Rohan P. and {Tacchella}, Sandro and {Mason}, Charlotte A. and {Bose}, Sownak and {Oesch}, Pascal A. and {Conroy}, Charlie},
        title = "{Rapid Reionization by the Oligarchs: The Case for Massive, UV-bright, Star-forming Galaxies with High Escape Fractions}",
      journal = {\apj},
         year = 2020,
        month = apr,
       volume = {892},
       number = {2},
          eid = {109},
        pages = {109},
          doi = {10.3847/1538-4357/ab7cc9},
archivePrefix = {arXiv},
       eprint = {1907.13130},
 primaryClass = {astro-ph.GA},
       adsurl = {https://ui.adsabs.harvard.edu/abs/2020ApJ...892..109N}
}

@ARTICLE{2015ApJ...813L...8M,
       author = {{Madau}, Piero and {Haardt}, Francesco},
        title = "{Cosmic Reionization after Planck: Could Quasars Do It All?}",
      journal = {\apjl},
         year = 2015,
        month = nov,
       volume = {813},
       number = {1},
          eid = {L8},
        pages = {L8},
          doi = {10.1088/2041-8205/813/1/L8},
archivePrefix = {arXiv},
       eprint = {1507.07678},
 primaryClass = {astro-ph.CO},
       adsurl = {https://ui.adsabs.harvard.edu/abs/2015ApJ...813L...8M}
}

@ARTICLE{2010A&A...523A...4B,
       author = {{Baek}, S. and {Semelin}, B. and {Di Matteo}, P. and {Revaz}, Y. and {Combes}, F.},
        title = "{Reionization by UV or X-ray sources}",
      journal = {\aap},
         year = 2010,
        month = nov,
       volume = {523},
          eid = {A4},
        pages = {A4},
          doi = {10.1051/0004-6361/201014347},
archivePrefix = {arXiv},
       eprint = {1003.0834},
 primaryClass = {astro-ph.CO},
       adsurl = {https://ui.adsabs.harvard.edu/abs/2010A&A...523A...4B}
}

@ARTICLE{2020MNRAS.498.6083E,
       author = {{Eide}, Marius B. and {Ciardi}, Benedetta and {Graziani}, Luca and {Busch}, Philipp and {Feng}, Yu and {Di Matteo}, Tiziana},
        title = "{Large-scale simulations of H and He reionization and heating driven by stars and more energetic sources}",
      journal = {\mnras},
         year = 2020,
        month = nov,
       volume = {498},
       number = {4},
        pages = {6083-6099},
          doi = {10.1093/mnras/staa2774},
archivePrefix = {arXiv},
       eprint = {2009.06631},
 primaryClass = {astro-ph.CO},
       adsurl = {https://ui.adsabs.harvard.edu/abs/2020MNRAS.498.6083E}
}

@ARTICLE{2013MNRAS.431..621M,
       author = {{Mesinger}, Andrei and {Ferrara}, Andrea and {Spiegel}, David S.},
        title = "{Signatures of X-rays in the early Universe}",
      journal = {\mnras},
         year = 2013,
        month = may,
       volume = {431},
       number = {1},
        pages = {621-637},
          doi = {10.1093/mnras/stt198},
archivePrefix = {arXiv},
       eprint = {1210.7319},
 primaryClass = {astro-ph.CO},
       adsurl = {https://ui.adsabs.harvard.edu/abs/2013MNRAS.431..621M}
}

@ARTICLE{2013ApJ...764...41F,
       author = {{Fragos}, T. and {Lehmer}, B. and {Tremmel}, M. and {Tzanavaris}, P. and {Basu-Zych}, A. and {Belczynski}, K. and {Hornschemeier}, A. and {Jenkins}, L. and {Kalogera}, V. and {Ptak}, A. and {Zezas}, A.},
        title = "{X-Ray Binary Evolution Across Cosmic Time}",
      journal = {\apj},
         year = 2013,
        month = feb,
       volume = {764},
       number = {1},
          eid = {41},
        pages = {41},
          doi = {10.1088/0004-637X/764/1/41},
archivePrefix = {arXiv},
       eprint = {1206.2395},
 primaryClass = {astro-ph.HE},
       adsurl = {https://ui.adsabs.harvard.edu/abs/2013ApJ...764...41F}
}

@ARTICLE{2013RPPh...76k2901B,
       author = {{Bromm}, Volker},
        title = "{Formation of the first stars}",
      journal = {Reports on Progress in Physics},
         year = 2013,
        month = nov,
       volume = {76},
       number = {11},
          eid = {112901},
        pages = {112901},
          doi = {10.1088/0034-4885/76/11/112901},
archivePrefix = {arXiv},
       eprint = {1305.5178},
 primaryClass = {astro-ph.CO},
       adsurl = {https://ui.adsabs.harvard.edu/abs/2013RPPh...76k2901B}
}

@ARTICLE{2001PhR...349..125B,
       author = {{Barkana}, R. and {Loeb}, A.},
        title = "{In the beginning: the first sources of light and the reionization of the universe}",
      journal = {\physrep},
         year = 2001,
        month = jul,
       volume = {349},
       number = {2},
        pages = {125-238},
          doi = {10.1016/S0370-1573(01)00019-9},
archivePrefix = {arXiv},
       eprint = {astro-ph/0010468},
 primaryClass = {astro-ph},
       adsurl = {https://ui.adsabs.harvard.edu/abs/2001PhR...349..125B}
}

@ARTICLE{2022MNRAS.514...55B,
       author = {{Bosman}, Sarah E.~I. and {Davies}, Frederick B. and {Becker}, George D. and {Keating}, Laura C. and {Davies}, Rebecca L. and {Zhu}, Yongda and {Eilers}, Anna-Christina and {D'Odorico}, Valentina and {Bian}, Fuyan and {Bischetti}, Manuela and {Cristiani}, Stefano V. and {Fan}, Xiaohui and {Farina}, Emanuele P. and {Haehnelt}, Martin G. and {Hennawi}, Joseph F. and {Kulkarni}, Girish and {Mesinger}, Andrei and {Meyer}, Romain A. and {Onoue}, Masafusa and {Pallottini}, Andrea and {Qin}, Yuxiang and {Ryan-Weber}, Emma and {Schindler}, Jan-Torge and {Walter}, Fabian and {Wang}, Feige and {Yang}, Jinyi},
        title = "{Hydrogen reionization ends by z = 5.3: Lyman-{\ensuremath{\alpha}} optical depth measured by the XQR-30 sample}",
      journal = {\mnras},
         year = 2022,
        month = jul,
       volume = {514},
       number = {1},
        pages = {55-76},
          doi = {10.1093/mnras/stac1046},
archivePrefix = {arXiv},
       eprint = {2108.03699},
 primaryClass = {astro-ph.CO},
       adsurl = {https://ui.adsabs.harvard.edu/abs/2022MNRAS.514...55B}
}

@ARTICLE{2015ApJ...809...89T,
       author = {{Trainor}, Ryan F. and {Steidel}, Charles C. and {Strom}, Allison L. and {Rudie}, Gwen C.},
        title = "{The Spectroscopic Properties of Ly{\ensuremath{\alpha}}-Emitters at z {\ensuremath{\sim}}2.7: Escaping Gas and Photons from Faint Galaxies}",
      journal = {\apj},
         year = 2015,
        month = aug,
       volume = {809},
       number = {1},
          eid = {89},
        pages = {89},
          doi = {10.1088/0004-637X/809/1/89},
archivePrefix = {arXiv},
       eprint = {1506.08205},
 primaryClass = {astro-ph.GA},
       adsurl = {https://ui.adsabs.harvard.edu/abs/2015ApJ...809...89T}
}

@ARTICLE{2014ApJ...795...33E,
       author = {{Erb}, Dawn K. and {Steidel}, Charles C. and {Trainor}, Ryan F. and {Bogosavljevi{\'c}}, Milan and {Shapley}, Alice E. and {Nestor}, Daniel B. and {Kulas}, Kristin R. and {Law}, David R. and {Strom}, Allison L. and {Rudie}, Gwen C. and {Reddy}, Naveen A. and {Pettini}, Max and {Konidaris}, Nicholas P. and {Mace}, Gregory and {Matthews}, Keith and {McLean}, Ian S.},
        title = "{The Ly{\ensuremath{\alpha}} Properties of Faint Galaxies at z \raisebox{-0.5ex}\textasciitilde 2-3 with Systemic Redshifts and Velocity Dispersions from Keck-MOSFIRE}",
      journal = {\apj},
         year = 2014,
        month = nov,
       volume = {795},
       number = {1},
          eid = {33},
        pages = {33},
          doi = {10.1088/0004-637X/795/1/33},
archivePrefix = {arXiv},
       eprint = {1408.3638},
 primaryClass = {astro-ph.GA},
       adsurl = {https://ui.adsabs.harvard.edu/abs/2014ApJ...795...33E}
}

@ARTICLE{2018A&A...619A.136M,
       author = {{Matthee}, Jorryt and {Sobral}, David and {Gronke}, Max and {Paulino-Afonso}, Ana and {Stefanon}, Mauro and {R{\"o}ttgering}, Huub},
        title = "{Confirmation of double peaked Ly{\ensuremath{\alpha}} emission at z = 6.593. Witnessing a galaxy directly contributing to the reionisation of the Universe}",
      journal = {\aap},
         year = 2018,
        month = nov,
       volume = {619},
          eid = {A136},
        pages = {A136},
          doi = {10.1051/0004-6361/201833528},
archivePrefix = {arXiv},
       eprint = {1805.11621},
 primaryClass = {astro-ph.GA},
       adsurl = {https://ui.adsabs.harvard.edu/abs/2018A&A...619A.136M}
}

@ARTICLE{2015ApJ...802L..19R,
       author = {{Robertson}, Brant E. and {Ellis}, Richard S. and
         {Furlanetto}, Steven R. and {Dunlop}, James S.},
        title = "{Cosmic Reionization and Early Star-forming Galaxies: A Joint Analysis of New Constraints from Planck and the Hubble Space Telescope}",
      journal = {\apjl},
         year = 2015,
        month = apr,
       volume = {802},
       number = {2},
          eid = {L19},
        pages = {L19},
          doi = {10.1088/2041-8205/802/2/L19},
archivePrefix = {arXiv},
       eprint = {1502.02024},
 primaryClass = {astro-ph.CO},
       adsurl = {https://ui.adsabs.harvard.edu/abs/2015ApJ...802L..19R}
}

@ARTICLE{2019ApJ...879...36F,
       author = {{Finkelstein}, Steven L. and {D'Aloisio}, Anson and
         {Paardekooper}, Jan-Pieter and {Ryan}, Russell, Jr. and
         {Behroozi}, Peter and {Finlator}, Kristian and {Livermore}, Rachael and
         {Upton Sanderbeck}, Phoebe R. and {Dalla Vecchia}, Claudio and
         {Khochfar}, Sadegh},
        title = "{Conditions for Reionizing the Universe with a Low Galaxy Ionizing Photon Escape Fraction}",
      journal = {\apj},
         year = 2019,
        month = jul,
       volume = {879},
       number = {1},
          eid = {36},
        pages = {36},
          doi = {10.3847/1538-4357/ab1ea8},
archivePrefix = {arXiv},
       eprint = {1902.02792},
 primaryClass = {astro-ph.CO},
       adsurl = {https://ui.adsabs.harvard.edu/abs/2019ApJ...879...36F}
}

@ARTICLE{2016MNRAS.463.1462O,
       author = {{Ocvirk}, Pierre and {Gillet}, Nicolas and {Shapiro}, Paul R. and
         {Aubert}, Dominique and {Iliev}, Ilian T. and {Teyssier}, Romain and
         {Yepes}, Gustavo and {Choi}, Jun-Hwan and {Sullivan}, David and
         {Knebe}, Alexander and {Gottl{\"o}ber}, Stefan and {D'Aloisio}, Anson and
         {Park}, Hyunbae and {Hoffman}, Yehuda and {Stranex}, Timothy},
        title = "{Cosmic Dawn (CoDa): the First Radiation-Hydrodynamics Simulation of Reionization and Galaxy Formation in the Local Universe}",
      journal = {\mnras},
         year = 2016,
        month = dec,
       volume = {463},
       number = {2},
        pages = {1462-1485},
          doi = {10.1093/mnras/stw2036},
archivePrefix = {arXiv},
       eprint = {1511.00011},
 primaryClass = {astro-ph.GA},
       adsurl = {https://ui.adsabs.harvard.edu/abs/2016MNRAS.463.1462O}
}

@ARTICLE{2020MNRAS.496.4087O,
       author = {{Ocvirk}, Pierre and {Aubert}, Dominique and {Sorce}, Jenny G. and
         {Shapiro}, Paul R. and {Deparis}, Nicolas and {Dawoodbhoy}, Taha and
         {Lewis}, Joseph and {Teyssier}, Romain and {Yepes}, Gustavo and
         {Gottl{\"o}ber}, Stefan and {Ahn}, Kyungjin and {Iliev}, Ilian T. and
         {Hoffman}, Yehuda},
        title = "{Cosmic Dawn II (CoDa II): a new radiation-hydrodynamics simulation of the self-consistent coupling of galaxy formation and reionization}",
      journal = {\mnras},
         year = 2020,
        month = may,
       volume = {496},
       number = {4},
        pages = {4087-4107},
          doi = {10.1093/mnras/staa1266},
       adsurl = {https://ui.adsabs.harvard.edu/abs/2020MNRAS.496.4087O}
}

@ARTICLE{2008MNRAS.391...63I,
       author = {{Iliev}, Ilian T. and {Shapiro}, Paul R. and {McDonald}, Patrick and {Mellema}, Garrelt and {Pen}, Ue-Li},
        title = "{The effect of the intergalactic environment on the observability of Ly{\ensuremath{\alpha}} emitters during reionization}",
      journal = {\mnras},
         year = 2008,
        month = nov,
       volume = {391},
       number = {1},
        pages = {63-83},
          doi = {10.1111/j.1365-2966.2008.13879.x},
archivePrefix = {arXiv},
       eprint = {0711.2944},
 primaryClass = {astro-ph},
       adsurl = {https://ui.adsabs.harvard.edu/abs/2008MNRAS.391...63I}
}

@ARTICLE{2022LRCA....8....3G,
       author = {{Gnedin}, Nickolay Y. and {Madau}, Piero},
        title = "{Modeling cosmic reionization}",
      journal = {Living Reviews in Computational Astrophysics},
         year = 2022,
        month = dec,
       volume = {8},
       number = {1},
          eid = {3},
        pages = {3},
          doi = {10.1007/s41115-022-00015-5},
archivePrefix = {arXiv},
       eprint = {2208.02260},
 primaryClass = {astro-ph.CO},
       adsurl = {https://ui.adsabs.harvard.edu/abs/2022LRCA....8....3G}
}

@ARTICLE{2007ApJS..170..335P,
       author = {{Page}, L. and {Hinshaw}, G. and {Komatsu}, E. and {Nolta}, M.~R. and {Spergel}, D.~N. and {Bennett}, C.~L. and {Barnes}, C. and {Bean}, R. and {Dor{\'e}}, O. and {Dunkley}, J. and {Halpern}, M. and {Hill}, R.~S. and {Jarosik}, N. and {Kogut}, A. and {Limon}, M. and {Meyer}, S.~S. and {Odegard}, N. and {Peiris}, H.~V. and {Tucker}, G.~S. and {Verde}, L. and {Weiland}, J.~L. and {Wollack}, E. and {Wright}, E.~L.},
        title = "{Three-Year Wilkinson Microwave Anisotropy Probe (WMAP) Observations: Polarization Analysis}",
      journal = {\apjs},
         year = 2007,
        month = jun,
       volume = {170},
       number = {2},
        pages = {335-376},
          doi = {10.1086/513699},
archivePrefix = {arXiv},
       eprint = {astro-ph/0603450},
 primaryClass = {astro-ph},
       adsurl = {https://ui.adsabs.harvard.edu/abs/2007ApJS..170..335P}
}

@ARTICLE{2007ApJ...671....1T,
       author = {{Trac}, Hy and {Cen}, Renyue},
        title = "{Radiative Transfer Simulations of Cosmic Reionization. I. Methodology and Initial Results}",
      journal = {\apj},
         year = 2007,
        month = dec,
       volume = {671},
       number = {1},
        pages = {1-13},
          doi = {10.1086/522566},
archivePrefix = {arXiv},
       eprint = {astro-ph/0612406},
 primaryClass = {astro-ph},
       adsurl = {https://ui.adsabs.harvard.edu/abs/2007ApJ...671....1T}
}

@ARTICLE{2007MNRAS.377.1043M,
       author = {{McQuinn}, Matthew and {Lidz}, Adam and {Zahn}, Oliver and {Dutta}, Suvendra and {Hernquist}, Lars and {Zaldarriaga}, Matias},
        title = "{The morphology of HII regions during reionization}",
      journal = {\mnras},
         year = 2007,
        month = may,
       volume = {377},
       number = {3},
        pages = {1043-1063},
          doi = {10.1111/j.1365-2966.2007.11489.x},
archivePrefix = {arXiv},
       eprint = {astro-ph/0610094},
 primaryClass = {astro-ph},
       adsurl = {https://ui.adsabs.harvard.edu/abs/2007MNRAS.377.1043M}
}

@ARTICLE{2006MNRAS.369.1625I,
       author = {{Iliev}, I.~T. and {Mellema}, G. and {Pen}, U.-L. and {Merz}, H. and {Shapiro}, P.~R. and {Alvarez}, M.~A.},
        title = "{Simulating cosmic reionization at large scales - I. The geometry of reionization}",
      journal = {\mnras},
         year = 2006,
        month = jul,
       volume = {369},
       number = {4},
        pages = {1625-1638},
          doi = {10.1111/j.1365-2966.2006.10502.x},
archivePrefix = {arXiv},
       eprint = {astro-ph/0512187},
 primaryClass = {astro-ph},
       adsurl = {https://ui.adsabs.harvard.edu/abs/2006MNRAS.369.1625I}
}

@ARTICLE{2016ApJ...825L...7H,
       author = {{Hu}, E.~M. and {Cowie}, L.~L. and {Songaila}, A. and {Barger}, A.~J. and {Rosenwasser}, B. and {Wold}, I.~G.~B.},
        title = "{An Ultraluminous Ly{\ensuremath{\alpha}} Emitter with a Blue Wing at z = 6.6}",
      journal = {\apjl},
         year = 2016,
        month = jul,
       volume = {825},
       number = {1},
          eid = {L7},
        pages = {L7},
          doi = {10.3847/2041-8205/825/1/L7},
archivePrefix = {arXiv},
       eprint = {1606.03526},
 primaryClass = {astro-ph.GA},
       adsurl = {https://ui.adsabs.harvard.edu/abs/2016ApJ...825L...7H}
}

@ARTICLE{2018ApJ...855...96H,
       author = {{Henry}, Alaina and {Berg}, Danielle A. and {Scarlata}, Claudia and {Verhamme}, Anne and {Erb}, Dawn},
        title = "{A Close Relationship between Ly{\ensuremath{\alpha}} and Mg II in Green Pea Galaxies}",
      journal = {\apj},
         year = 2018,
        month = mar,
       volume = {855},
       number = {2},
          eid = {96},
        pages = {96},
          doi = {10.3847/1538-4357/aab099},
archivePrefix = {arXiv},
       eprint = {1803.10243},
 primaryClass = {astro-ph.GA},
       adsurl = {https://ui.adsabs.harvard.edu/abs/2018ApJ...855...96H}
}

@ARTICLE{2022MNRAS.517.5642E,
       author = {{Endsley}, Ryan and {Stark}, Daniel P. and {Bouwens}, Rychard J. and {Schouws}, Sander and {Smit}, Renske and {Stefanon}, Mauro and {Inami}, Hanae and {Bowler}, Rebecca A.~A. and {Oesch}, Pascal and {Gonzalez}, Valentino and {Aravena}, Manuel and {da Cunha}, Elisabete and {Dayal}, Pratika and {Ferrara}, Andrea and {Graziani}, Luca and {Nanayakkara}, Themiya and {Pallottini}, Andrea and {Schneider}, Raffaella and {Sommovigo}, Laura and {Topping}, Michael and {van der Werf}, Paul and {Hutter}, Anne},
        title = "{The REBELS ALMA Survey: efficient Ly {\ensuremath{\alpha}} transmission of UV-bright z ≃ 7 galaxies from large velocity offsets and broad line widths}",
      journal = {\mnras},
         year = 2022,
        month = dec,
       volume = {517},
       number = {4},
        pages = {5642-5659},
          doi = {10.1093/mnras/stac3064},
archivePrefix = {arXiv},
       eprint = {2202.01219},
 primaryClass = {astro-ph.GA},
       adsurl = {https://ui.adsabs.harvard.edu/abs/2022MNRAS.517.5642E}
}

@ARTICLE{2026MNRAS.545f1862D,
       author = {{Davies}, Frederick B. and {Bosman}, Sarah E.~I. and {D'Odorico}, Valentina and {Campo}, Sofia and {Mesinger}, Andrei and {Qin}, Yuxiang and {Becker}, George D. and {Ba{\~n}ados}, Eduardo and {Chen}, Huanqing and {Cristiani}, Stefano and {Fan}, Xiaohui and {Gallerani}, Simona and {Haehnelt}, Martin G. and {Keating}, Laura C. and {Lai}, Samuel and {Ryan-Weber}, Emma and {Wang}, Feige and {Yang}, Jinyi and {Zhu}, Yongda},
        title = "{Updated dark pixel fraction constraints on reionization's end from the Lyman-series forests of XQR{\ensuremath{-}}30}",
      journal = {\mnras},
         year = 2026,
        month = jan,
       volume = {545},
       number = {2},
          eid = {staf1862},
        pages = {staf1862},
          doi = {10.1093/mnras/staf1862},
archivePrefix = {arXiv},
       eprint = {2510.25829},
 primaryClass = {astro-ph.CO},
       adsurl = {https://ui.adsabs.harvard.edu/abs/2026MNRAS.545f1862D}
}

@ARTICLE{2025MNRAS.544.4246P,
       author = {{Park}, Hyunbae and {Song}, Hyunmi and {Byrohl}, Chris and {Smith}, Aaron and {Yajima}, Hidenobu and {Luki{\'c}}, Zarija},
        title = "{Analytical model for scattered Ly{\ensuremath{\alpha}} emission in the post-reionization intergalactic medium}",
      journal = {\mnras},
         year = 2025,
        month = dec,
       volume = {544},
       number = {4},
        pages = {4246-4255},
          doi = {10.1093/mnras/staf1971},
archivePrefix = {arXiv},
       eprint = {2509.11268},
 primaryClass = {astro-ph.CO},
       adsurl = {https://ui.adsabs.harvard.edu/abs/2025MNRAS.544.4246P}
}

@ARTICLE{2021ApJ...922..263P,
       author = {{Park}, Hyunbae and {Jung}, Intae and {Song}, Hyunmi and {Ocvirk}, Pierre and {Shapiro}, Paul R. and {Dawoodbhoy}, Taha and {Iliev}, Ilian T. and {Ahn}, Kyungjin and {Bianco}, Michele and {Kim}, Hyo Jeong},
        title = "{Crucial Factors for Ly{\ensuremath{\alpha}} Transmission in the Reionizing Intergalactic Medium: Infall Motion, H II Bubble Size, and Self-shielded Systems}",
      journal = {\apj},
         year = 2021,
        month = dec,
       volume = {922},
       number = {2},
          eid = {263},
        pages = {263},
          doi = {10.3847/1538-4357/ac2f4b},
archivePrefix = {arXiv},
       eprint = {2105.10770},
 primaryClass = {astro-ph.GA},
       adsurl = {https://ui.adsabs.harvard.edu/abs/2021ApJ...922..263P}
}

@ARTICLE{2022MNRAS.512.3243S,
       author = {{Smith}, A. and {Kannan}, R. and {Garaldi}, E. and {Vogelsberger}, M. and {Pakmor}, R. and {Springel}, V. and {Hernquist}, L.},
        title = "{The THESAN project: Lyman-{\ensuremath{\alpha}} emission and transmission during the Epoch of Reionization}",
      journal = {\mnras},
         year = 2022,
        month = may,
       volume = {512},
       number = {3},
        pages = {3243-3265},
          doi = {10.1093/mnras/stac713},
archivePrefix = {arXiv},
       eprint = {2110.02966},
 primaryClass = {astro-ph.CO},
       adsurl = {https://ui.adsabs.harvard.edu/abs/2022MNRAS.512.3243S}
}

@ARTICLE{2011MNRAS.412.2543C,
       author = {{Calverley}, Alexander P. and {Becker}, George D. and {Haehnelt}, Martin G. and {Bolton}, James S.},
        title = "{Measurements of the ultraviolet background at 4.6 < z < 6.4 using the quasar proximity effect}",
      journal = {\mnras},
         year = 2011,
        month = apr,
       volume = {412},
       number = {4},
        pages = {2543-2562},
          doi = {10.1111/j.1365-2966.2010.18072.x},
archivePrefix = {arXiv},
       eprint = {1011.5850},
 primaryClass = {astro-ph.CO},
       adsurl = {https://ui.adsabs.harvard.edu/abs/2011MNRAS.412.2543C}
}

@ARTICLE{2021MNRAS.507.6108O,
       author = {{Ocvirk}, Pierre and {Lewis}, Joseph S.~W. and {Gillet}, Nicolas and {Chardin}, Jonathan and {Aubert}, Dominique and {Deparis}, Nicolas and {Th{\'e}lie}, {\'E}milie},
        title = "{Lyman-alpha opacities at z = 4-6 require low mass, radiatively-suppressed galaxies to drive cosmic reionization}",
      journal = {\mnras},
         year = 2021,
        month = nov,
       volume = {507},
       number = {4},
        pages = {6108-6117},
          doi = {10.1093/mnras/stab2502},
archivePrefix = {arXiv},
       eprint = {2105.01663},
 primaryClass = {astro-ph.CO},
       adsurl = {https://ui.adsabs.harvard.edu/abs/2021MNRAS.507.6108O}
}

@ARTICLE{2022MNRAS.516.3389L,
       author = {{Lewis}, Joseph S.~W. and {Ocvirk}, Pierre and {Sorce}, Jenny G. and {Dubois}, Yohan and {Aubert}, Dominique and {Conaboy}, Luke and {Shapiro}, Paul R. and {Dawoodbhoy}, Taha and {Teyssier}, Romain and {Yepes}, Gustavo and {Gottl{\"o}ber}, Stefan and {Rasera}, Yann and {Ahn}, Kyungjin and {Iliev}, Ilian T. and {Park}, Hyunbae and {Th{\'e}lie}, {\'E}milie},
        title = "{The short ionizing photon mean free path at z = 6 in Cosmic Dawn III, a new fully coupled radiation-hydrodynamical simulation of the Epoch of Reionization}",
      journal = {\mnras},
         year = 2022,
        month = nov,
       volume = {516},
       number = {3},
        pages = {3389-3397},
          doi = {10.1093/mnras/stac2383},
archivePrefix = {arXiv},
       eprint = {2202.05869},
 primaryClass = {astro-ph.CO},
       adsurl = {https://ui.adsabs.harvard.edu/abs/2022MNRAS.516.3389L}
}

\end{document}